\documentclass[11pt,onecolumn]{IEEEtran}
\usepackage[tbtags]{amsmath}
\usepackage{amsfonts}
\usepackage{amssymb}
\usepackage{subfigure}
\usepackage{cite}
\usepackage{calc}
\usepackage{color}
\usepackage{epsfig}

\newtheorem{defn}{Definition}
\newtheorem{thm}{Theorem}[section]
\newtheorem{cor}[thm]{Corollary}
\newtheorem{prop}{Proposition}
\newtheorem{lem}[thm]{Lemma}
\newtheorem{conj}[thm]{Conjecture}
\newtheorem{constr}[thm]{Construction}
\newtheorem{note}{Remark}
\newcommand{\bit}{\begin{itemize}}
\newcommand{\eit}{\end{itemize}}
\newcommand{\bcor}{\begin{cor}}
\newcommand{\ecor}{\end{cor}}
\newcommand{\beq}{\begin{equation}}
\newcommand{\eeq}{\end{equation}}
\newcommand{\beqn}{\begin{equation*}}
\newcommand{\eeqn}{\end{equation*}}
\newcommand{\beqa}{\begin{eqnarray}}
\newcommand{\eeqa}{\end{eqnarray}}
\newcommand{\beqan}{\begin{eqnarray*}}
\newcommand{\eeqan}{\end{eqnarray*}}
\newcommand{\ben}{\begin{enumerate}}
\newcommand{\een}{\end{enumerate}}
\newcommand{\bdefn}{\begin{defn}}
\newcommand{\edefn}{\end{defn}}
\newcommand{\bnote}{\begin{note}}
\newcommand{\enote}{\end{note}}
\newcommand{\bprop}{\begin{prop}}
\newcommand{\eprop}{\end{prop}}
\newcommand{\blem}{\begin{lem}}
\newcommand{\elem}{\end{lem}}
\newcommand{\bthm}{\begin{thm}}
\newcommand{\ethm}{\end{thm}}
\newcommand{\bconj}{\begin{conj}}
\newcommand{\econj}{\end{conj}}
\newcommand{\bconstr}{\begin{constr}}
\newcommand{\econstr}{\end{constr}}
\newcommand{\bpf}{\begin{proof}}
\newcommand{\epf}{\end{proof}}

\begin{document}

\title{Multi-hop Cooperative Wireless Networks: Diversity Multiplexing Tradeoff and Optimal Code Design}
%\author{K. Sreeram, S. Birenjith, P. Vijay Kumar}

\author{\authorblockN{K. Sreeram, S. Birenjith and P. Vijay Kumar $\dagger$} \\
\authorblockA{Department of ECE\\
Indian Institute of Science\\
Bangalore, India \\
Email: \{sreeramkannan, biren, vijay\}@ece.iisc.ernet.in} \\
\thanks{$\dagger$ P.Vijay Kumar is on leave of absence from the
University of Southern California, Los Angeles, USA. }
\thanks{This work was supported in part by NSF-ITR Grant CCR-0326628, in part by the DRDO-IISc Program on Advanced
Mathematical Engineering and in part by Motorola's University
Research Partnership Program.}}

\date{}
\maketitle

\begin{abstract}

We consider single-source single-sink (ss-ss) multi-hop relay
networks, with slow-fading links and single-antenna half-duplex
relay nodes. While two-hop cooperative relay networks have been
studied in great detail in terms of the diversity-multiplexing
tradeoff (DMT), few results are available for more general
networks. In this paper, we identify two families of networks that
are multi-hop generalizations of the two-hop network:
$K$-Parallel-Path (KPP) networks and layered networks.

KPP networks, can be viewed as the union of $K$ node-disjoint
parallel relaying paths, each of length greater than one. KPP
networks are then generalized to KPP(I) networks, which permit
interference between paths and to KPP(D) networks, which possess a
direct link from source to sink. We characterize the DMT of these
families of networks completely for $K > 3$. Layered networks are
networks comprising of layers of relays with edges existing only
between adjacent layers, with more than one relay in each layer.
We prove that a linear DMT between the maximum diversity
$d_\text{max}$ and the maximum multiplexing gain of $1$ is
achievable for single-antenna fully-connected layered networks.
This is shown to be equal to the optimal DMT if the number of
relaying layers is less than $4$. For multiple-antenna KPP and
layered networks, we provide an achievable DMT, which is
significantly better than known lower bounds for half duplex
networks.

For arbitrary multi-terminal wireless networks with multiple
source-sink pairs, the maximum achievable diversity is shown to be
equal to the min-cut between the corresponding source and the
sink, irrespective of whether the network has half-duplex or
full-duplex relays. For arbitrary ss-ss single-antenna directed
acyclic networks with full-duplex relays, we prove that a linear
tradeoff between maximum diversity and maximum multiplexing gain
is achievable.

Along the way, we derive the optimal DMT of a generalized parallel
channel and derive lower bounds for the DMT of triangular channel
matrices, which are useful in DMT computation of various
protocols. We also give alternative and often simpler proofs of
several existing results and show that codes achieving full
diversity on a MIMO Rayleigh fading channel achieve full diversity
on arbitrary fading channels. All protocols in this paper are
explicit and use only amplify-and-forward (AF) relaying. We also
construct codes with short block-lengths based on cyclic division
algebras that achieve the optimal DMT for all the proposed
schemes.

Two key implications of the results in the paper are that the
half-duplex constraint does not entail any rate loss for a large
class of cooperative networks and that simple AF protocols are
often sufficient to attain the optimal DMT.

\end{abstract}

\newpage
\section{Introduction\label{sec:introduction}}

\subsection{Prior Work\label{sec:prior_work}}

\begin{figure}[h!]
\centering
\includegraphics[width=60mm]{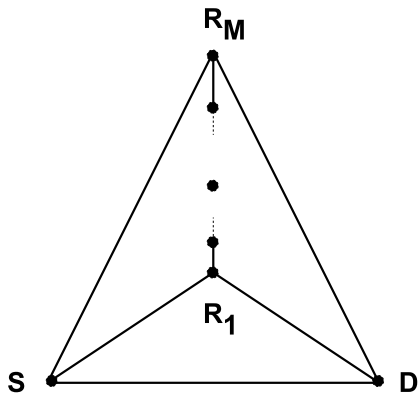}
\caption{Two Hop Cooperative Relay Network
\label{fig:classical_relay}}
\end{figure}

The concept of user cooperative diversity was introduced in
\cite{SenErkAaz1}. Cooperative diversity protocols were first
discussed in \cite{LanWor} for the two-hop relay network
(Fig.\ref{fig:classical_relay}) where the authors develop and
analyze the Orthogonal Amplify and Forward (OAF) protocol and the
Selection Decode and Forward (SDF) protocol for the case of a
single relay network.

Zheng and Tse \cite{ZheTse} proposed the Diversity-Multiplexing
gain Tradeoff (DMT) as a tool to evaluate point-to-point
multiple-antenna schemes in the context of slow fading channels.
The DMT was used as a tool to compare various protocols for half
duplex two-hop cooperative networks in \cite{LanTseWor,AzaGamSch}.
As noted in \cite{YukErk}, the DMT is a valuable tool in the study
of cooperative relay networks, because it is simple enough to be
analytically tractable and powerful enough to compare different
protocols.

In \cite{LanTseWor}, the SDF protocol is analyzed for an arbitrary
number of relays, where the authors give upper and lower bounds on
the DMT of the protocol. In these protocols, the relays and the
source node participate for equal time instants and the maximum
multiplexing gain $r$ that could be achieved was $0.5$.

For any network, an upper bound on the achievable DMT has been
given by the cut-set bound \cite{YukErk},\cite{CovTho}. A
fundamental question in this area is whether the two-hop
cooperative wireless system in Fig.\ref{fig:classical_relay} can
mimic a Multiple Input Single Output (MISO) system with $N+1$
transmit antennas and $1$ receive antenna and achieve the DMT
corresponding to the MISO system. This question still remains
open, see \cite{EliVinAnaKum}, \cite{PraVar} for a detailed
comparison of existing achievable regions.

 In \cite{AzaGamSch}, Azarian {\it et al.} analyze the class of
Non Orthogonal amplify and Forward (NAF) protocols, introduced
earlier by Nabar {\it et al.} in \cite{NabBolKne}. In
\cite{AzaGamSch}, the authors establish the improved DMT of the
NAF protocol in comparison to the class of OAF protocols
considered in \cite{LanTseWor}. However it has been shown in
\cite{EliVinAnaKum} that the DMT of the NAF protocol can be
obtained for the OAF protocols as well using appropriate unequal
slot lengths for source and relay transmissions.

The authors of \cite{AzaGamSch} also introduce the Dynamic Decode
and Forward (DDF) protocol wherein the time for which the relays
listen to the source depends on the source-relay channel gain.
They show that for the single relay case, the DMT of the DDF
protocol achieves the transmit diversity bound for $r \leq 0.5$,
beyond which the DMT falls below the transmit diversity bound.

Jing and Hassibi \cite{JinHas} consider cooperative communication
protocols where the relay nodes apply a linear transformation to
the received signal. The network model that they consider is the
same as the one shown in Fig.~\ref{fig:classical_relay} except
that there is no direct link between source and sink in their
model. The authors consider the case when both the source and the
relays transmit for an equal number of channel uses and the linear
transformation applied by the relays are restricted to the class
of unitary matrices. Rao and Hassibi \cite{RaoHas} consider
two-hop half-duplex multi-antenna cooperative networks without
direct link and analyze the DMT performance.

Yang and Belfiore consider a class of protocols called Slotted
Amplify And Forward (SAF) protocols in \cite{YanBelSaf}, and show
that these improve upon the performance of the NAF protocol
\cite{AzaGamSch} for the case of two relays. The authors also
provide an upper bound on the DMT of the SAF protocol with any
number of slots, and show that this upper bound tends towards the
transmit diversity bound as the number of slots increases. Under
the assumption of relay isolation and relay ordering, the naive
SAF scheme proposed in \cite{YanBelSaf} is shown to achieve the
SAF protocol upper bound.

Yuksel and Erkip in \cite{YukErk} have considered the DMT of the
DF and compress-and-forward (CF) protocols. They show that the CF
protocol achieves the transmit diversity bound for the case of a
single relay. We note however, that in the CF protocol, the relays
are assumed to know {\em all} the fading coefficients in the
system. The authors also translate cut-set upper bounds in
\cite{CovTho} for mutual information into the DMT framework for a
general multi-terminal network.

Yang and Belfiore in \cite{YanBelNew} consider AF protocols on a
family of MIMO multihop networks (termed as multi-antenna layered
networks in the current paper). They derive the optimal DMT for
the Rayleigh-product channel which they prove is equal to the DMT
of the AF protocol applied to this channel. They also propose AF
protocols to achieve the optimal diversity of these multi-antenna
layered networks.

Oggier and Hassibi \cite{OggHas} have proposed distributed space
time codes for multi-antenna layered networks that achieve a
diversity equal to the minimum number of relay nodes among the
hops. Recently, Vaze and Heath \cite{VazHea} have constructed
distributed space time codes based on orthogonal designs that
achieve the optimal diversity of the multi-antenna layered
network.

Borade, Zheng and Gallager in \cite{BorZheGal} consider AF schemes
on a class of multi-hop layered networks where each layer has the
same number of relays (termed as Regular networks in the current
paper). They show that AF strategies are optimal in terms of
multiplexing gain. They also compute lower bounds on the DMT of
the product Rayleigh channel.

From a capacity perspective as well, there have been some
investigations into single-source single-sink wireless networks.
Recently, Avestimehr, Diggavi and Tse \cite{AveDigTse} have
evaluated the capacity of deterministic wireless networks with
broadcast and interference constraints. They have also shown that
schemes from these deterministic networks can be lifted to
gaussian networks, to give achievable regions that are within a
constant away from outer-bounds. However, it must be noted that
they consider only full-duplex networks. The degrees of freedom of
arbitrary full-duplex ss-ss and multicast wireless networks is
established in \cite{SreBirKum3} using a connection with
deterministic wireless networks.

From the point of code design for multiple antenna systems,
Space-Time codes from Cyclic Division Algebra (CDA) was introduced
in \cite{SetRajSas}. Certain codes constructed from CDAs were
proved to be DMT optimal (in fact approximately universal - see
\cite{TavVis}) for the general MIMO channel in
\cite{EliRajPawKumLu}. These codes were tailored to suit the
structure of various static protocols for two-hop cooperation and
proved to be DMT optimal in \cite{EliVinAnaKum}. For the Dynamic
Decode and Forward protocol, DMT optimal codes were constructed
for arbitrary number of relays with multiple antennas in
\cite{EliVij}. Recently, in \cite{RajCaire}, codes for the single
relay single antenna DDF channel were constructed, which are not
only DMT optimal, but also have probability of error close to the
outage probability. In this paper, we present a DMT optimal code
design for all proposed protocols based on the approximately
universal codes in \cite{EliRajPawKumLu}.

Cooperative networks with asynchronous transmissions have also
been studied in the literature
\cite{Wei},\cite{LiXia},\cite{RajRajDist}. However, we consider
networks in which relays are synchronized. Codes for two-hop
cooperative networks having low decoding complexity and full
diversity are studied in \cite{JinJaf}, \cite{RajRajDist} and
\cite{YiKim}. While decoding complexity is not the primary focus
of the present paper, we do provide a
successive-interference-cancellation technique to reduce the code
length and therefore the complexity.

\subsection{Classification of Networks \label{sec:network_taxonomy}}

In this section, we define the classes of networks under
consideration here. Unless otherwise stated, all networks
considered possess a single source and a single sink and we will
apply the abbreviation ss-ss to these networks.

A cooperative wireless network can be built out of a collection of
spatially distributed nodes in many ways. For instance, we can
identify paths connecting source to the sink through a series of
nodes in such a manner that any two adjacent nodes fall in the
Rayleigh zone\cite{YukErk}. This process can be continued barring
those nodes which are already chosen. Such a construction will
result in a set of paths from the source to the sink. In the
simplest model, we can further impose the constraint that these
paths do not interfere each other, see
Fig.\ref{fig:classical_relay} thus motivating the study of a class
of multi-hop network which we shall refer to as the set of
K-Parallel Path (KPP) networks.

\begin{figure}[h!]
\centering
\includegraphics[height=60mm]{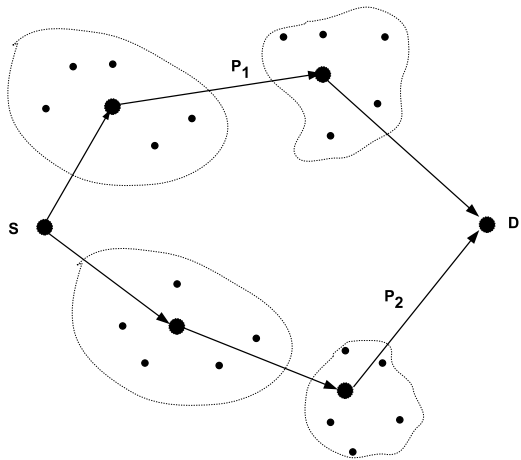}
\caption{Motivation for the KPP networks
\label{fig:kpp_network_motivation}}
\end{figure}

Alternatively, a layers of relays can be identified from a
collection of nodes between the source and the sink. This will
result in a layered network model, which is described in
\cite{BorZheGal}.

\subsubsection{Representation by a graph} Any wireless network can be associated with
a directed graph, with vertices representing nodes in the network
and edges representing connectivity between nodes. If an edge is
bidirectional, we will represent it by two edges one pointing in
either direction. An edge in a directed graph is said to be
\emph{live} at a particular time instant if the node at the head
of the edge is transmitting at that instant. An edge in a directed
graph is said to be \emph{active} at a particular time instant if
the node at the head of the edge is transmitting and the tail of
the edge is receiving at that instant.

\bnote Since most networks considered in this paper will have
bidirectional links, we will represent a bidirectional link by an
un-directed edge. Therefore, un-directed edges must be interpreted
as two directed edges, with one edge pointing in either direction.
\enote

A wireless network is characterized by broadcast and interference
constraints. Under the \emph{broadcast} constraint, all edges
connected to a transmitting node are simultaneously live and
transmit the same information. Under the \emph{interference}
constraint, the symbol received by a receiving end is equal to the
sum of the symbols transmitted on all incoming live edges. We say
a protocol avoids interference if only one incoming edge is live
for all receiving nodes.

In wireless networks, the relay nodes operate in either half or
full-duplex mode. In case of half duplex operation, a node cannot
simultaneously listen and transmit, i.e., an incoming edge and an
outgoing edge of a node cannot be simultaneously active.

\subsubsection{K-Parallel-Path Networks}

\begin{figure}[h!]
\centering
\includegraphics[width=80mm]{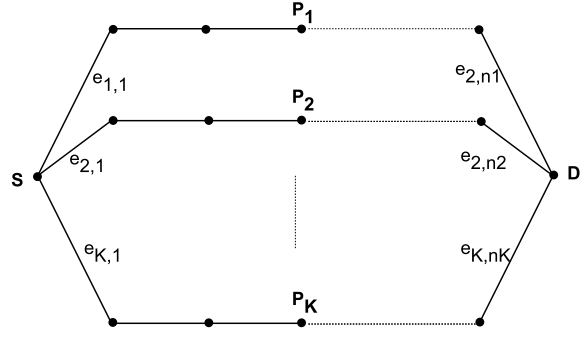}
\caption{The KPP network \label{fig:kpp_network}}
\end{figure}

One way of generalizing the two-hop relay network is to consider
this network as a collection of $K$ parallel, relaying paths from
the source to sink, each of length $>1$. This immediately leads to
a more general network that is comprised of $K$ parallel paths of
varying length, linking source and sink. More formally:

\bdefn \label{defn:kpp_defn} A set of edges $(v_1,v_2), (v_2,v_3),
\ldots, (v_{n-1},v_n)$ connecting the vertices $v_1$ to $v_n$ is
called a path. The length of a path is the number of edges in the
path. The K-parallel path (KPP) network  is defined as a ss-ss
network that can be expressed as the union of $K$ vertex-disjoint
paths, each of length greater than one, connecting the source to
the sink. Each of the node-disjoint paths is called a relaying
path. All edges in a KPP network are bidirectional (see
Fig.~\ref{fig:kpp_network}). \edefn

The communication between the source and the sink takes place in
$K$ parallel paths, labeled with the indices $P_1$, $P_2$,
$\ldots$, $P_K$. Along path $P_i$, the information is transmitted
from source to sink through multiple hops with the aid of $n_i-1$
intermediate relay nodes $\{R_{ij}\}_{j=1}^{n_i-1}$.

\bnote A network similar to the KPP network in
Definition~\ref{defn:kpp_defn} is considered in \cite{RibCaiGia},
albeit from a symbol error probability perspective. \enote

Definition~\ref{defn:kpp_defn} of KPP networks precludes the
possibility of either having a direct link between the source and
the sink, or of the existence of links connecting nodes lying on
distinct node-disjoint paths. We now expand the definition of KPP
networks to include both possibilities.

\bdefn \label{defn:kpp_general} If a given network is a union of a
KPP network and a direct link between the source and sink, then
the network is called a KPP network with direct link, denoted by
KPP(D). If a given network is a union of a KPP network and links
interconnecting relays in various paths, then the network is
called a KPP network with interference, denoted by KPP(I). If a
given network is a union of a KPP network, a direct link and links
interconnecting relays in various paths, then the network is
called a KPP network with interference and direct path, denoted by
KPP(I, D). \edefn

\bnote We adopt following terminology: For a KPP(D), KPP(I) or a
KPP(I,D) network, we consider the union of the $K$ node disjoint
paths as the \emph{backbone KPP network} (When there are several
choices for the K node-disjoint paths, we are free to choose any
one set of $K$ node-disjoint paths and refer to this collection of
$K$ paths as the backbone KPP network).  The K relaying paths in
these networks are referred to as the K \emph{backbone paths}. A
\emph{start node} and \emph{end node} of a backbone path are the
first and the last relays respectively in the path. \enote

\begin{figure}[h]
  \centering
  \subfigure[A KPP network]{\label{fig:kpp_eg}\includegraphics[width=50mm]{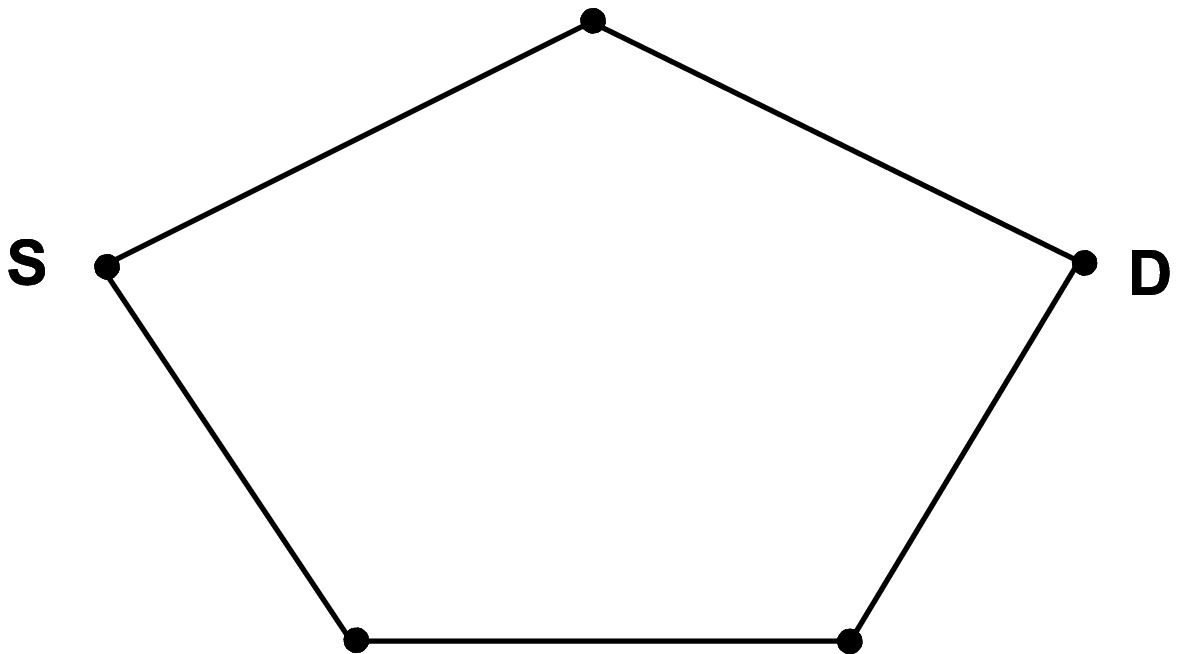}}
  \subfigure[A KPP(D) network]{\label{fig:kppd_eg}\includegraphics[width=50mm]{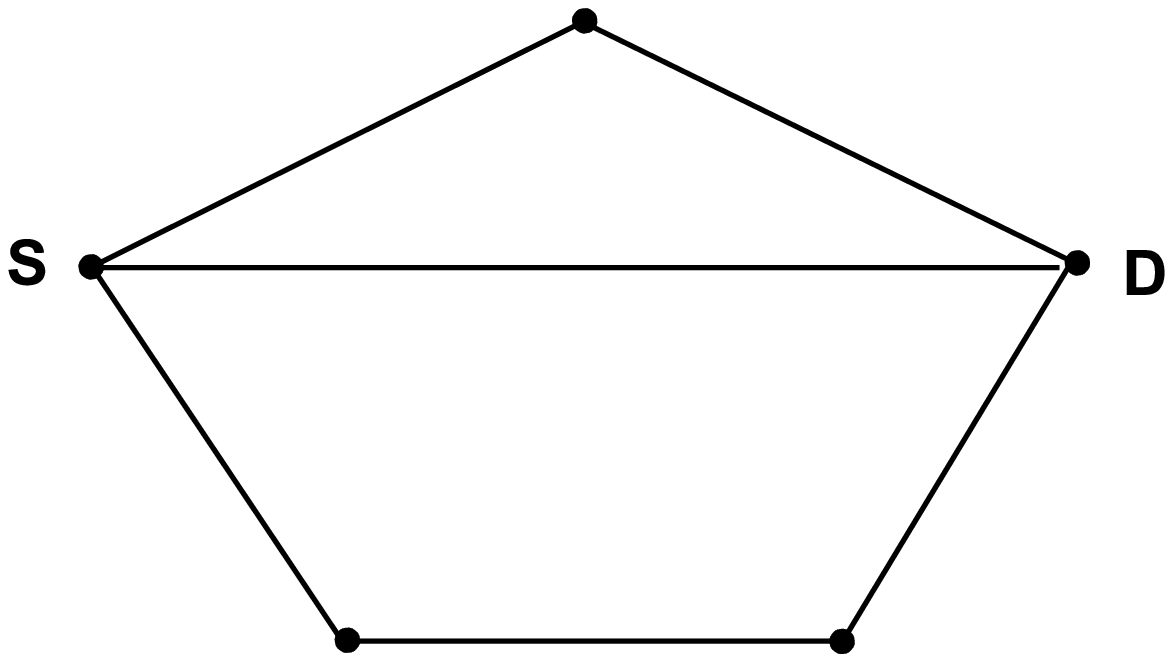}}
  \subfigure[A KPP(I) network]{\label{fig:kppi_eg}\includegraphics[width=50mm]{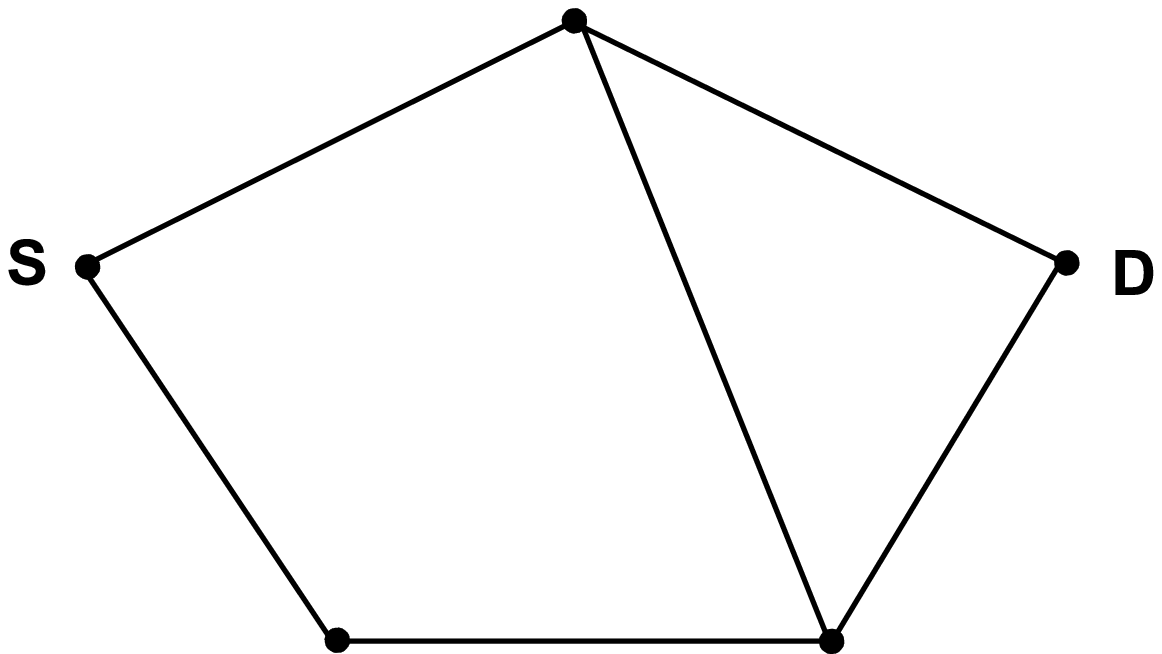}}
  \subfigure[A KPP(I, D) network]{\label{fig:kppid_eg}\includegraphics[width=50mm]{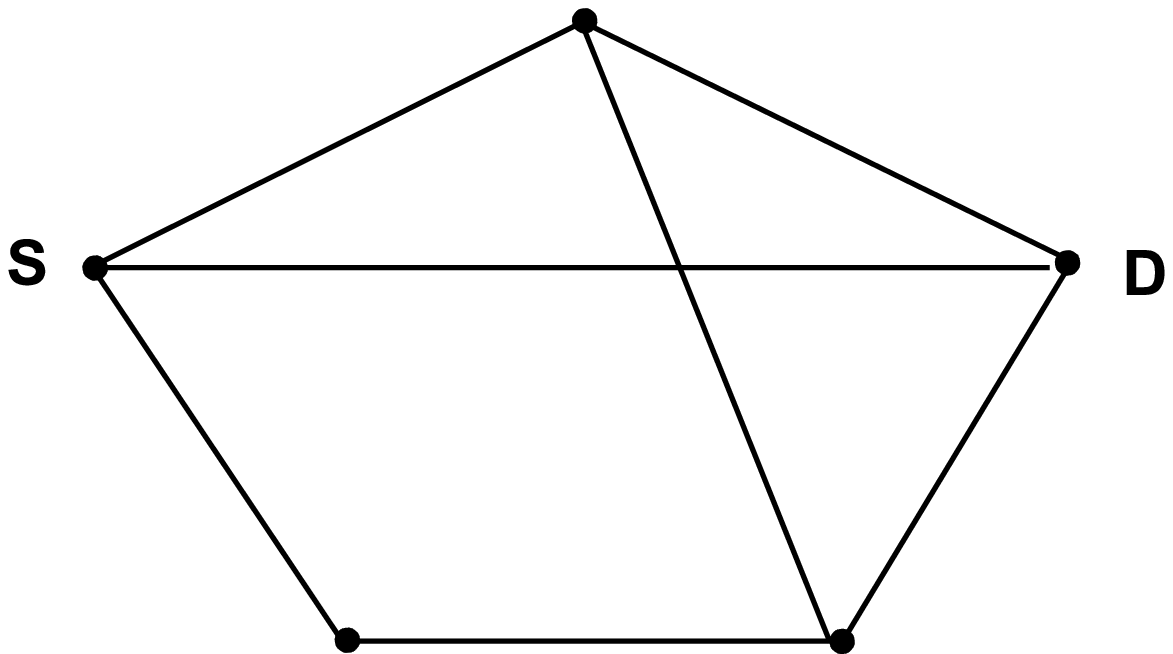}}
  \caption{Examples of KPP networks with K = 2}
  \label{fig:kpp_examples}
\end{figure}

Fig.~\ref{fig:kpp_examples} below provides examples of all four
variants of KPP networks.

For a KPP(D), KPP(I) or a KPP(I, D) network, we consider the union
of the $K$ node disjoint paths as the backbone KPP network. While
there may be many choices for the K node disjoint paths, we can
choose any one such choice and call that the backbone KPP network.
These K relaying paths in these networks are referred to as the K
\emph{backbone} paths. A \emph{start node} and \emph{end node} of
a backbone path are the first and the last relays respectively in
the path.

In a general KPP network, let $P_i,i=1,2,...,K$ be the $K$
backbone paths. Let $P_i$ have $n_i$ edges. The $j$-th edge on the
$i$-th path $P_i$ will be denoted by $e_{ij}$ and the associated
fading coefficient by $g_{ij}$.

\subsubsection{Layered Network} A second way of generalizing  a two-hop relay
network is to view the two-hop network as a network comprising of
a single layer of relays. The immediate generalization is to allow
for more layers of relays between source and sink, with the
proviso that all links are either inside a layer or between
adjacent layers. We label this class of multi-hop relaying
networks as {\em layered networks}:

\bdefn \label{defn:layered} Consider a ss-ss single-antenna
bidirectional network. A network is said to be a layered network
if there exists a a partition of the vertex set $V$ into subsets
$V_0,V_1,...,V_L,V_{L+1}$, such that \bit \item $V_0,V_{L+1}$
denote the singleton sets corresponding to the source and sink
respectively. \item If there is an edge between a node in vertex
set $V_i$ and a node in $V_j$, then $|i-j| \leq 1$. We assume
$|V_i| > 1, i=1,2,..,L$

\eit

We call $V_1,...,V_L$ as the relaying layers of the network. A
layered network is said to be fully connected if for any $i$, $v_1
\in V_i$ and $v_2 \in V_{i+1}$, then the $(v_1,v_2)$ is an edge in
the network. \edefn

\begin{figure}[h]
  \centering
  \subfigure[A layered network with with 4 relaying layers]{\label{fig:layered_eg.eps}\includegraphics[height=25mm]{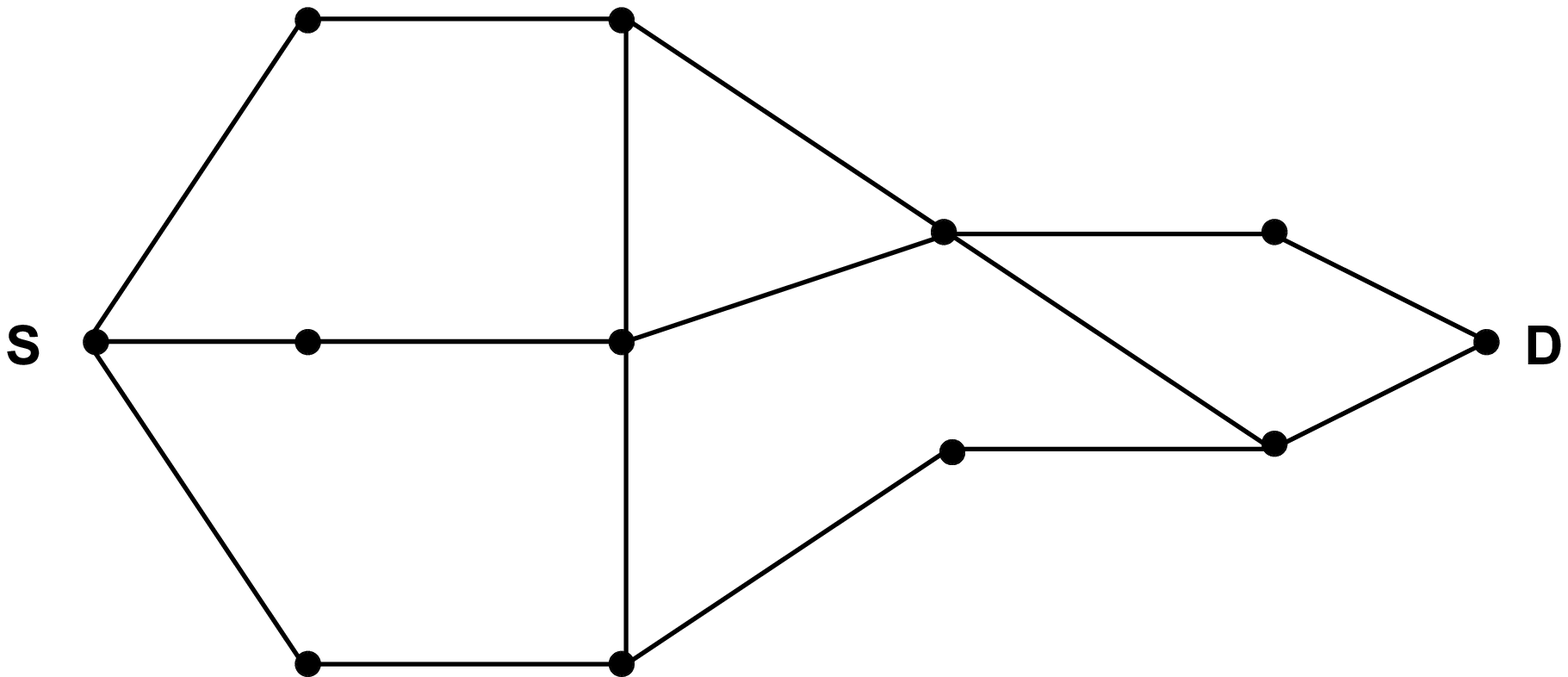}}
  \subfigure[A (3,4) regular network]{\label{fig:regular_eg}\includegraphics[height=25mm]{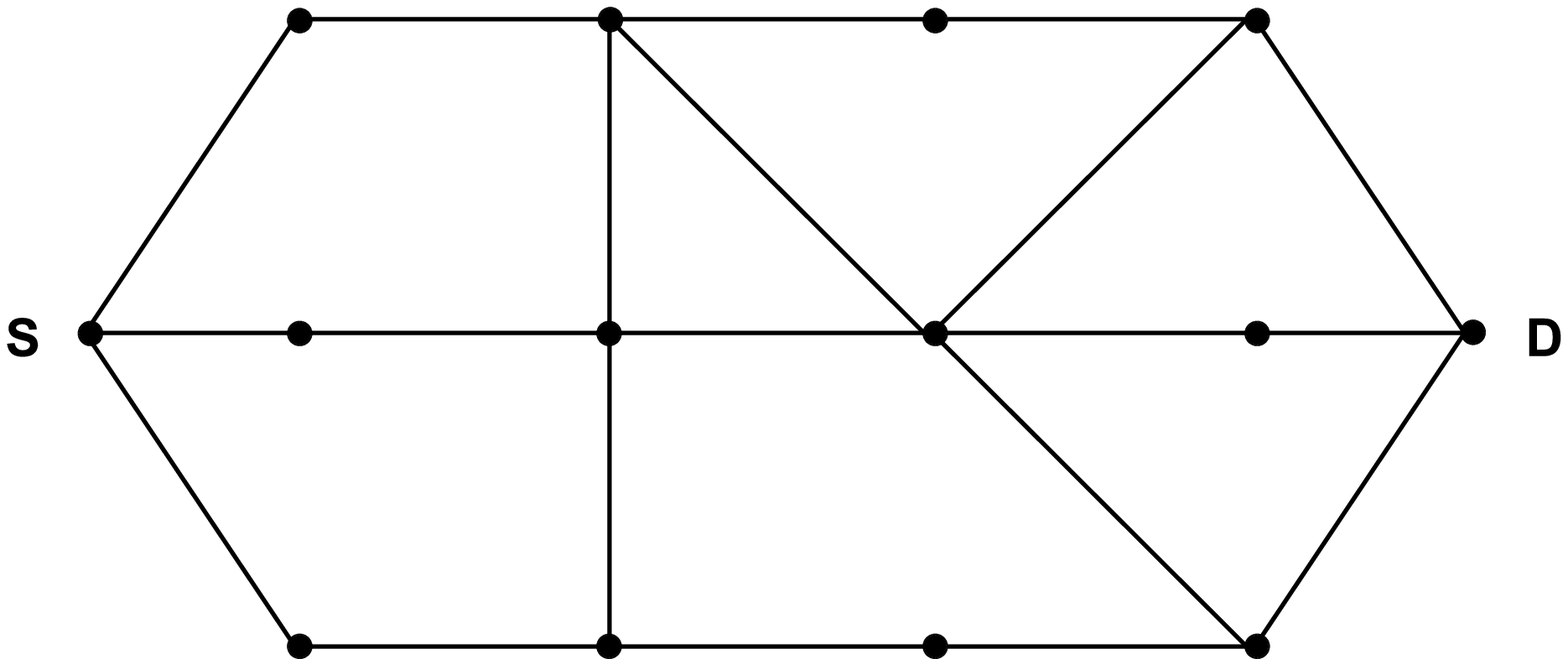}}
  \subfigure[A fully connected layered network]{\label{fig:layered_fc_eg}\includegraphics[height=25mm]{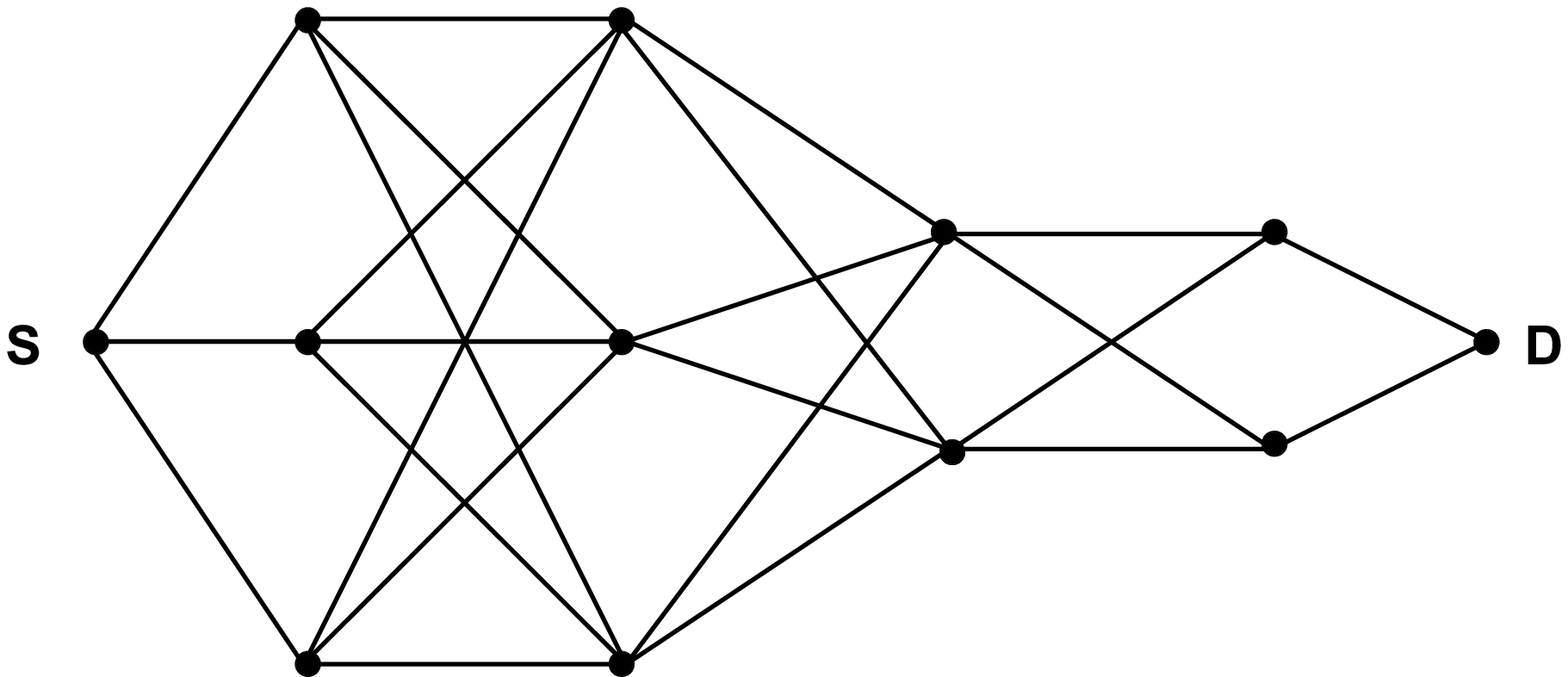}}
  \subfigure[A fully connected layered network with intra-layer links]{\label{fig:layered_fc_general}\includegraphics[height=25mm]{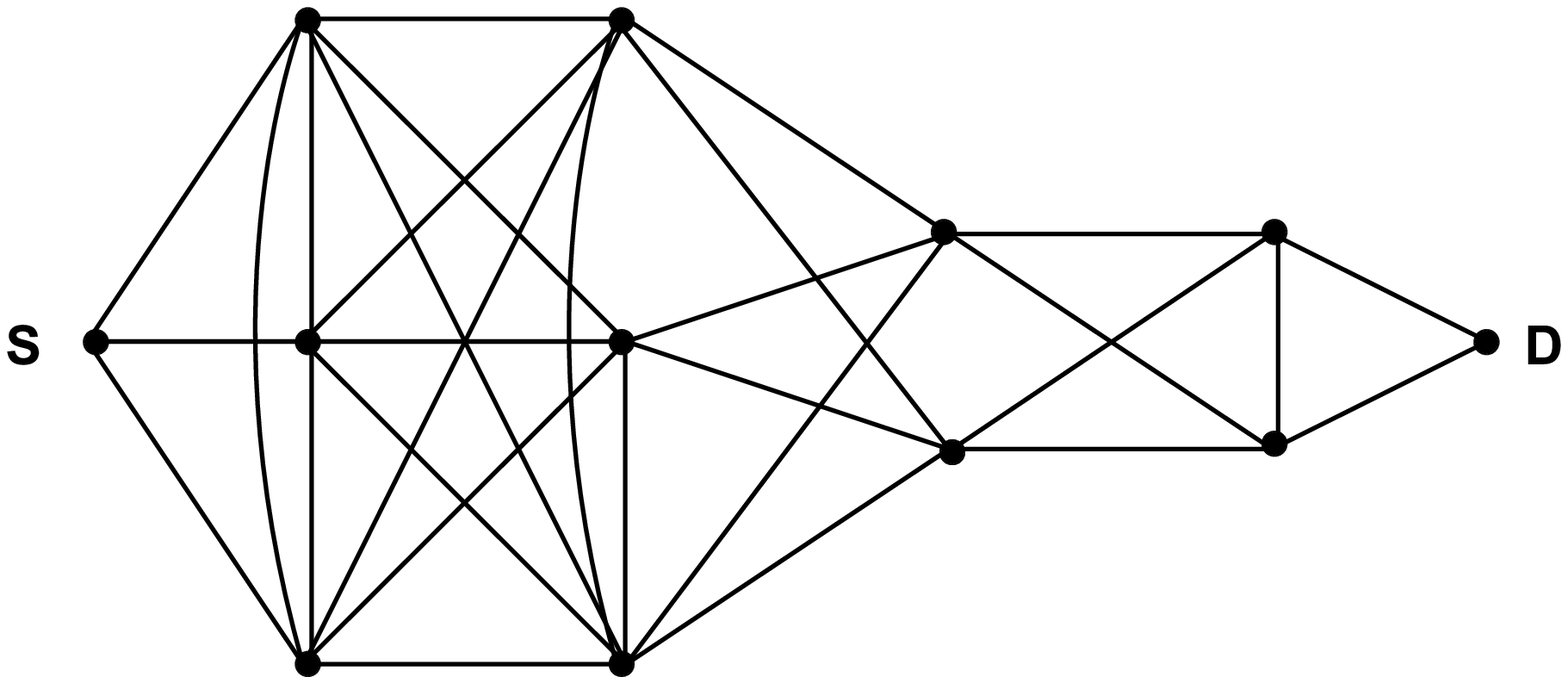}}
  \caption{Examples of Layered and Regular networks}
  \label{fig:layered_regular_examples}
\end{figure}

It must be noted that a fully connected layered network may or may
not have links inside of a layer. However, whenever we say fully
connected layered network, it applies to both networks that have
intra-layer links and those that do not have such links. Examples
of both these types of networks are shown in
Fig.~\ref{fig:layered_fc_eg} and
Fig.~\ref{fig:layered_fc_general}.

Every layered network will have a layer containing only the
source, and another layer containing only the sink. In
Fig.\ref{fig:layered_regular_examples}, examples of layered
networks are given. Layered networks were also considered in
\cite{YanBelNew} and \cite{BorZheGal}. In particular,
\cite{BorZheGal} considered layered networks with equal number of
relays on all layers. We refer to such layered networks as regular
networks.

\bnote In this remark, we characterize the intersection of KPP(I)
networks and layered networks. First we observe that one is not
contained in the other. Consider the subgraph of a given KPP(I)
network graph, consisting of all the nodes of the original network
except for the source and the sink. This subgraph will have the
property that the number of node-disjoint and edge-disjoint paths
is equal to the number of relay nodes immediately adjacent to the
source. This is a key property of KPP(I) networks, which in
general, does not hold for layered networks. On the other hand,
there can be cross links between the parallel paths in a KPP(I)
network in such a way that the network cannot be viewed as being
layered.  However, these two classes of networks are not mutually
exclusive and in fact, we term networks that lie in the
intersection of the two classes as regular networks. \enote

\bdefn \label{defn:regular} The $(K, L)$ Regular network is
defined as a KPP(I) network which is also a layered network
\cite{YanBelNew} with $L$ layers of relays (see
Fig.~\ref{fig:regular_eg}). \edefn

\bnote The two-hop relay network [Fig.\ref{fig:classical_relay}]
is a KPP(I,D) network with $K = M$, $M$ being the number of
relays. If we assume relay isolation, then it is a KPP(D) network
with $K = M$. If we exclude the direct link, then we have a ($M$,
1) regular network. \enote

\subsection{Setting and Channel Model \label{sec:channel_model}}

Between any two adjacent nodes $v_x$, $v_y$ of a wireless network,
we assume the following channel model.

\beq
    \bold{y} \ = \bold{ H {x} + {w}} \ ,
    \label{eq:channel_model}
\eeq where $\bold{y}$ corresponds to the received signal at node
$v_y$, $\bold{w}$ is the noise vector, $\bold{H}$ is a matrix and
$\bold{x}$ is the vector transmitted by the node $v_x$.

We follow the literature in making the assumptions listed below.
Our description is in terms of the equivalent complex-baseband,
discrete-time channel.

\ben \item All channels are assumed to be quasi-static and to
experience Rayleigh fading and hence all fade coefficients are
i.i.d., circularly-symmetric complex Gaussian $\mathbb{C}\mathcal
{N} (0,1)$ random variables. \item The additive noise at each
receiver is also modeled as possessing an i.i.d.,
circularly-symmetric complex Gaussian $\mathbb{C}\mathcal {N}
(0,1)$ distribution. \item Each receiver (but none of the
transmitters) is assumed to have perfect channel state information
of all the upstream channels in the network. \footnote{However,
for the protocols proposed in this paper, the CSIR is utilized
only at the sink, since all the relay nodes are required to simply
amplify and forward the received signal.} \een

An AF protocol $\wp$ i.e., a protocol $\wp$ in which each node in
the network operates in an amplify-and-forward fashion, induces
the following linear channel model between source and sink:  \beq
    \bold{y} \ = \bold{ H(\wp) {x} + {w} } \ ,
    \label{eq:channel_model}
\eeq where $\bold{y} \in \mathbb{C}^m$ denotes the signal received
at the sink, $\bold{w}$ is the noise vector, $\bold{H(\wp)}$ is
the $(m \times n)$ induced channel matrix and $\bold{x} \in
\mathbb{C}^n$ is the vector transmitted by the source. The
components of the $n$-tuple $\bold{x}$ are the $n$ symbols
transmitted by the source and similarly, the components of the
$m$-tuple $\bold{y}$ represent the symbols received at the sink.
Typically $m$ equals $n$. We impose the following energy
constraint on the transmitted vector $\bold{x}$ \beqan
\text{Tr}(\Sigma_x) \ := \ \text{Tr}(\mathbb{E}\{\bold{x}
\bold{x}^{\dagger}\}) & \leq & n \rho \eeqan where $\text{Tr}$
denote the trace operator, and we will regard $\rho$ as
representing the SNR on the network. We will assume a symmetric
power constraint on the relays and the source. However it will
turn out that given our high SNR perspective here, the exact power
constraint is not of significant importance. We consider both half
and full-duplex operation at the relay nodes.

\subsubsection{Diversity-Multiplexing Gain Tradeoff \label{sec:dmt}}

Let $R$ denote the rate of communication across the network in
bits per network use. Let $\wp$ denote the protocol used across
the network, not necessarily an AF protocol.  Let $r$ denote the
multiplexing gain associated to rate $R$ defined by \beqan R & = &
r \log (\rho) .\eeqan  The probability of outage for the network
operating under protocol $\wp$, i.e., the probability of the
induced channel in \eqref{eq:channel_model} is then given by \beqn
    P_{\text{out}}(\wp,R) = \inf_{\Sigma_x \ \geq \ 0, \
    \text{Tr}(\Sigma_x) \ \leq \ n \rho }
    \text{Pr}( I(\bold{x};\bold{y}) \ \leq \ n R | \bold{H(\wp)} = H(\wp) ).
\eeqn  Let the outage exponent $d_{\text{out}}(\wp,r)$ be defined
by \beqan d_{\text{out}}(\wp,r) & = & - \lim_{\rho \rightarrow
\infty} \frac{ P_{\text{out}}(\wp,R)}{\log( \rho) } \eeqan and we
will indicate this by writing \beqan \rho^{-d_{\text{out}}(\wp,r)}
& \doteq & P_{\text{out}}(\wp,R). \eeqan  The symbols $\dot \geq$,
$\dot \leq$ are similarly defined.

The outage $d_{\text{out}}(r)$ of the network associated to
multiplexing gain $r$ is then defined as the supremum of the
outages taken over all possible protocols, i.e., \beqan
d_{\text{out}}(r) & = & \sup_{\wp} d_{\text{out}}(\wp,r). \eeqan

A distributed space-time code (more simply a code) operating under
a protocol $\wp$ is said to achieve a diversity gain $d(\wp,r)$ if
\beqn P_{e}(\wp,\rho) \doteq \rho^{-d(\wp,r)} \ , \eeqn where
$P_e(\rho)$ is the average error probability of the code $C(\rho)$
under maximum likelihood decoding. Using Fano's inequality, it can
be shown (see \cite{ZheTse}) that for a given protocol, \beqan
    d(\wp,r) & \leq & d_{\text{out}}(\wp,r).
\eeqan

We will refer to the outage exponent $d_{\text{out}}(r)$ as the
DMT $d(r)$ of the corresponding channel since for every protocol
discussed in this paper we shall identify a corresponding coding
strategy in Section~\ref{sec:code_design} whose diversity gain
$d(\wp,r)$ equals $d_{\text{out}}(r)$.

For each of the networks described in this paper, we can get an
upper bound on the DMT, based on the cut-set upper bound on mutual
information \cite{CovTho}. This was formalized in \cite{YukErk} as
follows:

\blem \label{lem:CutsetUpperBound} Given a cut
$\mathcal{C}_i,i=1,2,..,M$ between any source and sink, let
$r^{(\mathcal{C}_i)} \log(\rho)$ be the rate of information flow
across the cut. Given a cut, there is a $H$ matrix connecting the
input terminals of the cut to the output terminals. Let us call
the DMT corresponding to this $H$ matrix as the DMT of the cut,
$d_{\mathcal{C}_i}(r^{(\mathcal{C}_i)})$. Then the DMT between the
source and the sink is upper bounded by \[ d({r}) \leq \min_{i }
\{d_{\mathcal{C}_i}(r^{(\mathcal{C}_i)})\}.\] \elem

\bdefn \label{defn:DMT_Matrix} Given a random matrix $\bold{H}$ of
size $m \times n$, we define the \emph{DMT of the matrix}
$\bold{H}$ as the DMT of the associated channel $\bold{y = Hx +
w}$ where $\bold{y}$ is a $m$ length received column vector,
$\bold{x}$ is a $n$ length transmitted column vector and
$\bold{w}$ is a $\mathcal{CN}(0,I)$ column vector. We denote the
DMT by $d_H(.)$ \edefn

\subsection{Results \label{sec:results}}

The principal results of this paper are tabulated in Table I. Some
of these results were presented in conference versions of this
paper \cite{SreBirKumITA}, \cite{WPMC}. We have characterized
achievable DMT/diversity for many classes of networks as given in
the table. When compared against the cut-set upper bound, in many
cases, the optimal DMT is achieved. In other cases, we prove that
a linear DMT between the maximum multiplexing gain and maximum
diversity is achievable, while the cut-set upper bound can be
concave in general.  Explicit schemes and code design is
established for all the achievable DMT. In the table, $M$ refers
to the min-cut of the network of interest.

\begin{table*}
\label{tab:summary} \caption{Principal Results Summary}
\begin{center}
\begin{tabular}{||c|c|c|c|c|c|c|c|c||}
\hline \hline &&&&&&&&\\
Network  & No of    & No of     & FD/ & Direct & Upper bound on  & Achievable & Is upper bound& Reference  \\
         & sources/ & antennas  & HD  &  Link  & Diversity/DMT   & Diversity/DMT& achieved? &\\
         &  sinks   & in nodes  &        &        & $d_\text{bound}(r)$  & $d_\text{achieved}(r)$& &\\
\hline \hline &&&&&&&&\\
Arbitrary      & Multiple     & Multiple     & FD/HD & $\checkmark$  & $d(0)=M$ & $d(0)=M$ & $\checkmark$&Theorem~\ref{thm:mincut}\\
&&&&&&&($d_{max}$ achieved)&\\
&&&&&&&&\\
\hline
&&&&&&&&\\
Arbitrary     & Multiple & Multiple   & FD/HD & $\times$ & $d(0)=M$ & $d(0)=M$ & $\checkmark$&Theorem~\ref{thm:mincut}\\
&&&&&&&($d_{max}$ achieved)&\\
&&&&&&&&\\
\hline
&&&&&&&&\\
Arbitrary & Single  & Single    & FD & $\checkmark$ & Concave & $M(1-r)^+$ & A linear DMT&Theorem~\ref{thm:FD_No_Direct_Path}\\
Directed &&&&&in general&& between $d_{max}$ and&\\
Acyclic Networks&&&&&&& $r_{max}$ is achieved&\\
&&&&&&&&\\
\hline
&&&&&&&&\\
KPP(K $\geq$ 3) & Single & Single & HD & $\times$ & $K(1-r)^+$ & $K(1-r)^+$ & $\checkmark$& Theorem~\ref{thm:KPP}\\
&&&&&&&&\\
\hline
&&&&&&&&\\
KPP(D)(K $\geq$ 3) & Single & Single & HD & $\checkmark$ & $(K+1)(1-r)^+$ & $(K+1)(1-r)^+$ & $\checkmark$& Theorem~\ref{thm:KPP_D}\\
&&&&&&&&\\
\hline
&&&&&&&&\\
KPP(I)(K $\geq$ 3) & Single & Single & HD & $\times$ & $K(1-r)^+$ & $K(1-r)^+$ & $\checkmark$ & Theorem~\ref{thm:KPP_I}\\
&&&&&&&&\\
\hline
&&&&&&&&\\
Fully & Single & Single & HD & $\times$ & Concave & $M(1-r)^+$ &  A linear DMT& Theorem~\ref{thm:fully_connected_layered}\\
Connected&&&&&in general&& between $d_{max}$ and &\\
Layered&&&&&&& $r_{max}$ is achieved. &\\
&&&&&&& $\checkmark$ for $L<4$ & Corollary~\ref{cor:optimal_layered} \\
\hline
&&&&&&&&\\
General & Single & Single & HD & $\times$ & Concave & $M(1-r)^+$ &  A linear DMT &Lemma~\ref{lem:General_layered_network}\\
Layered &&&&&in general&& between $d_{max}$ and &\\
(satisfying&&&&&&& $r_{max}$ is achieved &\\
Lemma~\ref{lem:General_layered_network})&&&&&&&&\\
\hline
&&&&&&&&\\
$(K,L)$ Regular & Single & Single & HD & $\times$ & $K(1-r)^+$ & $K(1-r)^+$ & $\checkmark$ & Theorem~\ref{thm:knregular_dmt}\\
&&&&&&&&\\
\hline \hline
\end{tabular}
\end{center}
\end{table*}

 For arbitrary
co-operative networks with multiple sources and sinks, each
potentially equipped with multiple antennas, we characterize the
maximum achievable diversity gain and give a scheme that achieves
this maximum diversity using an amplify-and-forward protocol in
Section.~\ref{sec:mincut_diversity}. For arbitrary ss-ss networks
with full duplex operation, we prove that a linear tradeoff
between maximum diversity and maximum multiplexing gain is
achievable using an amplify and forward protocol in
Section.~\ref{sec:full_duplex}.

For both KPP and layered networks, we propose an explicit protocol
that achieves a diversity multiplexing trade-off that is linear
between the maximum diversity and maximum multiplexing gain points
in Section.~\ref{sec:half_duplex}. For KPP networks, this
coincides with the upper-bound on the DMT as given by the cut-set
bound, thus characterizing the DMT of this entire family of
networks completely. For layered networks, the cut-set bound turns
out to be concave in the general case and does not coincide with
the achievable region. For general layered networks,  we give a
sufficient condition for the achievability of a linear DMT between
the maximum diversity and the maximum multiplexing gain in
Lemma~\ref{lem:General_layered_network}.

Along the way, we derive the optimal DMT of parallel channel in
Lemma~\ref{lem:parallel_channel}, provide alternative and often
simpler proofs of several existing results and in
Section.~\ref{sec:univ_full_div}, prove that codes achieving full
diversity on a MIMO Rayleigh fading channel achieve full diversity
on arbitrary fading channels.

In Section.~\ref{sec:code_design} we give explicit codes with
short block-lengths based on cyclic division algebras that achieve
the best possible DMT for all the schemes proposed above. We also
prove (Section.~\ref{sec:univ_full_div}) that full diversity codes
for all networks in this paper can be obtained by using codes that
give full diversity on a Rayleigh fading MIMO channel.

For KPP and layered networks with multiple antenna nodes, we
examine certain protocols and establish achievable DMT for these
protocols in Section.~\ref{sec:half_duplex_multiple}.

\section{Relation to Existing Literature}

In this section, we present how the results in this paper relate
to other in this area. Certain results in this paper can be used
to recover existing results on cooperative communication in a
simpler, concise and more intuitive manner.

\ben\item \emph{Proof of Conjecture 1 in the paper by Rao and
Hassibi \cite{RaoHas} and \cite{Rao}:}

The general NAF protocol considered in \emph{Example 3} in
Section~\ref{sec:examples_main_thm} of the present paper is the
same as that considered by Rao and Hassibi.  The results here
proves Conjecture 1 given in \cite{RaoHas} and \cite{Rao}.

\item \emph{The lower bound on the DMT of various AF Protocols:}
We prove lower bounds on the DMT of various AF protocols. While
most are previously known, the new method employed here presents a
simpler derivation. As it turns out, all lower bounds for single
antenna systems provided here are tight.

\emph{NAF Protocol:} The DMT of the NAF protocol was computed in
\cite{AzaGamSch}. We prove a lower bound on the DMT which turns
out to be tight.

\emph{SAF Protocol:} The Slotted Amplify and Forward protocol is
proposed in \cite{YanBelSaf} and upper and lower bounds on its DMT
under relay isolation is evaluated and shown to be equal. For
doing so, matrix theoretic techniques are employed in
\cite{YanBelSaf}. In the current paper, in \emph{Example 2} of
Section~\ref{sec:examples_main_thm}, the lower bound for the same
is developed using information theoretic techniques, which lends
insight into the form of the DMT.

\emph{N-Relay MIMO NAF Channel given in \cite{YanBelMimoAf}:}

In \cite{YanBelMimoAf}, the authors consider a two-hop relay
network with a direct link and $N$ relays. We prove an improved
lower bound on the DMT for the MIMO NAF protocol considered in
that paper (See \emph{Example 4} of
Section~\ref{sec:examples_main_thm}) .

\item \emph{The diversity of arbitrary cooperative networks.}

We characterize completely the maximum diversity order attainable
for arbitrary cooperative networks and it is shown that an amplify
and forward scheme is sufficient to achieve this. Special cases of
these were derived for the MIMO two-hop relay channel in
\cite{YanBelMimoAf}, under a certain condition on the number of
antennas (See \emph{Corollary 1} in that paper). Also, the
diversity order of layered networks using amplify and forward
networks is characterized in \cite{YanBelNew}. In
\cite{BoyFalYan}, upper bounds on the diversity order of an
arbitrary single-source single-sink network under the two cases of
common and independent code-books was derived. However, no
achievability results are given there.

\item \emph{The optimal DMT of the two-hop cooperative channel
without direct link.}

The optimal DMT of a (K,L) regular network is derived in
Theorem~\ref{thm:knregular_dmt} in Section \ref{sec:half_duplex}
of this paper. In an independent (parallel) work by Gharan,
Bayesteh and Khandani \cite{GhaBayKha}, the optimal DMT of a
two-hop network, which is a special case of a regular network (in
particular it is a (K,1) Network), is derived to be $d(r) =
L(1-r)$. The protocol they propose is the same as the protocol
employed in the present paper. In fact, both these protocols are
simply the SAF (Slotted Amplify and Forward) protocol
\cite{YanBelSaf} applied in the situation when there is no direct
link between source and sink. It must be noted however, that the
proof techniques used in this paper are entirely different from
those used in \cite{GhaBayKha}.

\item The DMT of the parallel channel in closed form is obtained
in Lemma.~\ref{lem:parallel_channel}. A special case of this
result is derived in \cite{YanBelNew} where the authors
characterize the parallel channel DMT when all the individual
channels have the same DMT.

\item For an arbitrary full-duplex networks, it is shown in the
present paper, that a linear DMT between the maximum diversity and
the maximum multiplexing gain is achievable. A special case of
this result is proved for the case of layered networks in
\cite{YanBelNew}. \een

\subsection{Outline \label{sec:outline}}

In Section \ref{sec:general_theory}, we present techniques and
general results which will of use in later sections. In this
section, we introduce the Information Flow diagram (i-f diagram),
and prove the result that min-cut equals diversity. In Section
\ref{sec:full_duplex}, we consider the case with full duplex
relays. We present schemes achieving optimal DMT for KPP(I,D)
networks. In Section \ref{sec:half_duplex_isolated}, we focus on
half-duplex KPP networks and present protocols achieving optimal
DMT for $K \geq 3$. In Section \ref{sec:half_duplex}, KPP(I)
networks with half-duplex relays are considered, and schemes
achieving optimal DMT are presented for KPP(I) networks allowing
certain types of interference. In Section \ref{sec:layered}, we
consider layered networks and show that a linear DMT between max
multiplexing of $1$ and diversity of $d_\text{max}$ is obtained,
which is indeed optimal if the number of layers is lesser than
$4$. In Section \ref{sec:multiple_antenna}, we consider
multi-antenna layered and KPP networks and give an achievable DMT,
which improves significantly on known bounds. Finally, in  Section
\ref{sec:code_design}, we give explicit CDA based codes of low
complexity for all the DMT optimal protocols.

\section{Techniques and General Results for Cooperative Networks \label{sec:general_theory}}

\subsection{Amplify and Forward Protocols}

We consider only amplify-and-forward (AF) protocols in this paper
by which we mean that relays are allowed to perform only linear
processing on their received signals prior to transmission.  In
particular, they are not permitted to decode and then re-encode.

In all of our protocols, we assume that the relays perform the
simplest form of linear processing; transmission upon scaling the
incoming by an appropriate constant to meet a transmit-power
constraint. \footnote{More sophisticated linear processing
techniques would include matrix transformations of the incoming
signal.}  Furthermore, it is known \cite{AzaGamSch}, that this
constant does not matter in the scale of interest. Therefore,
without loss of accuracy, we will assume that this constant is
indeed $1$.

It follows that, for any given network, we only need specify the
schedule to completely specify the protocol. Once the schedule is
specified, each node transmits the last received signal in the
next time instant in accordance with the schedule. This will
create a transfer matrix between the signal transmitted from the
source and the sink, with the noise being no longer white. To
compute the DMT offered by the protocol, we need to compute the
DMT of the equivalent channel $y = Hx + w$, where $H$ is the
effective transfer matrix and $w$ is the noise vector, which is
potentially colored.

In this section, we will develop techniques to handle non-white
noise and a general method to compute lower bounds on the DMT of
matrices with certain structure.

\subsection{The Information Flow Diagram \label{sec:info_flow_diagram}}

We begin by introducing the notion of an information-flow (i-f)
diagram as a means of characterizing the the mutual information
between the source and the sink in a ss-ss relay network. A ss-ss
relay network will have many paths between the source and the
sink, including a direct link.  Protocols employed in a wireless
network need to take into account the half-duplex, interference
and broadcast constraints at each of the nodes. Due to the
complexity of the network graph, it is in general difficult to
characterize the network information-theoretically under the
wireless constraints. The i-f diagram, that we propose, is an
attempt to abstract out the details of network graph, and to focus
our attention only on the mutual information between source and
sink, given a protocol.

As will be seen, the i-f diagram is well suited to studying
amplify and forward relay networks.

\begin{figure}[h!]
\centering
\includegraphics[width=60mm]{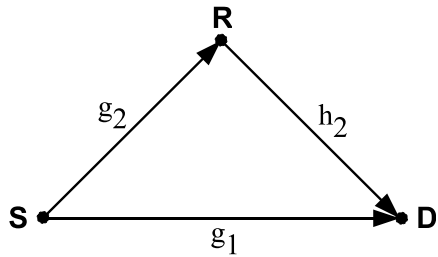}
\caption{Single Relay Channel \label{fig:one_relay}}
\end{figure}

\emph{Example 1} Consider a ss-ss, single-relay scenario,
operating under the Non-orthogonal Amplify and Forward (NAF)
protocol of \cite{AzaGamSch} (Fig.\ref{fig:one_relay}).  This is a
two slot protocol, wherein during the first slot, the source
transmits to both relay and sink.  During the second slot, the
relay re-transmits the information that it received during the
first time slot, while the source transmits new information at
this time. Let us represent the random vectors associated to
source transmissions at time slot one and two by $\bold{x_1,x_2}$
and the corresponding data received by sink in the two time slots
by $\bold{y_1,y_2}$.

Then the input-output relation takes on the following form

\beq
    \bold{y} = \bold{H} \bold{x} + \bold{n} , \label{eq:simple_relay_channel}
\eeq where \beqan
    \bold{n} & = & \left[ \begin{array}{c} \bold{w}_1 \\
     \bold{h_2}  \bold{v} \  +  \ \bold{w}_2
    \end{array} \right] \\
    \bold{H} & = & \left[ \begin{array}{cc} \bold{g_1} & \bold{0}\\
     \bold{g_2h_2}  & \bold{g_1} \end{array} \right] \\
    \bold{x} & = & \left[ \begin{array}{cc} \bold{x}_1 \\
    \bold{x}_2 \end{array} \right] , \\
     \bold{y} & = & \left[ \begin{array}{cc} \bold{y}_1 \\
    \bold{y}_2 \end{array} \right].
\eeqan Given that $\bold{H}$ is known, the covariance matrices of
the noise and signal vector are denoted by \beqan
    \Sigma_n & := & \mathbb{E} (\bold{n}
    \bold{n}^{\dagger}) \\
    & = & \left[ \begin{array}{cc} \sigma_w^2 & 0 \\
    0 & \sigma_w^2 \ + \ |h_2|^2\sigma_{v}^2 \end{array} \right]
\eeqan and \beqn
    \Sigma_x := \mathbb{E} (\bold{x} \bold{x}^{ \dagger}),
\eeqn where $\sigma_{v}^2,\sigma_{w}^2$ denote the variances of
the corresponding noise vectors. We will assume $\sigma_{v}^2 =
\sigma_{w}^2 = 1$ without loss of generality since the exact value
does not matter in the scale of interest.

\begin{figure}[h!]
\centering
\includegraphics[width=50mm]{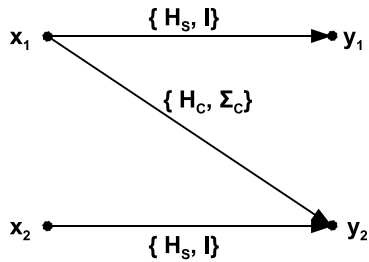}
\caption{Information flow diagram of single relay
channel\label{fig:flow_model}}
\end{figure}

We represent the induced channel by the i-f diagram in
Fig.\ref{fig:flow_model}.

In the i-f diagram in Fig.\ref{fig:flow_model}, we have used the
subscript $s$ denoting straight coupling, and subscript $c$
denoting cross coupling. So the following equivalence holds.

\beqan \bold{H_d} & := & \bold{g_1} \\
\bold{H_c} & := & \bold{g_2h_2} \\
\bold{\Sigma_c} & := & 1+|\bold{h_2}|^2 \eeqan

\begin{figure}[h]
  \centering
  \subfigure[A single link in i-f diagram]{\label{fig:equivalence_1_1}\includegraphics[width=50mm]{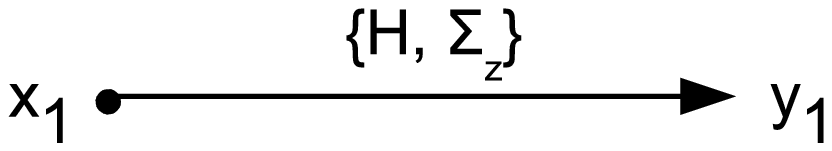}}
  \subfigure[Equivalent channel model]{\label{fig:equivalence_1_2}\includegraphics[width=50mm]{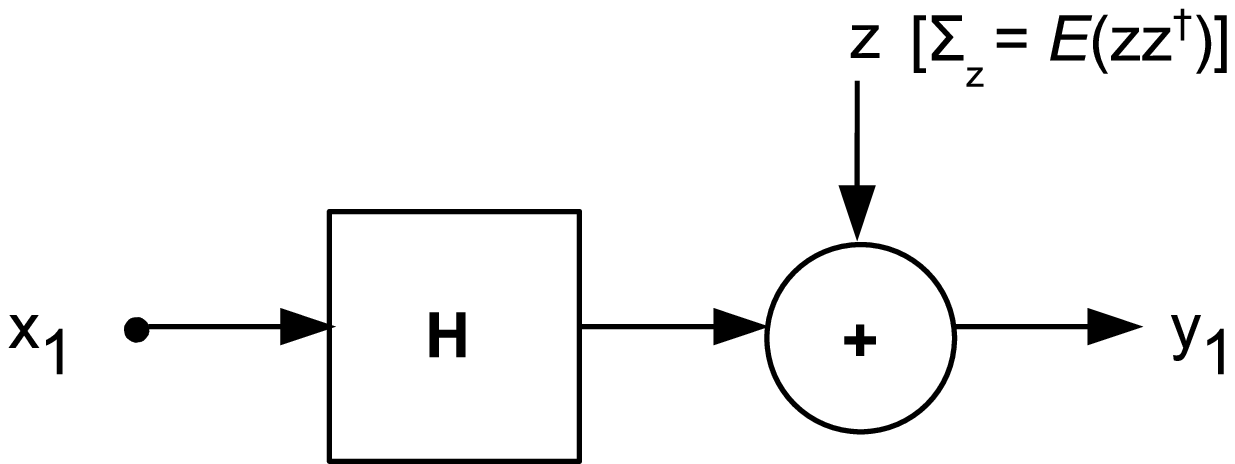}}
  \caption{Equivalent channel model of a single link in i-f diagram. }
  \label{fig:equivalence_1}
\end{figure}

\begin{figure}[h]
  \centering
  \subfigure[Multiple access links in i-f diagram]{\label{fig:equivalence_2_1}\includegraphics[width=50mm]{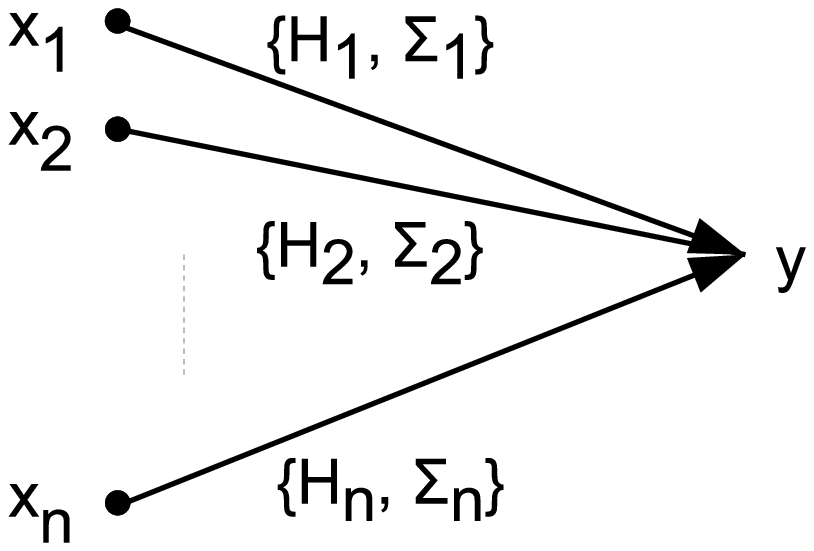}}
  \subfigure[Equivalent channel model]{\label{fig:equivalence_2_2}\includegraphics[width=50mm]{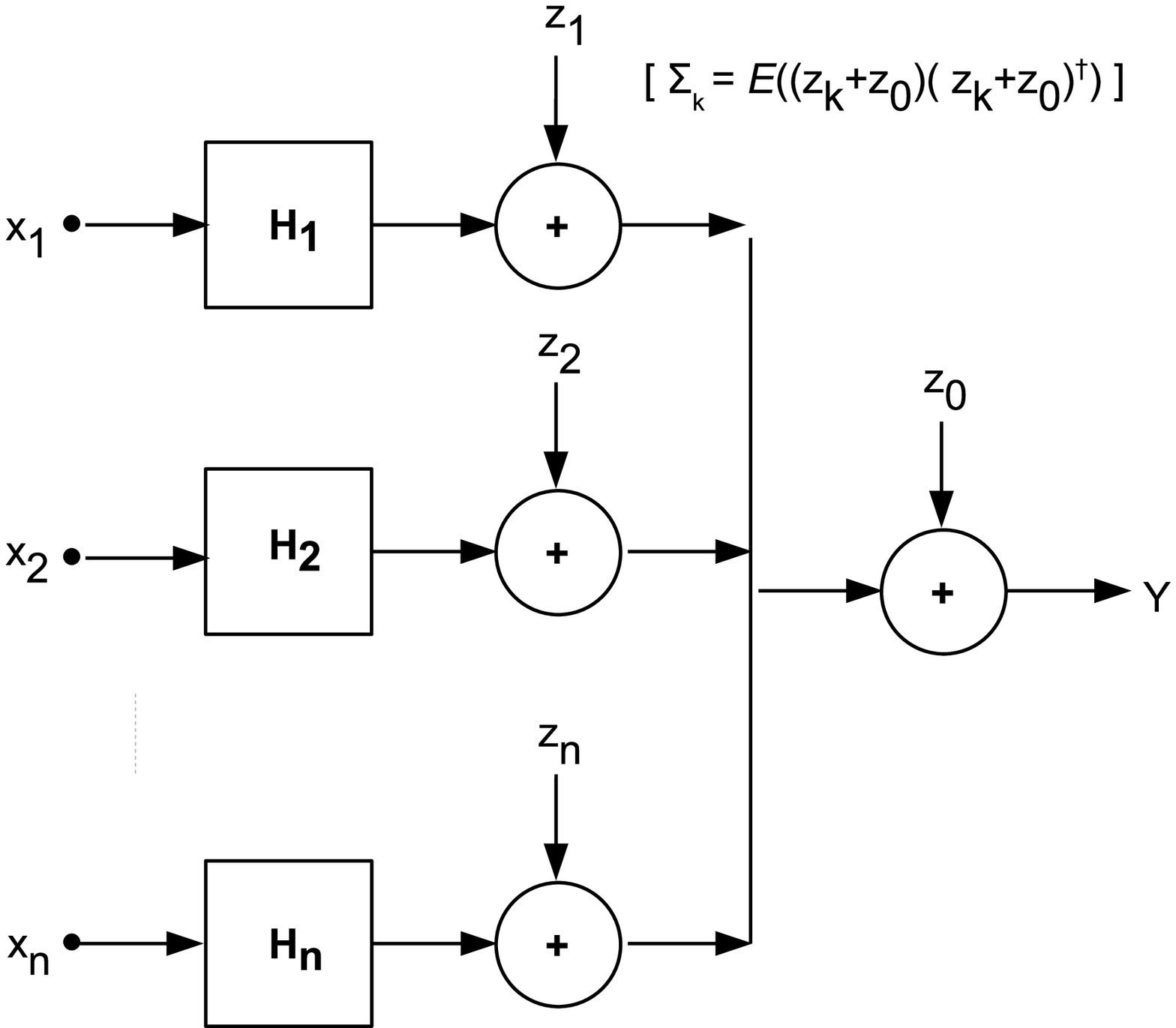}}
  \caption{Equivalent channel model of multiple-access links in i-f diagram. }
  \label{fig:equivalence_2}
\end{figure}

The interpretation of the arrows in the i-f diagram is illustrated
in Fig.\ref{fig:equivalence_1} and Fig.\ref{fig:equivalence_2}.


The two tuple notation $(\bold{H},\Sigma)$ for each link is used
to specify the channel matrix for the signal and the noise
covariance matrix. $\bold{H}$ is the channel matrix and therefore
the transmitted signal $x$ is multiplied by $\bold{H}$ to give
$\bold{H}x$ at the receiver. The noise is potentially correlated
because it is accumulated over multiple links, in such a way that
the noise added on one link gets multiplied by the channel matrix
of the next link.

The input output relation for the single link in the i-f diagram
Fig.\ref{fig:equivalence_1} is explained as follows:

\beqan \bold{y_1} & = & \bold{H{x_1} + z} \eeqan where $\bold{z}$
is a complex gaussian random variable, with $\Sigma_z =
\mathbb{E}(\bold{z} \bold{z}^{ \dagger})$.

The input output relation for the multiple links terminating in a
given node in the i-f diagram Fig.\ref{fig:equivalence_2} is
explained as follows:

\beqan \bold{y} & = & \bold{ \sum_{i=1}^N \ H_i{x_i} +
\sum_{i=1}^N \ {z_i} + z_0} \eeqan

where $\bold{z_i}$ and $\bold{z_0}$ are independent complex
gaussian random variables, $\Sigma_k = \mathbb{E}((\bold{z_k+z_0})
(\bold{z_k+z_0})^{ \dagger})$.

\subsection{White in the scale of interest\label{sec:noise}}

In this section, we provide two lemmas that will be extensively
used in all future sections: Lemma~\ref{lem:noise_white}, which
states that noise, even though correlated can be treated as white
in the scale of interest and Lemma~\ref{lem:signal_white}, which
proves that i.i.d. gaussian inputs are sufficient to attain the
outage exponent of any channel of the form $y=Hx+w$.

\blem \label{lem:noise_white} Consider a channel of the form
$\bold{y} = \bold{Hx} + \bold{z}$. Let $\bold{H}$, $\bold{F}_{j},
j=1,2,..,L$ be $n\times n$ independent random matrices, with
entries in each of the matrices being i.i.d. random variables with
complex Gaussian $\mathbb{C}\mathcal {N} (0,1)$ distribution. Let
$\bold{G}_i,i=1,2,..,M$ comprise of finite products of various
matrices from the set of $\bold{F}_j$. Let $\bold{z} = \bold{z}_0
+ \sum_{i=1}^{M} \bold{G}_i\bold{z}_i$. Let $\{\bold{z}_i\}$ be
i.i.d. circularly symmetric $n$-dimensional complex Gaussian
$\mathbb{C}\mathcal {N} (\underline{0},I)$ random vectors.

Then $\bold{z}$ is white in the scale of interest, i.e.,

\ben \item ${\lambda_i}  \doteq  \rho^{0} \ \ \forall i $ with
probability one, where $\lambda_i$ are eigenvalues of the noise
covariance matrix $\Sigma$.

\item $\log\det(I+ \rho \bold{H}\bold{H}^{\dagger}\Sigma^{-1})
\doteq \log\det(I+ \rho \bold{H}\bold{H}^{\dagger})$ with
probability one.

\item $Pr(\log\det(I+ \rho \bold{H}\bold{H}^{\dagger}\Sigma^{-1})
\leq r\log\rho) \doteq Pr(\log\det(I+ \rho
\bold{H}\bold{H}^{\dagger}) \leq r\log\rho) $

\een \elem

\bpf For a fixed set of values of $F_i$ and $H$, the noise
covariance matrix is given by,
\beqa \Sigma & = & \mathcal {E}[z{z^\dagger}] \nonumber \\
& = & I + \sum_{i=1}^{M} G_iG_i^\dagger \label{eq:sigma}\eeqa

Let $\lambda_i(A)$, $\lambda_{max}(A)$ and $\lambda_{min}(A)$
denote the $i$th, maximum and minimum eigenvalues of the positive
semi-definite matrix $A$.  If the context is clear, we may avoid
specifying the matrix, and just use $\lambda_i$, $\lambda_{max}$
and $\lambda_{min}$ respectively.

By Theorem 6.1.1 in \cite{HorJoh} due to Gersgorin, each
eigenvalue of $\Sigma$, when properly ordered, is bounded within
the interval

\beqa \Sigma_{ii} - R_i(\Sigma) \leq & \lambda_i(\Sigma) & \leq
\Sigma_{ii} - R_i(\Sigma) \ \text{where,} \label{eq:eigen_ineq} \\
R_i(\Sigma) & := &  \Sigma_{j=1 j \neq i}^{n}|\Sigma_{ij}|
\nonumber \eeqa

For $i=1,2,\ldots,M$, let $G_i$ be a product of $n_i$ matrices
from the set $\{F_j:j=1,2,\ldots,L\}$, and let them be labeled as
$F_{ij}, j=1,2,\ldots,n_i$. Let $F_{ij}(k, l)$ denote the
$(k,l)$th entry of the matrix $F_{ij}$. Note that each of
$F_{ij}(k, l)$ $\sim$ $\mathbb{C}\mathcal {N} ({0},1)$. Also, let
$G_i(k,l)$ denote the the $(k,l)$th entry of $G_i$. Then,

%Also, \beqan \Sigma_{ii} & = & 1 + \mid G_{ii}(f_0, f_1,\hdots,f_N)
%\mid^2 \\
%|\Sigma_{ij}| & = & \mid g_{ij}(f_0, f_1,\hdots,f_N) \mid \eeqan

\beqa \Sigma_{ii} & = & 1 + \sum_{\ell} \sum_{k}  \mid G_{\ell}(i,k)\mid^2 \label{eq:diagonal_entry}\\
|\Sigma_{ij}| & = & \mid \sum_{\ell} \sum_{k}G_{\ell}(i,k)G_{\ell}^{\dagger}(k,j) \mid \nonumber \\
& = & \mid \sum_{\ell} \sum_{k}G_{\ell}(i,k)(G_{\ell}(j,k))^{*} \mid \nonumber \\
& \leq & \sum_{\ell} \sum_{k} \mid G_{\ell}(i,k) \mid\mid
G_{\ell}(j,k) \label{eq:non_diagonal_entry}\mid\eeqa

Now every $G_{\ell}(i,j)$ is a polynomial function of
$\mathbb{C}\mathcal {N} (0,I)$ entries of $F_{\ell m}$, $m = 1, 2,
\ldots, n_{\ell}$. Define a random variable $\bold{v}$ such that
$\mid \bold{G}_{\ell}(i,j) \mid^2 \doteq \rho^{\bold{-v}}$. Now we
will prove that $\bold{v} \geq 0$ with probability one, for every
$\ell$, $i$ and $j$. Let $v$ denote a realization of the random
variable $\bold{v}$. It can be proved that polynomial functions of
independent random variables that have finite mean and variance
have finite mean and variance. Therefore $\mathbb{E} ( \mid
\bold{G}_{\ell}(i,j) \mid^2 )$ is finite.

Let the pdf of $\bold{v}$ be $p_{\bold{v}}(v)$. We have to prove
that $P(\bold{v}<0)=0$. Suppose we have proved that
$P(\bold{v}<-\frac{1}{n})=0$, for all $n \in \mathcal{N}$, then we
have:
\beqan P(\bold{v}<0) & = & P \{ \cup_{n=1}^{\infty} \{ \bold{v} < -\frac{1}{n} \} \}  \\
& \leq & \sum_{n=1}^{\infty} P(\bold{v} <-\frac{1}{n}) \label{white_eq} \\
& = & \sum_{n=1}^{\infty} 0 \\
& = & 0 \eeqan

Now we will prove that indeed $P(\bold{v}<-\frac{1}{n})=0$,
$\forall n$. Now, for any given $n$ and $\rho$,

\beqan \infty & > & \mathbb{E} ( \mid \bold{G}_{\ell}(i,j) \mid^2 ) \\
& = &  \mathbb{E} (\rho^{-\bold{v}}) \\
& = & \int_{-\infty}^{+\infty} \rho^{-{v}} p_{\bold{v}}(v) dv \\
& \geq & \int_{-\infty}^{-\frac{1}{n}} \rho^{-{v}} p_{\bold{v}}(v) dv \\
& \geq & \rho^{\frac{1}{n}} \int_{-\infty}^{-\frac{1}{n}} p_{\bold{v}}(v) dv \\
& = & \rho^{\frac{1}{n}} P({\bold{v}}<-\frac{1}{n}) \eeqan

Taking limit as $\rho$ tends to infinity on both sides \beqan
\infty & > & \lim_{\rho->\infty} \rho^{\frac{1}{n}}
P(\bold{v}<-\frac{1}{n}) \eeqan

This can only imply that $P(\bold{v}<-\frac{1}{n}) = 0$ since
otherwise, the RHS will grow to infinity as $\rho$ tends to
infinity. Hence with probability 1, \beqa \mid
\bold{G}_{\ell}(i,j) \mid^2 \doteq \rho^{-\bold{v}} \ \
\text{with} \ \ \bold{vv}
> 0. \label{eq:rho} \eeqa

By equations \eqref{eq:eigen_ineq}, \eqref{eq:rho},
\eqref{eq:non_diagonal_entry} and \eqref{eq:diagonal_entry}, it
follows that with probability one, the following equations are
true:
\beqa \lambda_i(\Sigma) & \dot \leq & 1 + \rho^{-v} \nonumber \\
& \doteq & \rho^{0} \ \ \ \forall \ i \nonumber \\
\Rightarrow  \lambda_{max} & \dot \leq & \rho^{0}
\label{eq:eigen_max} \eeqa

We now provide a lower bound for each $\lambda_i(\Sigma)$. Let
$e_i$ be the eigen vector corresponding to $\lambda_i(\Sigma)$.
Then,

\beqa {\lambda_i}\parallel{e_i}\parallel^{2} & = & {e_i}^\dagger\Sigma{e_i} \nonumber \\
 & = & {e_i}^\dagger\left(I + \sum_{i=1}^{M} G_iG_i^\dagger\right){e_i} \nonumber \\
 & = & \parallel{e_i}\parallel^2 + {e_i}^\dagger{\left(\sum_{i=1}^{M} G_iG_i^\dagger\right)}{e_i} \nonumber \\
 & \geq & \parallel{e_i}\parallel^2 \nonumber \\
\Rightarrow {\lambda_i} & \geq & 1 \ \ \forall i \nonumber \\
\Rightarrow \lambda_{min} & \geq & \rho^{0} \label{eq:eigen_lower}
\eeqa

By \eqref{eq:eigen_max}, \eqref{eq:eigen_lower}, we have that with
probability one: \beqa {\lambda_i} & \doteq & \rho^{0} \ \ \forall
i  \label{eq:sigma_dot} \eeqa

To prove the second assertion of the lemma, we use the Amir-Moez
bound on the eigen values of the product of Hermitian,
positive-definite matrices \cite{AmiMoe}. By this bound, for any
two positive definite $n \times n$ Hermitian matrices $A, B$:

\beqa \lambda_i(A)\lambda_{min}(B) \leq & \lambda_i(AB) & \leq
\lambda_i(A)\lambda_{max}(B) \nonumber \eeqa

So we get,  \beqa \det (I + \rho A B) & = &
\prod_{i}(1+\rho\lambda_i({AB}))
\nonumber \\
 & \leq & \prod_{i}(1+\rho\lambda_i(A)\lambda_{\text{max}}(B)) \nonumber \\
 & = & \det (I + \rho  \lambda_{\text{max}}(B)A )  \nonumber \eeqa

Similarly, \beqan \det (I + \rho A B) & \geq & \det (I + \rho
\lambda_{\text{min}}(B) A) \eeqan

Therefore, \beqa \det (I + \rho  \lambda_{\text{min}}(B) A) \ \leq
& \det (I + \rho A B) & \leq \ \det (I + \rho
\lambda_{\text{max}}(B) A) \label{eq:reduced_Amir_Moez} \eeqa

Applying \eqref{eq:reduced_Amir_Moez} to $A ={H}{H}^\dagger$ and
$B = \Sigma^{-1}$, we get

\beqa \Rightarrow \det(I+\rho{H}{H}^\dagger \lambda_{min}
(\Sigma^{-1}))
& \leq & \det(I+\rho{H}{H}^\dagger\Sigma^{-1}) \label{eq:amir_moez_lower_bound} \\
 & \leq & \det(I+\rho{H}{H}^\dagger \lambda_{max}
(\Sigma^{-1})) \label{eq:amir_moez_upper_bound} \eeqa

Since the eigenvalue of $\Sigma$ and of $\Sigma^{-1}$ are
reciprocals, it follows that $\lambda_{max}(\Sigma^{-1}) =
\lambda_{min}(\Sigma) \doteq \rho^{0}$, and
$\lambda_{min}(\Sigma^{-1}) = \lambda_{max}(\Sigma^{-1}) \doteq
\rho^{0}$ with probability one. Hence, we have with probability
one, \beqa \det(I+\rho{H}{H}^\dagger\Sigma^{-1}) & \doteq &
\det(I+\rho{H}{H}^\dagger) \eeqa

This proves the second assertion of the lemma.

Continuing from \eqref{eq:amir_moez_lower_bound} and
\eqref{eq:amir_moez_upper_bound}, we have

\beqa Pr\{\log(\det(I+\rho{H}{H}^\dagger \lambda_{min}
(\Sigma^{-1}))) < r\log\rho\} & \geq &
Pr\{\log\det(I+\rho{H}{H}^\dagger\Sigma^{-1})
< r\log\rho\} \nonumber \\
 & \geq &
Pr\{\log(\det(I+\rho{H}{H}^\dagger \lambda_{max} (\Sigma^{-1}))) <
r\log\rho\} \label{eq:outage_bound} \eeqa

In the following, we will prove that both the bounds coincide as
$\rho$ $\rightarrow$ $\infty$. We begin with the bounds on
$\lambda_{min}(\Sigma)$ and $\lambda_{max}(\Sigma)$. By
\eqref{eq:sigma}, we know that
\beqa \lambda_{min}(\Sigma) & \geq & 1 \nonumber \\
\lambda_{max}(\Sigma^{-1}) & \leq & 1 \nonumber \\
\text{Hence, } Pr\{\log\det(I+\rho{H}{H}^\dagger\Sigma^{-1}) <
r\log\rho\} & \geq &
Pr\{\log\det(I+\rho{H}{H}^\dagger\lambda_{max}(\Sigma^{-1})) < r\log\rho\} \nonumber \\
& \geq &  Pr\{\log\det(I+\rho{H}{H}^\dagger) < r\log\rho\}
\label{eq:outage_lower_bound}\eeqa

Now bounding $\lambda_{max}(\Sigma)$,

\beqa \lambda_{max}(\Sigma) & = & \lambda_{max}(I + \sum_{i=1}^{M} G_iG_i^\dagger) \nonumber \\
& = & 1 + \lambda_{max}(\sum_{i=1}^{M}(G_iG_i^\dagger) \nonumber \\
& \leq & 1 + Tr(\sum_{i=1}^{M}G_iG_i^\dagger) \nonumber \\
& = & 1 + \sum_{i=1}^{M}Tr(G_iG_i^\dagger) \nonumber \\
& = & 1 + \sum_{i=1}^{M}{||G_i||^2}_{F} \nonumber \\
& \leq & 1 + \sum_{i=1}^{M} \prod_{j=1}^{n_i}{||F_{ij}||^2}_{F}
\label{eq:max_lambda_bound_1} \\
& \leq &  f(u_1,u_2, \ldots, u_S) \eeqa

Now, it follows that RHS of \eqref{eq:max_lambda_bound_1} is a
multinomial in random variables $u_1, u_2, \ldots, u_S$ with
constant term 1 and non-negative integer coefficients. Here, each
$u_i$ is the squared norm of a $\mathbb{CN}(0,1)$ random variable,
and therefore has a exponentially distribution.

\beqan f(u_1,u_2, \ldots, u_S) & = & \sum_{\underline{e} \in E} c_{\underline{e}} \underline{u}^{\underline{e}} \\
\text{where }  \underline{e} & = & (e_1, e_2, \ldots, e_S) \in E
\subset {\mathbb{Z}_+}^S,  \ \ \mid E \mid \ < \infty \\
c_e & \in & \mathbb{Z}_+ \eeqan

Clearly, \beqa f(u_1,u_2, \ldots, u_S) > {\rho}^{\epsilon} &
\Rightarrow & \exists \ \underline{e} \ \text{ s.t. }
{\underline{u}}^{\underline{e}} >
\frac{{\rho}^{\epsilon}}{T_{\underline{e}}} \text{ , where }
\nonumber \\
& & T_{\underline{e}} \text{ is the number of terms in the
multinomial.} \nonumber \\
\Rightarrow Pr\{ f(u_1,u_2, \ldots, u_S)
> {\rho}^{\epsilon} \} & \leq &  Pr\left\{ \bigcup_{\underline{e}}
\left( \underline{u}^{\underline{e}} > \frac{{\rho}^{\epsilon}}
{T_{\underline{e}}} \right) \right\} \label{eq:union_multinomial}
\eeqa

Now we evaluate a single term in the RHS of
\eqref{eq:union_multinomial}. Define $T := \max_{\underline{e}}
T_{\underline{e}}$.

\beqan Pr\left( \underline{u}^{\underline{e}}
> \frac{{\rho}^{\epsilon}}{T_{\underline{e}}} \right)  & \leq & S
\ Pr \left( u_i > \frac{{\rho}^{\epsilon}}{S G T_{\underline{e}}}
\right)  \\
& \dot \leq & Pr \left( u_i > \frac{{\rho}^{\epsilon}}{S G T}
\right)  \\
& = & Pr \left( u_i > a {\rho}^{\epsilon} \right)  \\
& = & \exp(- a {\rho}^{\epsilon}) \\
\Rightarrow Pr\left( \underline{u}^{\underline{e}}
> \frac{{\rho}^{\epsilon}}{T_{\underline{e}}} \right) & \dot \leq &  \exp(- a {\rho}^{\epsilon}) \eeqan

where $G$ is the maximum degree of $f$ in any variable and $a$ is
a constant.

Continuing with \eqref{eq:union_multinomial},

\beqan Pr\{ f(u_1,u_2, \ldots, u_S)
> {\rho}^{\epsilon} \} & \dot\leq & \sum_{\underline{e}} \exp \left (-a {{\rho}^{\epsilon}} \right ) \\
& = & |E| \exp \left (-a {{\rho}^{\epsilon}} \right ) \\
& \doteq & \exp \left (-a {{\rho}^{\epsilon}} \right ) \eeqan

So we have, \beqan Pr\{ f(u_1,u_2, \ldots, u_S) > {\rho}^{\epsilon} \} & \dot \leq &  \exp(-a{\rho}^{\epsilon}) \\
\Rightarrow Pr\{ \lambda_{max}(\Sigma) > {\rho}^{\epsilon} \} &
\dot\leq & \exp(-a{\rho}^{\epsilon}) \eeqan

Let $\mathcal{H}$ denote the support of all the fading
coefficients in the network, and let $\hbar \in \mathcal{H}$
denote a realization of the fading coefficients. Clearly, once a
$\hbar$ is given, the values of the matrices $H,G_i$ and $F_i$ are
all well defined.

Let $A = \{ \hbar \in \mathcal{H} \mid \log\det( I +
{\rho}HH^{\dagger}\Sigma^{-1} ) < {\rho}^r \}$ and $B = \{ \hbar
\in \mathcal{H} \mid \lambda_{max}(\Sigma) > {\rho}^{\epsilon}
\}$. Then, \beqa
Pr(A) & = & Pr(A \cap B^c) + Pr(A \cap B) \nonumber \\
& \leq & Pr(A \cap B^c) + Pr(B) \nonumber \\
& \dot \leq & Pr(A \cap B^c) + \exp(-a\rho^\epsilon) \label{eq:prob_bound_1} \\
\text{Now, } A & \subset & \{ \hbar \in \mathcal{H} \mid \log\det(
I + {\rho}HH^{\dagger}\lambda_{min}(\Sigma^{-1}) ) < {\rho}^r \} \nonumber \\
& = & \{ \hbar \in \mathcal{H} \mid \log\det( I +
{\rho}HH^{\dagger}{(\lambda_{max}(\Sigma))}^{-1} ) < {\rho}^r \} \nonumber \\
A \cap B^c & \subset & \{ \hbar \in \mathcal{H} \mid \log\det( I +
{\rho}^{1-\epsilon}HH^{\dagger} ) < {\rho}^r \}
\label{eq:prob_bound_2} \eeqa

\beqa
\frac{\log Pr(A)}{\log\rho} & \leq & \frac{\log\{ Pr(A \cap B^c) + Pr(B) \}}{\log\rho} \nonumber \\
& \dot \leq & \frac{\log\{ Pr(\hbar \in \mathcal{H} \mid \log\det(
I + {\rho}^{1-\epsilon}HH^{\dagger} ) < {\rho}^r)
+ \exp(-{\rho}^{\epsilon}) \}}{\log\rho} \nonumber  \\
\lim_{\rho\rightarrow\infty} \frac{\log Pr(A)}{\log\rho} & \dot
\leq & \lim_{\rho\rightarrow\infty} \frac{\log\{ Pr(\hbar \in
\mathcal{H} \mid \log\det( I + {\rho}^{1-\epsilon}HH^{\dagger} ) <
{\rho}^r) \}}{\log\rho} \eeqa

The last equation follows since the first term in the RHS is
polynomial in $\rho$ whereas the second term is exponential and
therefore the sum is dominated by the first term.

After doing the variable change, $\rho^{'} = \rho^{1-\epsilon}$
and using the variable $\rho$ itself in place of $\rho^{'}$, \beqa
\lim_{\rho\rightarrow\infty} \frac{\log Pr(A)}{\log\rho} & \dot
\leq & (1-\epsilon) \lim_{\rho\rightarrow\infty} \frac{\log\{
Pr(\hbar \in \mathcal{H} \mid \log\det( I + {\rho}HH^{\dagger} ) <
{\rho}^{(\frac{r}{1-\epsilon})} \}}{\log\rho}
\label{eq:outage_upper_bound} \eeqa

In \eqref{eq:outage_upper_bound}, $\epsilon$ is arbitrary, and we
tend it to zero. Hence, by \eqref{eq:outage_upper_bound} and
\eqref{eq:outage_lower_bound}, the exponents for both the bounds
in \eqref{eq:outage_bound} coincide and hence we get, \beqa
Pr\{\log\det(I+\rho{H}{H}^\dagger\Sigma^{-1}) < r\log\rho\} &
\doteq & Pr\{\log\det(I+\rho{H}{H}^\dagger) < r\log\rho\}
\nonumber \eeqa

This proves the third assertion of the lemma. \epf

\blem \label{lem:signal_white} \cite{ZheTse} For any channel that
is of the form $y = \bold{H}x + \bold{w}$ with $w$ being white
gaussian noise, i.i.d. gaussian inputs are sufficient to attain
the best possible outage exponent of the channel.\elem

\bpf Proof is available in \cite{ZheTse}. We sketch the outline of
the same proof for completeness. The outage probability is given
by,

\beqan P_{\text{out}}(R) & = & \inf_{\Sigma_x: \ Tr(\Sigma_x) \leq
\mathbb{P}} Pr\{I(\bold{x};\bold{y} \mid \bold{H}
= H) \leq R\} \\
& = & \inf_{\Sigma_x: \ Tr(\Sigma_x) \leq \mathbb{P}} Pr\{\log\det(I+\rho{H}\Sigma_x{H}^\dagger) \leq R\} \\
\eeqan If $x$, $y$ $\in$ $\mathbb{C}^m$, then the outage
probability can be bounded below and above as,

\beqan \lefteqn{Pr\{\log\det(I+\frac{\rho}{m}{H}{H}^\dagger) \leq R\}} \\
  & \geq & P_{\text{out}}(R) \geq Pr\{\log\det(I+ \rho{H}{H}^\dagger) \leq R\} \eeqan
As $\rho$ $\rightarrow$ $\infty$, it can be shown that the bounds
are tight and hence we get (Equation (9) in \cite{ZheTse}), \beqa
P_{\text{out}}(R) & \doteq & P(\log\det(I+\rho HH^\dagger)<R)
\eeqa \epf

\bnote \label{rem:signal_white} Because of
Lemma~\ref{lem:signal_white}, it is sufficient to consider i.i.d.
gaussian input distribution for characterizing the outage
exponent.  Also, for characterizing outage exponent, we are
allowed to assume that the noise is white in the scale of interest
(see Lemma~\ref{lem:noise_white}). It can be verified that noise
that we deal with in this paper is always satisfies the conditions
in Lemma~\ref{lem:noise_white}. Hence we will make these two
assumptions throughout the paper \bit \item Signal is distributed
as i.i.d gaussian. \item Noise is white in the scale of interest.
\eit \enote

\subsection{A DMT Lower Bound}

\bdefn \label{defn:Block_Lower_Triangular}Consider a set of $N_i
\times N_j$ matrices $A_{ij},j=1,2,...,N, i \geq j$. Let $A$ be a
matrix comprised of the block matrices $A_{ij}$ in the $(i,j)$th
position, i.e., \beqan A & = & \left[\begin{array}{cccc}
    A_{11} & 0 & \ldots & 0\\
    A_{21} & A_{22} & \ldots & 0\\
    \vdots & & \ddots &\vdots\\
    A_{N1} & A_{N2} & \ldots & A_{NN} \\
    \end{array}\right].
\eeqan We will call $A$ as a block lower-triangular matrix. Define
the $l$-th sub-diagonal matrix, $A_{\ell}$ of a block lower
triangular matrix $A$ as the block lower triangular matrix
comprising of entries
$A_{{\ell}1},A_{({\ell}+1)2},...,A_{({\ell}+N-1)N}$ and zeros
everywhere else i.e.,

\beqa (A_{\ell})_{ij} & = & A_{ij} \ \text{ if } \ i-j=l-1, \
\text{ else } \ (A_{\ell})_{ij}=0_{N_i \times N_j}. \eeqa The last
sub-diagonal matrix of $A$ is defined as the sub-diagonal matrix
$A_{\ell}$ of $A$, with the maximum $l$ such that $A_{\ell}$ is a
\emph{non-zero} matrix. \edefn

\bthm \label{thm:main_theorem} Consider a block lower triangular
random matrix $\bold{H}$ made of matrices $\bold{H_{ij}}$ of size
$N_i \times N_j$. Let $M := \sum_{i=1}^{N} N_i$ be the size of the
square matrix $\bold{H}$. Consider a channel of the form $\bold{y
= Hx + w}$, where $\bold{H}$ is the $M \times M$ block lower
triangular random matrix, $\bold{x}, \bold{y}, \bold{w}$ are $M
\times 1$ vectors. Let $\bold{w}$ be a noise vector, which is
white in the scale of interest. Let $\bold{x_i}, \bold{y_i},
\bold{w_i}$ be vectors of length $N_i$ such that $\bold{x} =
[\bold{x_1}, \bold{x_2},\ldots,\bold{x_N}]^{T}$, $\bold{y} =
[\bold{y_1}, \bold{y_2},\ldots, \bold{y_N}]^{T}$ and $\bold{w} =
[\bold{w_1}, \bold{w_2},\ldots, \bold{w_N}]^{T}$.

Let $\bold{H_d}$ be the block-diagonal part of the matrix
$\bold{H}$ and $\bold{H_{\ell}}$ denote the last sub-diagonal
matrix of $\bold{H}$, as per
Definition~\ref{defn:Block_Lower_Triangular}. Then

\ben \item $d_H(r) \geq d_{H_d}(r)$. \item $d_H(r) \geq
d_{H_{\ell}}(r)$.

\item In addition, if the entries of $H_{\ell}$ are independent of
the entries in $H_d$, then $d_H(r) \geq d_{H_d}(r) +
d_{H_{\ell}}(r)$ \een \ethm

\bpf The channel is given by $\bold{y} = \bold{H}\bold{x} +
\bold{w}$. Since the noise is white in the scale of interest, by
Lemma~\ref{lem:noise_white}, the DMT of this channel is the same
as that of a channel with the noise distributed as
$\mathbb{CN}(0,I)$. Therefore, without loss of generality, we
assume that $\bold{w}$ is distributed as $\mathbb{CN}(0,I)$.

%Let us first consider the case when the noise is white, i.e., $W$ is distributed as
%$\mathcal{CN}(0,I)$.

We have the block-diagonal part of $\bold{H}$, $\bold{H_d} = diag
\{ \bold{H_{11}},\bold{H_{22}},\ldots,\bold{H_{NN}} \}$ and the
last sub-diagonal matrix $H_{\ell}$ contains $N-l+1$ non-zero
entries $\{ H_{l1},H_{(l+1)2},...,H_{N(N-l+1)} \}$ in the $l$-th
sub-diagonal.

The outage probability exponent\cite{ZheTse} is given by \beqan
\rho^{-d(r)} & \doteq & \inf_{\Sigma_x: Tr\Sigma_x \leq
\mathbb{P}} Pr\{I(\bold{x};\bold{y}:\bold{H} = H) \leq r\log\rho\}
\label{eq:outage_exponent} \eeqan In order to evaluate this
exponent, we first evaluate the mutual information. Let us assume
that the input $x$ is distributed as $\mathcal{CN}(0,I)$. By
Lemma~\ref{lem:signal_white}, this input distribution is indeed
DMT optimal. We will compute the mutual information terms under
this assumption that the inputs are iid gaussian.

\begin{figure}[h!]
\centering
\includegraphics[height=50mm]{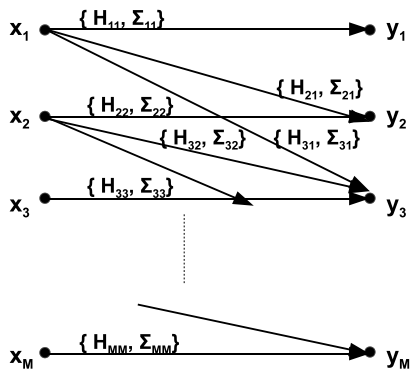}
\caption{The i-f diagram for the block lower triangular channel
matrix.\label{fig:if_block_lt}}
\end{figure}

See Fig.\ref{fig:if_block_lt} for the i-f diagram. Now, we proceed
to find a lower bound on the DMT of the protocol.

Consider the following series of inequalities for all $i=1,...,N$.
\beqa I(\bold{x_i};\bold{y} | \bold{H} = H, \bold{x_1^{i-1}}) & \geq & I(\bold{x_i};\bold{y_i} | \bold{H} = H, \bold{x_1^{i-1}}) \nonumber \\
& = & I(\bold{x_i}; \bold{H_{ii} x_i + H_{i(i-1)} x_{i-1} + ... + H_{i(i-{\ell})} x_{i-{\ell}} +  w_i} | \bold{H} = H, \bold{x_1^{i-1}}) \nonumber \\
& = & I(\bold{x_i}; H_{ii} \bold{ x_i} + H_{i(i-1)} \bold{x_{i-1}} + ... + H_{i(i-{\ell})} \bold{x_{i-{\ell}} +  w_i} | \bold{H} = H, \bold{x_1^{i-1}}) \nonumber \\
& = & I(\bold{x_i}; H_{ii} \bold{ x_i} + H_{i(i-1)} \bold{x_{i-1}} + ... + H_{i(i-{\ell})} \bold{x_{i-{\ell}} +  w_i} | \bold{x_1^{i-1}}) \nonumber \\
& = & I(\bold{x_i}; H_{ii} \bold{ x_i  +  w_i} | \bold{x_1^{i-1}}) \nonumber \\
& = & I(\bold{x_i}; H_{ii} \bold{ x_i  +  w_i} ) \nonumber \eeqa
The last step follows since $\{ \bold{ x_i } \}$ are independent.

\beqa \Rightarrow I(\bold{x};\bold{y} |\bold{H}=H) & = &  \sum_{i=1}^{M} I(\bold{x_i};\bold{y} | \bold{H} = H, \bold{x_1^{i-1}} ) \nonumber \\
& \geq & \sum_{i=1}^{M} I(\bold{x_i}; H_{ii} \bold{ x_i  +  w_i}) \nonumber \\
& \geq & I(\bold{x};  \bold{H_dx + w} | \bold{H_d}=H_d)
\label{eq:main_diagonal_bound} \eeqa

In the above, whenever the index of a variable is not positive, we
assume that the variable is not present in the conditioning, in
order to simplify the notation.

Now by equation \eqref{eq:outage_exponent}, \beqa \rho^{-d_H(r)} & = &  Pr\{I(\bold{x};\bold{y}|\bold{H} = H) \leq r\log\rho\} \\
& \leq & Pr\{ I(\bold{x};  \bold{H_dx + w} |\bold{H_d} = H_d)  \leq r\log\rho \}\nonumber \\
& = & \rho^{-d_{H_d}(r)} \nonumber \\
d_H(r) & \geq & d_{H_d}(r) \eeqa

We have another series of inequalities for all $i=1,...,M-1$.

\beqa I(\bold{x_{i-{\ell}}};\bold{y} | \bold{H} = H, \bold{x_{i-{\ell}+1}^N}) & \geq & I(\bold{x_{i-{\ell}}};\bold{y_{i}} | \bold{H} = H, \bold{x_{i-{\ell}+1}^N}) \nonumber \\
& = & I(\bold{x_{i-{\ell}}}; \bold{H_{ii} x_i + H_{i(i-1)} x_{i-1} + ... + H_{i(i-{\ell})} x_{i-{\ell}} +  w_{i}}  | \bold{H} = H, \bold{x_{i-{\ell}+1}^N}) \nonumber \\
& = & I(\bold{x_{i-{\ell}}}; H_{ii} \bold{ x_i} + H_{i(i-1)} \bold{ x_{i-1}} + ... + H_{i(i-{\ell})} \bold{ x_{i-{\ell}} +  w_{i}}  | \bold{H} = H, \bold{x_{i-{\ell}+1}^N}) \nonumber \\
& = & I(\bold{x_{i-{\ell}}}; H_{ii} \bold{ x_i} + H_{i(i-1)} \bold{ x_{i-1}} + ... + H_{i(i-{\ell})} \bold{ x_{i-{\ell}} +  w_{i}}  | \bold{x_{i-{\ell}+1}^N}) \nonumber \\
& = & I(\bold{x_{i-{\ell}}}; H_{i(i-{\ell})} \bold{ x_{i-{\ell}} +  w_{i}}  | \bold{x_{i-{\ell}+1}^N}) \nonumber \\
& = & I(\bold{x_{i-{\ell}}}; H_{i(i-{\ell})} \bold{ x_{i-{\ell}} +  w_{i}}) \nonumber \\
\Rightarrow I(\bold{x};\bold{y} |\bold{H}=H) & = &  \sum_{i=N}^{1} I(\bold{x_i};\bold{y} | \bold{H} = H, \bold{x_{i+1}^N} ) \nonumber \\
& \geq &  \sum_{i=N}^{l+1} I(\bold{x_{i-{\ell}}};\bold{y} | \bold{H} = H, \bold{x_{i-{\ell}+1}^N} ) \nonumber \\
& \geq &  \sum_{i=N}^{l+1} I(\bold{x_{i-{\ell}}}; H_{i(i-{\ell})} \bold{ x_{i-{\ell}} +  w_{i}})\nonumber \\
& = & I(\bold{x}; \bold{H_{\ell}x + w} | \bold{H_{\ell}}=H_{\ell})
\label{eq:sub_diagonal_bound} \eeqa

Now by equation \eqref{eq:outage_exponent}, \beqa \rho^{-d_H(r)} & = &  Pr\{I(\bold{x};\bold{y}|\bold{H} = H) \leq r\log\rho\} \\
& \leq & Pr\{ I(\bold{x};  \bold{H_{\ell}x + w} |\bold{H_{\ell}} = H_{\ell})  \leq r\log\rho \}\nonumber \\
& = & \rho^{-d_{H_d}(r)} \nonumber \\
d_H(r) & \geq & d_{H_d}(r) \eeqa

Therefore, \beqa I(\bold{x};\bold{y} |\bold{H}=H) & \geq &
\max(I(\bold{x};  \bold{H_dx + w} | \bold{H_d}=H_d), I(\bold{x};
\bold{H_{\ell}x + w} | \bold{H_{\ell}}=H_{\ell}))
\label{eq:max_bound}\eeqa The outage probability
exponent\cite{ZheTse} is given by \beqan
 \rho^{-d(r)} & \doteq &
\inf_{\Sigma_x: Tr\Sigma_x \leq \mathbb{P}}
Pr\{I(\bold{x};\bold{y} \mid \bold{H} = H) \leq r\log\rho\} \eeqan

Now by equation \eqref{eq:max_bound},

\beqa \rho^{-d_H(r)} & = &  Pr\{I(\bold{x};\bold{y}:\bold{H} = H) \leq r\log\rho\} \\
& \leq & Pr\{ \max(I(\bold{x};  \bold{H_dx  +  w} | \bold{H_d}=H_d), I(\bold{x}; \bold{H_{\ell}x + w} | \bold{H_{\ell}}=H_{\ell})) \leq r\log\rho \}\nonumber \\
& = & Pr\{ I(\bold{x};  \bold{H_dx  +  w} | \bold{H_d}=H_d)\leq r\log\rho, \nonumber\\
&& I(\bold{x}; \bold{H_{\ell}x + w} | \bold{H_{\ell}}=H_{\ell})) \leq r\log\rho \}\nonumber \\
& = & Pr\{ I(\bold{x};  \bold{H_dx  +  w} | \bold{H_d}=H_d)\leq r\log\rho \} \nonumber\\
& & \times Pr\{ I(\bold{x};  \bold{H_{\ell}x  +  w} | \bold{H_{\ell}}=H_{\ell})\leq r\log\rho \} \nonumber\\
& = & \rho^{-d_{H_d}(r)} \rho^{-d_{H_{\ell}}(r)} \nonumber \\
\ & = & \rho^{-d_{H_d}(r)+d_{H_{\ell}}(r)}  \nonumber \\
d_H(r) & \geq & d_{H_d}(r)+d_{H_{\ell}}(r) \eeqa

where the first step comes about because of the independence of
the entries in $\bold{H_d}$ and $\bold{H_{\ell}}$, which is indeed
the case because of the assumption that all the fading
coefficients in the system are independent. The second step is
because iid complex gaussian inputs are optimal in the scale of
interest.

\epf

\bcor \label{cor:Upper_Triangular} Theorem~\ref{thm:main_theorem}
holds even for the case when the matrix $H$ is block
upper-triangular instead of block lower-triangular. \ecor

\bpf Follows from the proof of Theorem~\ref{thm:main_theorem}
since the DMT of a matrix $H$ and its transpose $H^T$ are the
same. \epf

\bnote \label{rem:Matrix_Inequality} The following two matrix
inequalities can be deduced from the proof of
Theorem~\ref{thm:main_theorem}, with $H_d$ and $H_{\ell}$ defined
as in the theorem:

\beqan \det ( I  + \rho H H^\dagger ) & \geq &  \det ( I  + \rho
H_d H_d^\dagger ) \\
\text{and } \det ( I  + \rho H H^\dagger ) & \geq &  \det ( I  +
\rho H_{\ell} H_{\ell}^\dagger ) \eeqan

\enote

\bnote \label{rem:Diagonal_Matrix_DMT} The DMT of a matrix $H$ is
greater than or equal to the DMT of the block diagonal matrix
$H_d$. This bound will be most frequently used whenever we recall
Theorem~\ref{thm:main_theorem} \enote

\subsection{Example Applications of the Main Theorem}
\label{sec:examples_main_thm} In this section, we recover lower
bounds on DMT of various existing amplify and forward protocols.
While these are already known, the derivations presented here are
surprisingly simple and they lead to intuitive explanation of how
these protocols achieve the DMT.

\vspace{0.05in}

\emph{Example 1: Single Source, Single Sink, Single relay, NAF
protocol}

\vspace{0.05in}

Consider the relay network in Fig.\ref{fig:one_relay}, considered
in Section \ref{sec:info_flow_diagram}. The i-f diagram is given
in Fig.\ref{fig:flow_model}. \beq
    \bold{y} = \bold{H} \bold{x} + \bold{n} , \label{eqref:naf_model}
\eeq where \beqan
    \bold{H} & = & \left[ \begin{array}{cc} \bold{g_1} & 0\\
     \bold{g_2h_2}  & \bold{g_1} \end{array} \right] \\
    \bold{n} & = & \left[ \begin{array}{c} \bold{w}_1 \\
    \bold{w}_2  \  +  \ h_2  \bold{v}
    \end{array} \right]
\eeqan

Since two time instants are used in order to obtain the equivalent
channel matrix, we have a rate loss by a factor of 2, and hence
$d(r) = d_H(2r)$. It can be checked that the noise vector
$\bold{n}$ satisfies the conditions in Lemma~\ref{lem:noise_white}
and therefore is white in the scale of interest. Now it is
sufficient to study the DMT of the matrix $H$. Let $H_d = H
\bigotimes I$, where $ \bigotimes $ denotes the Hadamard product
(entry-wise product) of matrices. Let $H_{\ell}$ denote the matrix
with only the lower triangular entry and set all other entries to
zero, i.e.,
\beqan \bold{H_d} & := & \left[ \begin{array}{cc}  \bold{g_1} & 0\\
     0  & \bold{g_1} \end{array} \right] \\
      \bold{H_{\ell}} & := & \left[ \begin{array}{cc} 0 & 0\\
     \bold{g_2h_2}  & 0 \end{array} \right] \eeqan

The fading coefficients $\bold{g_1,g_2,h_2}$ are independent and
therefore $\bold{H_d}$ is independent of $\bold{H_{\ell}}$. We use
Theorem~\ref{thm:main_theorem} and we get that:

\beqan d_{H}(r) & \geq & d_{H_d}(r) + d_{H_{\ell}}(r) \eeqan

It is easy to evaluate $d_{H_d}(r)$ and $d_{H_{\ell}}(r)$:

\beqan d_{H_d}(r) & = & (1 - \frac{r}{2})^{+} \\
d_{H_{\ell}}(r) & = & (1 - {r})^{+}\\
\Rightarrow d_H(r) & \geq & (1 - \frac{r}{2})^{+} + (1 - {r})^{+}
\eeqan

We can get the DMT of the protocol as
\beqan  d(r) & = & d_H(2r)\\
\Rightarrow d(r) & \geq & (1 - {r})^{+} + (1 - 2r)^{+} \eeqan

From \cite{AzaGamSch} we know that this bound is indeed tight.
However, we will not proceed to find an upper-bound here.

\vspace{0.05in}

\emph{Example 2: Single source, Single sink, Multiple relays, SAF}

\vspace{0.05in}

Consider the network in Fig.\ref{fig:classical_relay} with $N$
relays. We employ an M-slot amplify-and-forward protocol termed
Slotted Amplify-and-Forward (SAF) introduced in \cite{YanBelSaf}.
Each of symbols transmitted by the source reach the sink through
the direct link, and through a relayed path. For the case when
relays are isolated from each other (see \cite{YanBelSaf} for a
description), the induced channel matrix for a $M$ slot protocol
is given by a $M \times M$ channel matrix, with $g_d$, the fading
coefficient of the direct link, along the diagonal and $g_1,
\ldots, g_N$, the product coefficients on relay paths, repeating
cyclically along the second sub-diagonal. Let $M = kN + 1$ be the
slot length, with $k$ a positive integer.

For example, for $M=5$, $N=2$, $k=2$ case, the induced channel
matrix is given by: \beqan \bold{H} & := & \left[
\begin{array}{ccccc}
\bold{g_d} & 0 & 0 & 0 & 0 \\
\bold{g_1} & \bold{g_d} & 0 & 0 & 0  \\
0 & \bold{g_2} & \bold{g_d} & 0 & 0  \\
0 & 0 & \bold{g_1} & \bold{g_d} & 0 \\
0 & 0 & 0 & \bold{g_2} & \bold{g_d}  \\ \end{array} \right] \eeqan

\begin{figure}[h!]
\centering
\includegraphics[height=60mm]{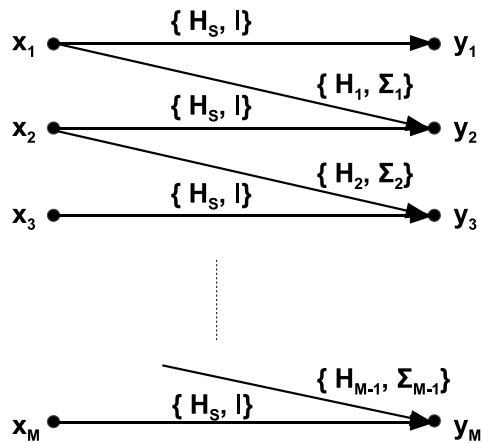}
\caption{The i-f diagram for M-slot SAF protocol.
\label{fig:if_saf_fixed_delay}}
\end{figure}

See Fig.\ref{fig:if_saf_fixed_delay} for the i-f diagram, where
$H_d := g_d, H_i := g_{(i-1 \mod 2) + 1}, \Sigma_i = 1 + |f_i|^2$.
Since the channel is used for $M$ time slots, we have the relation
$d(r) = d_H(Mr)$ between the DMT of the protocol, $d(r)$, and the
DMT of the matrix $d_H(r)$ . Now, we proceed to find a lower bound
on the DMT of the matrix.

Let $\bold{H_d} = \bold{g_d}I$ be the diagonal matrix
corresponding to $H$. Let $H_{\ell}$ be the second sub-diagonal
matrix corresponding to $H$. It contains
$\bold{g_1},...,\bold{g_M}$ each for $k$ times in the second
sub-diagonal. From Theorem~\ref{thm:main_theorem}, the DMT of
$\bold{H}$ can be lower bounded as: \beqa
d_H(r) & \geq & d_{H_d}(r) + d_{H_{\ell}}(r) \\
\text{ We already have  } \ \ \ \ d(r) & = & d_{H}(Mr)\\
\Rightarrow d(r) & \geq & d_{H_d}(Mr) + d_{H_{\ell}}(Mr) \eeqa

Now the DMT of the matrices $\bold{H_d}$ and $\bold{H_{\ell}}$
can be easily derived as: $ d_{H_d}(r) = (1 - \frac{r}{M})^{+} $
and $ d_{H_{\ell}}(r) = N(1 - \frac{r}{M-1})^{+}$

\beqa d(r) & \geq & (1-r)^{+} + N(1 - \frac{M}{M-1}r)^{+} \eeqa

The right hand side is infact shown to be equal to the DMT of the SAF
protocol in \cite{YanBelSaf}.

\vspace{0.05in} \emph{Example 3: Single Source, Single Sink,
Multiple Antenna, Single relay, NAF protocol} \vspace{0.05in}

Let us first consider a single relay network with the source, the
relay and sink equipped with multiple antennas $n_s$, $n_r$,
$n_d$. Let us use the NAF protocol \cite{AzaGamSch} in this
scenario, as is done in \cite{YanBelMimoAf}. The channel matrix
turns out to be

\beqa \bold{H} & := & \left[ \begin{array}{cc}  \bold{H_d} & 0\\
     \bold{H_{\ell}}  & \bold{H_d} \end{array} \right] \label{eq:NAF}\eeqa

where $\bold{H_d}$ is the $n_d \times n_s$ fading matrix between
source and the sink, $\bold{H_{\ell}}$ is the product fading
matrix of an $n_r \times n_s$ matrix between the source and the
relay and an $n_d \times n_r$ matrix between relay and sink.
Proceeding in the same manner as in \emph{Example 1}, we can get
that $d(r) \geq d_{H_d}(r) + d_{H_{\ell}}(2r)$, where $d_{H_d}(r)$
is the DMT of the direct link matrix $H_d$, and $d_{H_{\ell}}(r)$
is the DMT of the product matrix $H_{\ell}$. This lower bound was
derived as Theorem 1 of \cite{YanBelMimoAf}.

Let us now consider a generalized NAF protocol (see \cite{RaoHas})
where, for the first $T$ time instants, the source transmits to
the relays and then the relays transmit a linear transformation of
the received vector over the $T$ time instants. Even in this case,
the input output transformation can be represented using a
equation of the form \eqref{eq:NAF}. However $H$ is now a $2Tn_d
\times 2Tn_s$ matrix, $H_{d}$ is a $Tn_s \times Tn_d$ block
diagonal matrix with the direct link fading matrix repeated $T$
times and $H_{\ell}$ is any $Tn_d \times Tn_s$ matrix (which
depends on the linear transformations used at the relays) relating
the inputs to the output at the sink due to the relaying path. Let
$d_C(r) := d_{H_{\ell}}(Tr)$ denote the DMT of the same scheme
used without the direct link and with full duplex relays. Let
$d_D(r) := d_{H_d}(Tr)$ denote the DMT of the direct path fading
matrix.

Then Theorem~\ref{thm:main_theorem} can be used to get the
following inequality for the DMT of this generalized NAF scheme:
\beqan d(r) \geq d_{D}(r) + d_C(2r) \eeqan

This proves Conjecture $1$ of \cite{RaoHas}.

\vspace{0.05in} \emph{Example 4: Single Source, Single Sink,
Multiple Antenna, Multiple relays, NAF protocol} \vspace{0.05in}

In \cite{YanBelMimoAf}, the authors consider a two-hop relay
network with a direct link and $N$ relays. Consider the NAF
protocol for the $N$ relay case suggested in \cite{YanBelMimoAf}
in which each path is used for equal duration. Here we consider a
general version of the NAF Protocol, where different relaying
paths are activated for different fractions of time. Let the
relaying path through relay $i$ be used for $f_i$ fraction of the
time. For this protocol, let us derive the DMT. The matrix
connecting the input and the output is a block lower-triangular
matrix with the direct-link fading matrix $\bold{H_d}$ repeated on
the block-diagonal. The second sub-diagonal contains entries
matrices $R_1, R_2, \ldots, R_N$, where $R_i$ is the product
matrix along the $i$th relay. We can bound the DMT of resulting
matrix using Theorem~\ref{thm:main_theorem}:

\beqa d(r) \geq d_{H_d}(r) + d_C(2r) \label{eq:N_Relay_NAF} \eeqa

where $d_C(r)$ is the DMT of a parallel channel with entries $R_i$
occurring for a fraction $f_i$ of the time. We can evaluate
$d_C(r)$ explicitly from the DMT $d_i(r)$ of the product channel
$R_i$.

The DMT of this channel can be computed using the parallel channel
formula given in equation
Equation~\eqref{eq:parallel_repeated_coeffs} in
Lemma~\ref{lem:parallel_correlated_channel} and it is given
by,\beqa d_{C}(r) & = & \sup_{(f_1,f_2,\cdots,f_K)}
\inf_{(r_1,r_2,\cdots,r_K): \ \sum_{i=1}^{K} \ f_i r_i = r }\
\sum_{i=1}^{K} {d_i(r_i)}  \eeqa where $d_i(r)$ is the DMT of the
product channel in the $i$th channel and corresponds to the DMT of
the product matrix $G_iH_i$.

Therefore the overall DMT is given by \beqa d(r) & \geq &
d_{H_d}(r) + \sup_{(f_1,f_2,\cdots,f_K)} \ \
\inf_{(r_1,r_2,\cdots,r_K): \ \sum_{i=1}^{K} \ f_i r_i = r }\
\sum_{i=1}^{K} {d_i(r_i)} \label{eq:N_Relay_Better_Bound} \eeqa

As a particular choice, if $f_i = 1/N$ for all $i$, then \beqa
d_{C}(r) & = & \inf_{(r_1,r_2,\cdots,r_K): \ \sum_{i=1}^{K} \ r_i
= N r }\ \sum_{i=1}^{K} {d_i(r_i)} \eeqa Let $\theta_i :=
\frac{r_i}{Nr}$. Then we have

\beqa d_{C}(r) & = &  \inf_{(\theta_1,\theta_2,\cdots,\theta_K): \
\sum_{i=1}^{K} \ \theta_i = 1 }\ \sum_{i=1}^{K} {d_i( N \theta_i
r)}  \eeqa

We plug this equation into \eqref{eq:N_Relay_NAF} and get \beqa
d(r) & \geq & d_{H_d}(r) +
\inf_{(\theta_1,\theta_2,\cdots,\theta_K): \ \sum_{i=1}^{K} \
\theta_i = 1 }\ \sum_{i=1}^{K} {d_i( 2 N \theta_i r)} \eeqa which
is indeed the formula in \emph{Theorem 2} of \cite{YanBelMimoAf}.
However the lower bound on DMT that we have in
Equation~\eqref{eq:N_Relay_Better_Bound} is better than the lower
bound in \emph{Theorem 2} of \cite{YanBelMimoAf} since we allow
for arbitrary periods of activation which is a more general
approach.

\bnote In the notation of \cite{YanBelMimoAf}, $d_{H_d}(r) =
d_F(r)$ since $F$ is the matrix of transformation between source
and sink through the direct link. Also $G_i$ is the matrix between
source to relay $i$ and $H_i$ matrix between relay $i$ to sink.
According to notation of \cite{YanBelMimoAf}, $d_{G_iH_i}(r)$ is
the DMT corresponding to the product matrix $G_iH_i$. \enote

\subsection{DMT of elementary network connections\label{sec:elementary_nw}}

\subsubsection{Parallel Network \label{sec:parallel} }

\blem \label{lem:parallel_channel} Consider a parallel channel
with $M$ links, the each link being represented by $y_i =
\bold{H_i}x_i + \bold{w_i}$, and let the optimal DMT of the $i$th
link be $d_i(.)$. Then the optimal DMT of the parallel channel is
given by \beq d(r) = \inf_{(r_1,r_2,\cdots,r_M): \ \sum_{i=1}^{M}
r_i = r} \ \sum_{i=1}^{M} {d_i(r_i)} \label{eq:parallel_dmt} \eeq
\elem

\vspace{0.1in} \bpf The input-output relation of the parallel
channel is given by

\beq \left[ \begin{array}{c}
\bold{y_1}\\
\bold{y_2}\\
\vdots\\
\bold{y_M}\\
\end{array} \right]
 =  \left[\begin{array}{cccc}
    \bold{H_1} & & & \\
     & \bold{H_2} & &\\
     &  & \ddots &\\
     &  &  & \bold{H_M} \\
    \end{array}\right]
    \left[ \begin{array}{c}
             \bold{x_1}\\
             \bold{x_2}\\
             \vdots\\
             \bold{x_M}\\
            \end{array} \right] + \bold{n} \label{eq:parallel_relation}
\eeq

\begin{figure}[h!]
\centering
\includegraphics[height=50mm]{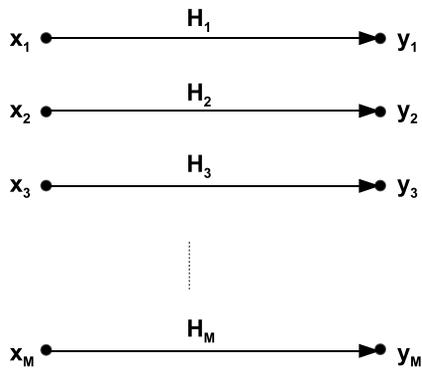}
\caption{The Parallel channel with M
sub-channels\label{fig:parallel_network}}
\end{figure}

\beqa {I(\bold{x};\bold{y} | \bold{H} = H )} & = & h(\bold{y} | \bold{H} = H )- \sum_{i=1}^{M} h(\bold{y_i} | \bold{y_1^{i-1}}, \bold{x}, \bold{H} = H ) \nonumber \\
& = & h(\bold{y} | \bold{H} = H )- \sum_{i=1}^{M} h(\bold{y_i} |
\bold{x_{i}} ,
\bold{H} = H ) \nonumber \\
& \leq & \sum_{i=1}^{M} h(\bold{y_i} | \bold{H} = H )-
\sum_{i=1}^{M} h(\bold{y_i} |
\bold{x_{i}} , \bold{H} = H ) \nonumber \\
& = & \sum_{i=1}^{M} [ h(\bold{y_i} | \bold{H} = H )-
h(\bold{y_i} | \bold{x_{i}} ,
\bold{H} = H ) ] \nonumber \\
& = & \sum_{i=1}^{M} I(\bold{x_i};\bold{y_i} | \bold{H} = H ) \nonumber \\
& = & \sum_{i=1}^{M} I(\bold{x_i};\bold{y_i} | \bold{H_i} = H_i )
\eeqa

\beqan \Rightarrow Pr\{I(\bold{x};\bold{y} \mid \bold{H} = H) \leq
r\log\rho\} & \leq & Pr\{ \sum_{i=1}^{M} I(\bold{x_i};\bold{y_i} |
\bold{H_i} = H_i ) \leq r\log\rho \} \eeqan

The equality in the last equation occurs if all the $\bold{x_i}$
are independent. So we will choose the $\bold{x_i}$ to
independent, for the rest of the discussion, since this maximizes
the mutual information and hence minimizes the error probability.
Define $\bold{Z_i} := I(\bold{x_i};\bold{y_i} | \bold{H_i} = H_i
)$. Now $\bold{Z_i}$ is a random variable which depends on the
realization of the channel. Since $\{ \bold{H_i} \}$ are
independent, $ \{ \bold{Z_i} \} $ are also independent. Let $R_i =
r_i \ log(\rho)$ and $R = r \ log(\rho)$ for $i=1,2$.

Now our goal is to evaluate $P \{ \ \sum_{i=1}^{M} \bold{Z_i} \leq
r log(\rho) \ \}$. To do this, first we consider the case when
$M=2$ and we evaluate $P \{ \ \bold{Z_1 + Z_2} \leq r log(\rho) \
\}$. Then we extend this to general $M$ by induction.
\beqan F_{Z_i} (R_i) & := & P \{\bold{Z_i} < R_i \} \\
f_{Z_i} (R_i) & := & \frac{d}{d{R_i}} F_{Z_i} (R_i) \\
\text{Let } F_{Z_i} (R_i) &  \doteq & \rho^{-d_i(r_i)} \\
\text{Then } f_{Z_i} (R_i) &  \doteq & \frac{d}{d{r_i \ log(\rho)}} \rho^{-d_i(r_i)}\\
&  \doteq  & \rho^{-d_1(r_1)} \frac{d}{d{r_i}} d_i(r_i) \\
&  \doteq  & \rho^{-d_1(r_1)} \eeqan

\beqan P(\bold{Z_1+Z_2} \leq R) & = & \rho^{-d(r)}\\
 & = & \int_{0}^{\infty} \ {f_{Z_1} (R_1) F_{Z_2} (R-R_1) d{R_1} } \\
 & \doteq & \int_{0}^{\infty} \ \rho^{-d_1(r_1)} \ \rho^{-d_2(r-r_1)} ln (\rho) d(r_1) \eeqan

By Varadhan's Lemma\cite{DemZei}, the SNR exponent integral can be
evaluated in the scale of interest as:

\beqan d(r) & = & \inf_{r_1 \geq 0} {d_1(r_1)} + d_2(r-r_1) \\
& = & \inf_{(r_1,r_2): \ r_1+r_2=r} \ \sum_{i=1}^{2} {d_i(r_i)}
\eeqan Now, consider the general case with $M$ parallel channels
\beqan \rho^{-d(r)} & \doteq & P \{ \ \sum_{i=1}^{M} Z_i \leq r
log(\rho) \ \} \eeqan

Proceeding by induction, we get:  \beqan d(r) & = &
\inf_{(r_1,r_2,\cdots,r_M): \ \sum_{i=1}^{M} r_i = r} \
\sum_{i=1}^{M} {d_i(r_i)} \eeqan \epf

\bnote The following lower and upper bounds on the outage exponent
are immediate from Equation \eqref{eq:parallel_dmt}:
\beqa d(r) & \leq &  \sum_{i=1}^{M} {d_i(\frac{r}{K})} \label{eq:dmt_parallel_upper} \\
d(r) & \geq &  \sum_{i=1}^{M} {d_i(r)}
\label{eq:dmt_parallel_lower} \eeqa \enote

We recall the following Lemma from the theory of majorization
\cite{MarOlk}:

\blem \label{lem:majorization} \cite{MarOlk}  If $f(.)$ is a
symmetric function in variables $r_1, r_2,\ldots, r_N$ and is
convex in each of the variables $r_i, i=1, 2,\ldots, N$, then,
\beq \inf_{(r_1,r_2,\cdots,r_N): \ \sum_{i=1}^{N} r_i = r} \
f(r_1,r_2,...,r_N) = f \left (
\frac{r}{N},\frac{r}{N},...,\frac{r}{N} \right ) \eeq \elem

\vspace{0.1in} \blem The DMT of a parallel channel with all the
individual channels being identical and having a convex DMT is
given by: \beqa d(r) & = & Md_1 \left ( \frac{r}{M} \right )
\label{eq:dmt_parallel} \eeqa \elem

\bpf Consider $\sum_{i=1}^{M} {d_i(r_i)}$ as a function of the
variables $r_i$. Then the function satisfies the conditions of
Lemma~\ref{lem:majorization}. Therefore,
 \beqan d(r) & = &
\inf_{(r_1,r_2,\cdots,r_M): \ \sum_{i=1}^{M} r_i = r} \
\sum_{i=1}^{M} {d_i(r_i)} \\
& = &  \sum_{i=1}^{M} {d_i(\frac{r}{M})} \\
& = &  M {d_1(\frac{r}{M})} \eeqan \epf

\subsubsection{Parallel Channel with Repeated Coefficients\label{sec:parallel_dependent}}

\vspace{0.1in} \blem \label{lem:parallel_correlated_channel}
Consider a parallel channel with $M$ links with repeated channel
matrices. Let there be $N$ distinct channel matrices
$H^{(1)},H^{(2)},...,H^{(N)}$, with $H^{(i)}$ repeating in $n_i$
sub-channels, such that $\sum_{i=1}^{N}n_i = M$. Let $f_i =
\frac{n_i}{M}$. Then the DMT of the parallel channel is given by,
\beq d(r) = \inf_{(r_1,r_2,\cdots,r_M): \ \sum_{i=1}^{N} \ f_i r_i
= \frac{r}{M}} \ \sum_{i=1}^{N} {d_i(r_i)}
\label{eq:parallel_repeated_coeffs} \eeq \elem

\begin{figure}[h!]
\centering
\includegraphics[height=60mm]{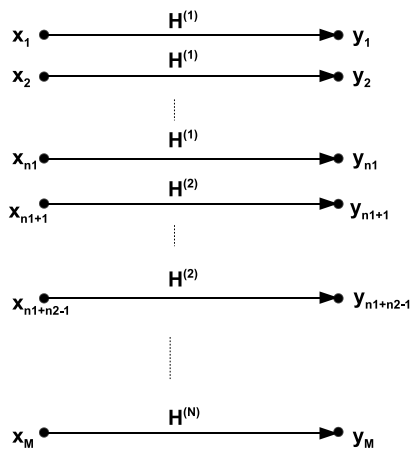}
\caption{The Parallel Network with repeated
coefficients\label{fig:parallel_network_repeat}}
\end{figure}

\vspace{0.1in} \bpf Following the same line of arguments in the
proof of Lemma~\ref{lem:parallel_channel}, choose $\bold{x_i}$ to
be independent. For computing the DMT, we know from
Lemma~\ref{lem:signal_white} that the inputs can in fact be
independent and identically distributed with a $\mathbb{CN}(0,I)$
distribution. So we have \beqan I(\bold{x};\bold{y} | \bold{H} = H
)
& = & \sum_{i=1}^{M}I(\bold{x_i};\bold{y_i} | \bold{H_i} = H_i ) \\
P \{ I(\bold{x};\bold{y} | \bold{H} = H ) \leq r log{\rho} \}
& = & P \{\sum_{i=1}^{M} I(\bold{x_i};\bold{y_i} | \bold{H_i} = H_i ) \leq r log{\rho} \} \\
& = & P \{ \sum_{i=1}^N \ n_iI(\bold{x_i};\bold{y_i} | \bold{H_i}
= H_i ) \leq r log{\rho} \} \eeqan

%Here we are assuming that the input distribution $p_{X_i} = p_x$ is the same for all $i=1, 2,\ldots, M$. In
%particular, we can assume that $p_X = \mathcal{C} \mathcal{N} (0,I)$, which is optimal according to the argument
%above.

Now, define $Z_i := n_iI(\bold{x_i};\bold{y_i} | \bold{H_i} = H_i
)$. Also let
\beqan \rho^{-d_i^{'}(r)} & \doteq & P \{Z_i < {r}log(\rho) \} \\
& = & P \{I(\bold{x_i};\bold{y_i} | \bold{H_i} = H_i ) < (\frac{r}{n_i})log(\rho) \} \\
& = & \rho^{-d_i(\frac{r}{n_i})} \\
\text{where, }\rho^{-d_i(r)} & \doteq & P \{  \mu (H^{(i)},p_x) <
r log(\rho) \}  \eeqan

Using the same convolution argument in the proof of
Lemma~\ref{lem:parallel_channel}, \beqan d(r) & = &
\inf_{(r_1,r_2,\cdots,r_N): \ \sum_{i=1}^{N} r_i = r} \
\sum_{i=1}^{N} {d_i^{'}(r_i)} \\
& = & \inf_{(r_1,r_2,\cdots,r_N): \ \sum_{i=1}^{N} r_i = r} \ \sum_{i=1}^{N} {d_i(\frac{r_i}{n_i})} \\
& = & \inf_{(r_1,r_2,\cdots,r_N): \ \sum_{i=1}^{N} \ f_i r_i =
\frac{r}{M}} \ \sum_{i=1}^{N} {d_i(r_i)} \eeqan \epf

\subsection{Achievability of outage exponent} In all the above derivations, it was assumed that the outage exponent was equal to the DMT.
It needs to be shown that the outage exponent can indeed be
achieved. We first give a simple compound channel argument for the
achievability, similar to the argument in \cite{TavVis}. Consider
a compound channel, where a channel, $s$ is chosen from a set of
possible channels $\mathcal{S}$ and the channel remains fixed.
Then the capacity of the compound channel is given by

\beqa C & = & \sup_{p_X(x)} \ \inf_{s \in \mathcal(S)}
I(\bold{X};\bold{Y} | \bold{S} = s ) \eeqa

If the maximizing input distribution $p_X^{*}(x)$ is the same for
all possible channels $s \in \mathcal{S}$, then \beqan C & = &
\inf_{s \in \mathcal{S}} \ C_s \text{, where}\\  C_s & := &
I(\bold{X};\bold{Y} | \bold{S} = s ) \eeqan evaluated for
$p_X^{*}(x)$, which is indeed the capacity of the channel $s$.

Consider the set of all channels not in outage, $\mathcal{H}$.
Then $\mathcal{H}$ is defined as \beqa \mathcal{H} = \{ H:
I(\bold{X};\bold{Y} | \bold{H} = H )> r log(\rho) \} \eeqa

If the optimizing distribution is independent of $H$ in
$\mathcal{H}$, then the capacity of the compound channel
$\mathcal{H}$ is given by $C = rlog{\rho}$.

This means that there exists a code for this compound channel,
whose probability of error is less than $\epsilon$ for any given
$\epsilon
> 0$. The probability of error of this code when used on the slow
fading channel is given by

\beqa P_e & = & P_{\text{out}} P_{\text{e/out}} + P_{\text{out}^{c}} P_{ {e / {out^c}}} \\
& \leq & P_{\text{out}} + P_{ {e / {out^c}}}  \\
& \leq & P_{\text{out}} + \epsilon \\
&  \dot \leq & P_{\text{out}} \eeqa

where $P_{\text{out}}$ is the probability of the channel being in
outage and $P_{\text{out}^{c}}$ is the probability of the channel
not being in outage. $P_{\text{e/out}}$ is the probability of
error of the code given the channel is in outage and $P_{ {e /
{out^c}}}$ is the probability of error of the code given the
channel is not in outage. Thus the outage probability is
achievable if the optimizing distribution is independent of $H$.

Since the outage exponent optimizing distribution is iid gaussian,
which is independent of $H$, as shown in
Lemma~\ref{lem:signal_white}, we can show that outage exponent is
achievable using universal codes. It should be pointed out here
that short approximately universal codes for the MIMO parallel
channel were given recently in \cite{EliVij}. These codes indeed
achieve the outage exponent of the parallel channels considered in
Section~\ref{sec:parallel}.

\section{Full Duplex Relay Networks\label{sec:full_duplex}}

In this section, we consider networks equipped with full duplex
(FD) relay nodes. First, we draw a general result on the optimum
diversity of a multi-terminal network. We also provide an
achievable DMT region for an ss-ss network with single antenna
nodes.

\subsection{Mincut equals Diversity \label{sec:mincut_diversity}}

\bthm \label{thm:mincut} Consider a multi-terminal fading network
with nodes having multiple antennas with each edge having iid
Rayleigh-fading coefficients. The maximum diversity achievable for
any flow is equal to the min-cut between the source and the sink
corresponding to the flow. Each flow can achieve its maximum
diversity simultaneously. \ethm

\bpf First we consider the case where there is only a single
source-sink pair. We will prove the theorem in two cases: the
single antenna antenna case and the multiple antenna case. We
shall assume that all the fade coefficients are independent.

\vspace{0.05in}

\emph{Case I: Network with single antenna nodes}

\vspace{0.05in}

Let the source be $S_i$ and sink be $D_j$. Let $\mathbb{C}_{ij}$
denote the set of all cuts between $S_i$ and $D_j$.

From cutset bound \cite{YukErk},

\beqan d(r) & \leq & \min_{C\in \mathbb{C}_{ij}}d_C(r) \\
\Rightarrow  d(0) & \leq & \min_{C\in \mathbb{C}_{ij}}d_C(0)
\\ & =: & \ m
\eeqan

where $m$ is the number of edges in the mincut between $S_i$ and
$D_j$.

Sufficient to prove that diversity order of $m$ is achievable. We
know that the number of edges in the mincut is the maximum number
of edge disjoint paths between source and the sink. Schedule the
network in such a way that each edge in a given edge disjoint path
is activated one by one. Same is repeated for all the edge
disjoint paths. Thus, the same data symbol is transmitted through
all the edge disjoint paths from $S_i$ to $D_j$. Let the number of
edges in the $i$th edge disjoint path be $n_i$. The $j$th edge in
the the $i$th edge disjoint path is denoted by $e_{ij}$ and the
associated fading coefficient be $h_{ij}$. So the activation
schedule will be as follows: $e_{11}, e_{12}, \cdots,
e_{1(n_1)},e_{21}, \cdots, e_{2(n_2)}, \cdots, e_{m1}, e_{m2},
\cdots, e_{m(n_m)}$. Now define $h_i := \prod_{j=1}^{n_i}h_{ij}$.
Let the total number of time slots required be $N =
\Sigma_{i=1}^{m}n_i$.

With this protocol in place, the equivalent channel seen by a
symbol is

\beqan H & = & \left[
\begin{array}{cccccc}
        h_1         &   0   & \hdots    & &&0\\
        0   & h_2   &   0   & &&0\\
        \vdots      &       &\vdots &\ddots &&\\
        0 & \hdots  & \hdots      & && h_m
        \end{array}
        \right]
\eeqan

If $d_e(r)$ is the outage exponent for this channel,

\beqan \rho^{-d_e(r)} & \doteq &  Pr\{ \Sigma_{i = 1}^{m}
\log(1+|h_i|^2) \leq rlog\rho\} \\
& = & Pr\{ \Sigma_{i = 1}^{m} \log(1+\prod_{j=1}^{n_i}|h_{ij}|^2)
\leq r\log\rho\} \\
& \doteq & Pr\{ \Sigma_{i = 1}^{m}
\log(1+\rho^{(1-\Sigma_{j=1}^{n_i}u_{ij})}) \leq r\log\rho\} \\
\text{where } |h_{ij}|^2 & = & \rho^{-u_{ij}} \\
& \doteq & Pr\{ \Sigma_{i = 1}^{m}
\log(1+\rho^{(1-\Sigma_{j=1}^{n_i}u_{ij})}) \leq \log\rho^r\} \\
& \doteq & Pr\{ \prod_{i = 1}^{m}
(\rho^{(1-\Sigma_{j=1}^{n_i}u_{ij})^+}) \leq \rho^r\} \eeqan

Following the same lines of arguments as in \cite{ZheTse},

\beq d(r) = \inf_{\mathcal{A}}
\Sigma_{i=1}^{m}\Sigma_{j=1}^{n_i}u_{ij} \eeq

where

\beq \mathcal{A} = \{ u_{ij}: \Sigma_{i=1}^{m}(1-
\Sigma_{j=1}^{n_i}u_{ij})^+ \leq r \} \eeq

Let $\Sigma_{j=1}^{n_i}u_{ij} = u_i$. Then,

\beqan
d(r) & = & \inf_{\mathcal{A}'} \Sigma_{i=1}^{m}u_i \\
\text{where } \mathcal{A}' & = & \{u_{i}: \Sigma_{i=1}^{m}(1-
u_{i})^+ \leq r \} \\
\Rightarrow d_e(r) & = & m-r \eeqan

Since we use N channel uses, the effective outage exponent is
given by,

\beqa d(r) & = & d_e(Nr) \nonumber \\
& = & m - Nr \label{eq:mincut_diversity} \eeqa

Hence the maximum achievable diversity is $m$.

\vspace{0.1in}

\emph{Case II: Network with multiples antenna nodes}

\vspace{0.1in}

In the multiple antenna case, we regard any link between a $n_t$
transmit and $n_r$ receive antenna as being composed of $n_tn_r$
links, with one link between each transmit and each receive
antenna. Note that it is possible to selectively activate
precisely one of the $n_tn_r$ Tx-antenna-Rx-antenna pairs by
appropriately transmitting from just one antenna and listening at
just one Rx antenna. The same strategy as in the single antenna
case can then be applied to achieve this diversity in the network.

\begin{figure}[h]
  \centering
  \subfigure[Original network with multiple antenna nodes]{\label{fig:mincut_1}\includegraphics[width=67mm,  height=50mm]{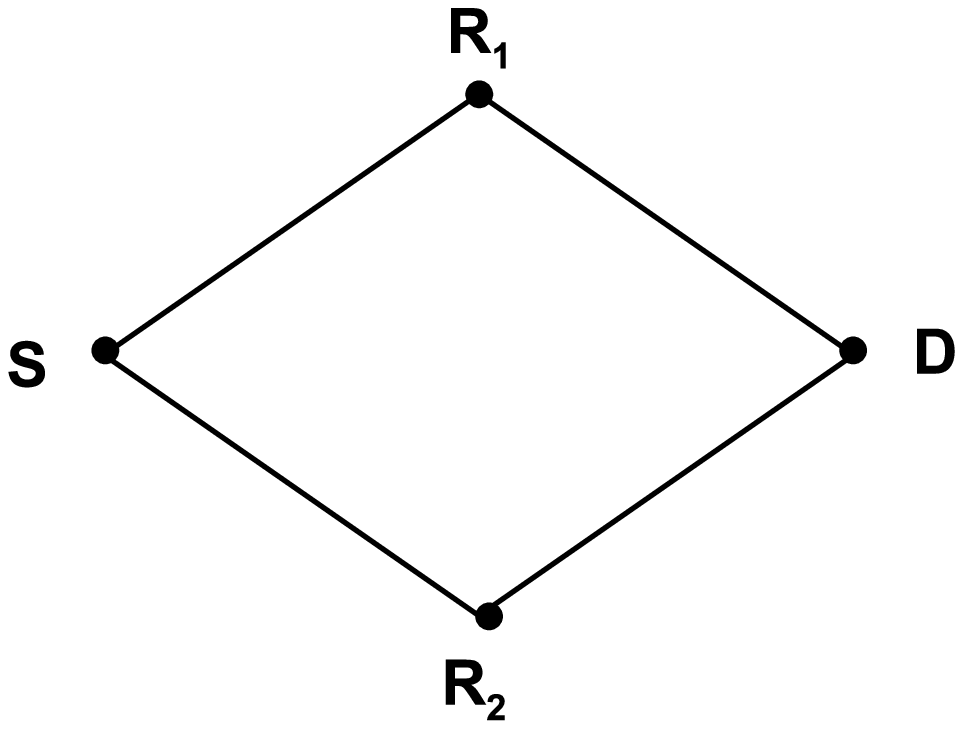}}
  \subfigure[Equivalent network with single antenna nodes]{\label{fig:mincut_2}\includegraphics[width=67mm, height=50mm]{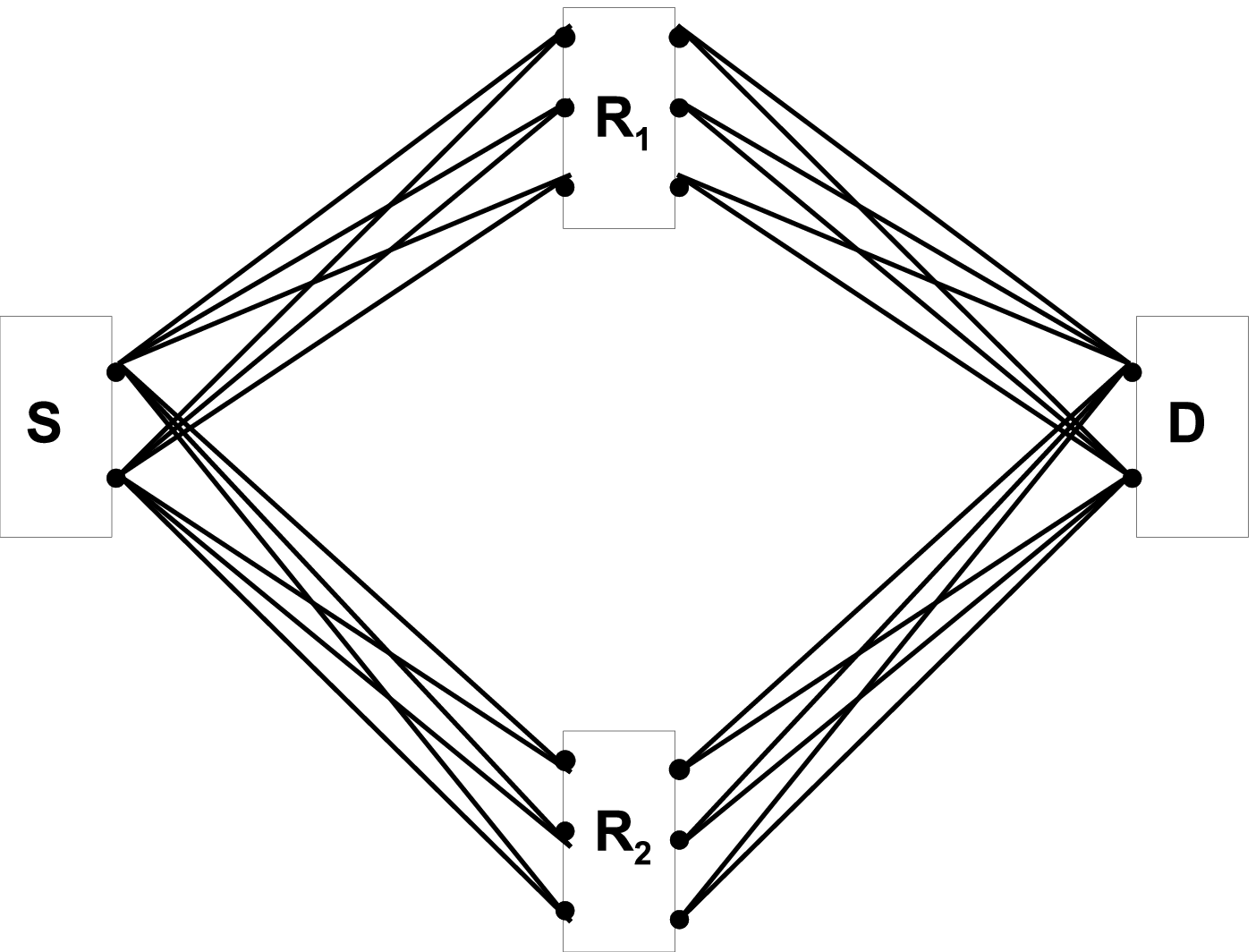}}
  \caption{Illustration: $n_S = n_D = 2, n_1 = n_2 = 3$}
  \label{fig:mincut_multiple_antennas}
\end{figure}

Fig. \ref{fig:mincut_multiple_antennas} illustrates this
conversion for the case of a single source $S$, two relays $R_1$
and $R_2$ and a sink $D$.  Having converted the multiple antenna
network into one with single antenna nodes, \emph{Case II} follows
from \emph{Case I}.

Thus the proof is complete for the single flow from $S_i$ to
$D_j$.

When there are multiple flows in the network, we simply schedule
the data of all the flows in a time-division manner. This will
entail a rate loss - however, since we are interested only in the
diversity, we can still achieve each flow's maximum diversity
simultaneously. \epf

\bdefn Consider a network $N$ and a path $P$ from source to sink.
This path $P$ is said to have an intermediate direct path if there
is a direct link in $N$ connecting two non-consecutive nodes in
$P$. \edefn

\bthm \label{thm:FD_No_Direct_Path} Consider a ss-ss full-duplex
network with single antenna nodes. Let the min-cut of the network
be $M = d_\text{max}$. Let the network satisfy \emph{either} of
the two conditions: \ben \item \emph{None} of the $M$ edge
disjoint paths between source and sink have intermediate direct
paths, or \item The directed graph representing the network has no
directed cycles. \een Then, a linear DMT $d(r) = M(1-r)^{+}$
between the maximum multiplexing gain of $1$ and maximum diversity
is achievable. \ethm

\bpf Given that the network has min-cut $M$, it means that there
are $M$ edge disjoint paths from source to sink. By the hypothesis
of the lemma, we have that these edge disjoint paths do not have
any intermediate direct paths. Let us call the edge disjoint paths
$e_1,e_2,...,e_M$. Let the product of the fading coefficients
along the path $e_i$ be $g_i$. Let $D_i$ be the delay of each
path. Let $D = \max D_i$.  Add delays $D-D_i$ to the path $e_i$
such that now all paths have equal delay. We follow the following
steps in order to activate the edges:

\ben \item \ben \item Activate edge disjoint path $e_1$ for a
period $T$, where $T > D$: activating all edges of the edge
disjoint path simultaneously. This will create a transfer matrix
from the source symbols to sink symbols as a diagonal matrix with
zeros on the first $D$ rows, and only one non-zero thread in the
matrix comprised of coefficients equal to $g_1$ which is the
product coefficient on path $e_1$. After this is done, the various
nodes in the network store the data that have not yet been passed
to the sink for future use. \item Repeat \emph{Step 1.a} for all
edge disjoint paths $e_1,...,e_M$. The net transfer matrix will
comprise $MD$ zero rows and one non-zero thread which contains
each $g_i$ for $T-D$ durations. \een

\item Activate all the edge disjoint paths each for time $T$. This
time, the net transfer matrix will comprise of a single non-zero
thread which contains each product coefficient $g_i$ for $T$
durations. There will be no zero rows since all nodes always have
information to transmit.

\item Repeat \emph{Step 2} for $L-2$ more times, thereby all edge
disjoint paths have been activated for $L$ times. \een

Now the induced channel matrix from source to sink will comprise
of $MD$ zeros initially and on removing these rows we get a
transfer matrix, $H$. $d(r) = d_H(LMTr)$. For $L$ large, we will
have $d(r) = d_H(LMTr)$.

This matrix $H$ will have each $g_i$ for $LT-D$ times along the
diagonal. This matrix will be lower triangular if none of the $M$
edge disjoint paths between source and sink have intermediate
direct paths. This matrix will be upper triangular if the directed
graph representing the network has no directed cycles. In either
case, we can use Theorem~\ref{thm:main_theorem} and
Corollary~\ref{cor:Upper_Triangular}, we get that $d_H(r) \geq
d_{H_d}(r)$, where $H_d$ is the diagonal matrix corresponding to
the matrix $H$. But $H_d$ contains $LT-D$ entries each of $g_i$,
therefore this matrix DMT is given by $d_{H_d}(r) =
d_{H_1}(\frac{1}{LT-D}r)$  where $H_1 = diag (g_1,...,g_M)$.
$\Rightarrow d(r) = d_H(LMTr) \geq d_{H_d}(LMT r) = d_{H_1}(
\frac{LMT}{LT-D}r)$. For $LT$ tending to $\infty$, we get $d(r)
\geq d_{H_1}(Mr)$. Now $d_{H_1}(r) = (M-r)^{+}$. Since $M =
d_{\text{max}}$, we get

\beqa \Rightarrow d(r) & \geq & d_{\text{max}}(1-r)^{+} \eeqa

\epf

\bcor \label{cor:FD_KPP_Layered} For the full duplex KPP networks
without direct link (i.e. KPP(I) networks) and full duplex layered
networks, a DMT of $M(1-r)^{+}$ which is a linear DMT between the
maximum diversity and maximum multiplexing gain can be achieved.
\ecor \bpf It can be easily shown that the $M$ edge disjoint paths
between source and sink for KPP(I) and layered networks do not
have any intermediate direct path. Therefore it satisfies
condition $(1)$ of Theorem~\ref{thm:FD_No_Direct_Path} and hence
proved. \epf

\section{Half duplex networks with isolated paths - KPP Networks \label{sec:half_duplex_isolated}}

In this section, we consider single-source single-sink(ss-ss) half
duplex networks in which relaying paths are isolated(i.e.,
interference between the paths is absent). Every node is equipped
with a single antenna. In general, it is assumed that half duplex
networks incur a loss in multiplexing gain by a factor of 2. But
we will establish that we can achieve the same performance in DMT
with half duplex relays as that of full duplex ones, in most of
the cases. We will show systematic ways of constructing protocols
for multi-hop networks with half duplex relays. We will show that
we can achieve the same optimal DMT of KPP networks with/without
direct link.

We first consider KPP networks in the absence of a direct link. At
the end of this section we extend the results to KPP(D) networks.

\subsection{Protocols for K-Parallel Path Networks} \label{sec:ortho_protocols}

We consider amplify-and-forward (AF) protocols in this paper. In
the class of AF protocols considered in this paper, the
communication takes place in a block of $N$ time instants, during
which the channel fading coefficients remain fixed. We assume that
the edge activations are periodic, and we refer to $N$ as the
cycle length of the protocol. We shall describe all our protocols
in a simple manner, as an edge coloring scheme. Let $C = \{c_1,
c_2, \ldots, c_N\}$ be the set of $N$ colors used in the scheme.
All the edges in the network are assigned a subset of colors from
the set $C$. The subset of colors assigned to the edge $e_{ij}$
will be denoted by $A_{ij}$. Each color in $A_{ij}$ represents the
time instants during which the edge $e_{ij}$ is active.
\footnote{We assume that the network is in operation for
sufficient amount of time, so that if an edge is active, the node
at beginning of the edge always has a symbol to transmit.}
However, due to the broadcast nature, a node will experience
interference if there is any other node connected to this one is
transmitting, apart from its intended transmitting node. A
protocol which avoids this interference is said to be an
interference free protocol, which will be of interest to us. Also,
in the class of AF protocols that we consider, we assume that
neither the source broadcasts simultaneously to different nodes
nor does the sink listen to simultaneous transmission by different
nodes. We will see later that imposing such a constraint on the
protocol is not restrictive, since we are able to achieve the best
possible DMT performance with such a protocol.

The upper bound on DMT for the class of KPP networks using the
cutset bound ( Lemma~\ref{lem:CutsetUpperBound} ) is given by:
\beqn
    d(r) \ \leq \ K(1 - r) .
\eeqn

Hence, for each of the KPP networks, we shall try to approach this
bound. Since this bound corresponds to a MISO channel, we refer to
this as the MISO bound. We shall prove, by constructing protocols
and computing their DMT, that this bound can be achieved for all
$K \geq 3$.

\subsection{Protocols achieving MISO bound \label{sec:miso_protocols}}

In this section we propose protocols for the $K$-parallel path
network and compute their DMT. For the case when $K \geq 3$ the
DMT of proposed protocols achieve the MISO bound. Also, for the
case $ K =2$ we find the maximum multiplexing gain that a protocol
can achieve among the class of AF protocols considered in this
paper.

\bdefn \label{defn:orth_protocol} A half duplex protocol is said
to be an orthogonal protocol if at any node, at a given time
instant, only one of the incoming or outgoing edges is active and
none of the nodes perform any processing of the symbols, but just
forwards the incoming packets. We put a further condition that an
orthogonal protocol for a KPP network has all edges on a given
parallel path activated equal number of times. \edefn

\bnote In networking literature \cite{KodNan}, a network is said
to have orthogonal channels if interference is avoided at all
nodes and each node can communicate with at most one other node at
any given time. While Definition~\ref{defn:orth_protocol} is
similar to this, the notion of orthogonal protocols will be
generalized to networks with interference as well in
Section~\ref{sec:half_duplex}. \enote

\bprop \label{prop:coloring} Let $C = \{ c_1, c_2,..., c_N\}$ be
the set of colors. An edge coloring is a map $\psi:E \to {\cal
P}_C$ which takes $e_{ij}$ to $A_{ij}$.

Every orthogonal protocol can be described as an edge coloring of
the network satisfying the following constraints. Similarly, every
edge coloring satisfying the following constraints describes an
orthogonal protocol.

\beqa  A_{i1} \cap A_{j1} & = & \phi, i  \neq  j. \\
A_{in_i} \cap A_{jn_j} & = & \phi, i \neq j. \\
A_{ij} \cap A_{i{j+1}} & = & \phi, j =1,2,...,{n_i}-1. \\
|A_{ij}| & = & m_i, j = 1,2,...,n_i. \eeqa

\eprop

Each color in $C$ represents a time slot and so the length of the
cycle for the protocol is $N$. Each color in $A_{ij}$ represents
the time slots during which the edge $e_{ij}$ is active.

The first constraint corresponds to the fact that for an
orthogonal protocol, only one outgoing edge is active at the
source. Similarly the second constraint corresponds to the fact
that for an orthogonal protocol, only one incoming edge is active
at the sink. The third constraint captures the half duplex nature
of the protocol. The last constraint indicates that all the edges
in a given path are active for equal duration of time so that all
the symbols transmitted by the source are forwarded to the sink.

\bdefn \label{defn:rate_orth_protocol} The rate, R of an
orthogonal protocol is defined as the ratio of the number of
symbols transmitted by the source to the total number of time
slots. In the notation above, we have

\[R = \frac{\sum_{i=1}^{K} m_i }{N}\]

\edefn

\begin{figure}[h!]
\centering
\includegraphics[width=60mm]{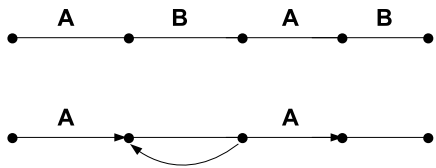}
\caption{Back-flow on a path \label{fig:backflow}}
\end{figure}

\bdefn Consider a KPP network. Let $v_1, v_2, v_3, v_4$ be four
consecutive vertices lying on one of the $K$ paths leading from
source to sink. Let $v_1$ and $v_3$ transmit, thereby causing the
edges $(v_1, v_2)$ and $(v_3, v_4)$ to be active. Due to the
broadcast and interference constraints, transmission from $v_3$
interferes with the reception at $v_2$. This is termed as
back-flow, and is illustrated in Fig.\ref{fig:backflow} \edefn

Back-flow can be avoided if we make sure that there is at least
two inactive edges between any two active edges. We formalize this
in the following remark:

\bnote \label{note:no_backflow} An orthogonal protocol avoids
back-flow if the corresponding coloring satisfies the following
condition: \beqan
    A_{ij} \cap A_{i{j+2}} = \phi, j = 1,2,...,{n_i}-2.
\eeqan \enote

\vspace*{0.1in}

By Remark \ref{note:no_backflow}, it is evident that any three
adjacent edges $e_{ij}, e_{i(j+1)},$ and $e_{i(j+2)}$ will map to
disjoint sets of colors when the coloring scheme corresponds to an
orthogonal protocol avoiding back-flow. Moreover, it remains
consistent with the constraints to repeat the same set of colors
in every third edge. This suggests an easy way of describing the
edge coloring. For a given path in the network, we will have three
sets of colors in order and they are cyclically associated to
edges starting from source to sink. For reasons that will become
apparent later, the last edge (edge connected to the sink) in the
given path may get associated to a different set of colors. So, to
describe an orthogonal protocol, we define a tuple of sets $G_i =
[ G_{i0},G_{i1},G_{i2} ] $ and a set $F_i$ for all $i$ such that,

 \beqa {A_{ij}} & = &
    \left \{\begin{array}{ccc}
        G_{i (j \text{ mod } 3)} ,& &  \ \mbox{$j \neq n_i$} \\
        F^i ,& &  \ \mbox{$j = n_i$} \\
    \end{array}
    \right.  \label{eqref:color_class}
\eeqa

Hereafter, we will use $G^i$ and $F^i$ for $i = 1,2,...,K$ to
completely describe an orthogonal protocol. Here, $G^i$ specifies
the colors that are repeated cyclically on the edges of the path
$P_i$ and $F^i$ specifies the color on the last edge $e_{i{n_i}}$
of path $P_i$.

\blem \label{lem:dmt_opt} Consider a KPP network. If an orthogonal
protocol satisfies the following constraints:

\ben \item The rate of the protocol is equal to one. \item In
every cycle, the sink receives equal number of symbols from each
one of the $K$ parallel paths. \item The protocol avoids
back-flow. \een

Then the protocol achieves the MISO bound\footnote{Throughout the
paper, keeping in mind that the number $N$ of symbols transmitted
can be made large, we ignore a rate-loss factor of $\frac{N}{N+D}$
arising from the presence of $D$ units of delay in the network.},
i.e., \beqn d(r) = K(1-r)^+  \eeqn \elem

\bpf The induced channel matrix for any orthogonal protocol for a
K-parallel path network can be split into block diagonal matrices.
This is by virtue of the fact that we are dealing with $K$
parallel paths and at any time instant, the sink receives a symbol
from only one of the $K$ paths. Further, the input symbols can be
reordered such that the matrices $H_i$ on the block-diagonal
contain fading coefficients corresponding to the $i$-th path.

So, the induced channel matrix $H$ between the source and sink,
considering $mK$ time instants of transmission, can be written in
terms of the channel matrices $H_i$, $i = 1,2, \cdots K$, where
$H_i$ is the $m \times m$ channel matrix for path $P_i$.

\begin{equation}
H = \left[ \begin{array}{c|c|c|c} H_1 & 0 & \cdots & 0 \\
\hline 0 & H_2 & \cdots & 0 \\ \hline \vdots  & \vdots  & \ddots &
\vdots  \\ \hline
0 & 0 & \cdots & H_K \\
\end{array}
\right].
\end{equation}

\beqa \Rightarrow \text{det}(I+ \rho HH^{\dagger}) & = &
\prod_{i=1}^{K} \text{det}(I+ \rho {H_i}{H_i}^{\dagger})
\label{eqref:det} \eeqa

For protocols which avoid back-flow and use all paths equally, the
channel matrix for path $ P_i$ is given by \beqa H_i & = &
{g_i}{I_m}, \ i = 1,2, \cdots, K. \eeqa

where $g_i =\prod_{j=1}^{n_i} g_{ij}$

Consider one cooperation frame of the protocol satisfying the
above constraints. Let $\bold{x_i}$ be the column vector  of $m$
symbols transmitted by the source to path $P_i$ and $\bold{y_i}$
be the column vector of $m$ symbols received by the sink from the
path $P_i, 1 \leq i \leq K$. Since $\bold{x_i}$ passes through all
the edges $e_{ij}, \ \  1 \leq j \leq n_i,$ before reaching the
sink, the channel model for one cooperation frame can be written
as

\beqa \left[ \begin{array}{c}
\bold{y_1}\\
\bold{y_2}\\
\vdots\\
\bold{y_K}\\
\end{array} \right]
 &=&  \left[\begin{array}{cccc}
    {\bold{g_1}}{I_m} & & & \\
    & {\bold{g_2}}{I_m} & &\\
    & & \ddots &\\
    & & & {\bold{g_K}}{I_m} \\
    \end{array}\right]
    \left[ \begin{array}{c}
             \bold{x_1}\\
             \bold{x_2}\\
             \vdots\\
             \bold{x_K}\\
            \end{array} \right] + \bold{n} \\
 \bold{y} &=& \bold{H x} +\bold{n}
\eeqa where $\bold{n}$ is the equivalent colored noise seen at the
sink and $H$ is the equivalent parallel channel. It can be easily
shown that the noise
becomes white, in the scale of interest \cite{EliVinAnaKum}.% in the high SNR regime.
 The DMT of the above channel, $H$, can be shown to be, \beqn
    d(r) = K (1- r)^+ \ ,
\eeqn which is the MISO bound.  Here, the notation $(1 - r)^+$
indicates that we must choose the maximum of $0$ and $1 - r$.

\epf

\bcor \label{cor:dmt_equal} If any orthogonal protocol has a
channel matrix $H$, with $H_i$ as the channel matrix for the path
$P_i$, such that det $(I_m + \rho H_iH_i^{{\dagger}}) =$ det $(I_m
+ \rho H_i^{'}H_i^{'{\dagger}})$, where $H_i^{'} = {g_i}{I_m} $
for $i=1,2, \cdots, K$, then that protocol achieves the MISO
bound. \ecor

\bpf The DMT depends only upon det $(I + \rho HH^{\dagger})$ which
remains the same as that in (\ref{eqref:det}). Therefore the DMT
remains same. \epf

\bthm \label{thm:K_geq_4} When $K \geq 4$, there exists a protocol
achieving MISO bound for KPP networks. \ethm

\bpf  We now establish an orthogonal protocol for the case when $K
\geq 4$. By Prop~\ref{prop:coloring}, it is sufficient to
establish a coloring of the edges. We will give the map $\psi$
explicitly for the given network by specifying $A_{ij} \ , \forall
i,j$.

We will be using the set of colors $C = \left\{ c_1, c_2, ...,
c_K\right\}$. In the following, whenever we refer to color $c_i$
assume $c_0 = c_K$ and for $i > K$, $c_i = c_{(i \ \text{mod} \
K)}$.

We will specify the coloring scheme by giving a tuple of sets $G_i
= [ G_{i0},G_{i1},G_{i2} ] $ and a set $F_i$ for all $i$.

\vspace*{0.1in}

$G_i = [ \{c_i\}, \{c_{i+1}\},\{c_{i+2}\} ]$

$F_i = \{ c_{i+3} \}$

It is easy to verify that the scheme described satisfies all the
constraints of Lemma~\ref{lem:dmt_opt}, and therefore will achieve
the MISO bound \epf

\subsection{Back-flow does not impair the DMT}

\blem \label{lem:back_flow_lower_bound} Consider a network running
an orthogonal protocol, which, in the absence of back-flow creates
a block-diagonal matrix as the transfer matrix between the input
and the output. For such a network, the DMT when back-flow is
present, is lower bounded by the DMT in the absence of back-flow.
\elem

\bpf  The presence of back-flow creates entries in the strictly
lower-triangular portion of the transfer matrix. Since the DMT of
a lower triangular matrix is lower bounded by the DMT of the
corresponding diagonal matrix (by Theorem~\ref{thm:main_theorem}),
we have that the system with back-flow will yield a better DMT
than the one without back-flow. \epf

\subsubsection{Back Flow does not alter DMT in the Single
Antenna Case}

Since we already have a lower bound on the DMT of the networks
with back-flow, it is sufficient to get an upper bound on the DMT,
which is the same as the lower bound.

\blem \label{lem:backflow_upper_bound} Consider a KPP network
running an orthogonal protocol with single antenna nodes, which in
the absence of back-flow creates a diagonal matrix as the transfer
matrix between the input and the output. For such a network, the
DMT when back-flow is present, is the \emph{same} as the DMT in
the absence of back-flow. \elem

\bpf If the network has back-flow, then the channel matrix would
be \beqan H & = & \left[
\begin{array}{ccccc}
        h_1                 &   0               & \hdots    & &0\\
        h_2(g_{21})   & h_2               & 0     & &0\\
        h_3(g_{31})   & h_3(g_{32}) & h_3   & &\\
        \vdots              &                   &\vdots &\ddots &\\
        h_n(g_{n1}) & \hdots        & \hdots      & &h_n
        \end{array}
        \right],
\eeqan

If the network did not have back-flow, then the channel matrix
would be

\beqan H_d & = & \left[
\begin{array}{ccccc}
        h_1                 &   0               & \hdots    & &0\\
        0   & h_2               & 0     & &0\\
        0   & 0 & h_3   & 0 &\\
        \vdots              &                   &\vdots &\ddots &\\
        0 & \hdots        & \hdots      & &h_n
        \end{array}
        \right].
\eeqan

$(I+\rho{H}{H}^\dagger)$ is a positive definite Hermitian matrix
and by invoking Theorem 16.8.2 of \cite{CovTho}, we have that the
determinant is upper bounded by the product of row-norms:

\beqa \det(I+\rho{H}{H}^\dagger) & \leq & (1 + \rho|h_1|^2)(1 +
\rho|h_2|^2 + \rho|g_{21}|^2|h_2|^2)\cdots \nonumber \\
& & (1 + \rho|h_n|^2 + \rho|g_{n(n-1)}|^2|h_n|^2 + \cdots
+ \rho|g_{n1}|^2|h_n|^2) \nonumber \\
& = & \prod_{i=1}^{n}(1 + \rho|h_i|^2(1 + |g_{i(i-1)}|^2 + \cdots + |g_{i1}|^2)) \nonumber \\
& \doteq & \prod_{i=1}^{n}(1 + \rho|h_i|^2) \label{eq:scalar_bound_doteq} \\
& = & \det(I+\rho{H_d}{H_d}^\dagger) \nonumber \eeqa

The dot equivalence (\ref{eq:scalar_bound_doteq}) follows from
equation (\ref{eq:rho}) in the proof of
Lemma~\ref{lem:noise_white}.

Already we have from Lemma~\ref{lem:back_flow_lower_bound},

\beqan \det(I+\rho{H}{H}^\dagger) & \geq &
\det(I+\rho{H_d}{H_d}^\dagger) \eeqan

Therefore we get \beqan \det(I+\rho{H}{H}^\dagger) & \doteq &
\det(I+\rho{H_d}{H_d}^\dagger). \eeqan Therefore, the DMT with
back-flow is the same as without back-flow. \epf

\bthm \label{thm:k3_miso} When $K = 3$, there exists a protocol
achieving MISO bound for KPP networks. \ethm

\bpf By Prop~\ref{prop:coloring}, it is sufficient to establish a
coloring of the edges. We will give the map $\psi$ explicitly for
the given network by specifying $A_{ij} \ \forall i,j$.  Define
   \beqa {a_i} & = &
    \left\{ \begin{array}{ccc}
        1 ,& &  \ \mbox{$n_i = 1$  mod $3$} \\
        0 ,& &  \ \mbox{$n_i \neq 1$ mod $3$} \\
    \end{array}
    \right.
    \eeqa

Without loss of generality we assume that the paths are ordered
such that for the first $l$ paths, $a_i=1$ followed by the paths
for which $a_i=0$. We give a protocol for various possibilities of
$l$. \bit

\item \emph{Case 1: ($l = 0, 1, \ or \ 3$)}

We will give a coloring scheme such that the corresponding
protocol avoids back-flow, uses all paths equally, and achieves
rate 1. By Lemma~\ref{lem:dmt_opt}, this protocol will achieve the
transmit diversity bound.

We will specify the coloring scheme by giving the tuple of sets
$G_i = [ G_{i0},G_{i1},G_{i2} ]$ for all $i$. $G_i$ is defined
exactly the same way how it is in the proof of
Theorem~\ref{thm:K_geq_4}.

The set of colors used is $C = \left\{ c_1, c_2, c_3\right\}$. In
the following, whenever we refer to color $c_i$, assume $c_0 =
c_3$ and for $i > 3$, $c_i = c_{(i \ \text{mod} \ 3)}$ .

\vspace*{0.1in}

For $l=0$,

$G_i =
    \left\{ \begin{array}{ccc}
        \mbox{$[ \{c_i\}, \{c_{i+2}\},\{c_{i+1}\} ]$} ,& &  \ \mbox{$n_i = 0$  mod $3$} \\
        \mbox{$[ \{c_i\}, \{c_{i+1}\},\{c_{i+2}\} ]$} ,& &  \ \mbox{$n_i = 2$  mod $3$} \\
    \end{array}
    \right.$

\vspace*{0.2in}

For $l=1$,

$G_1 = [ \{c_1\}, \{c_2\},\{c_3\} ]$

\vspace*{0.05in}

$G_2 =
    \left\{ \begin{array}{ccc}
        \mbox{$[ \{c_2\}, \{c_1\},\{c_3\} ]$} ,& &  \ \mbox{$n_2 = 0$  mod $3$} \\
        \mbox{$[ \{c_2\}, \{c_3\},\{c_1\} ]$} ,& &  \ \mbox{$n_2 = 2$  mod $3$} \\
    \end{array}
    \right.$

\vspace*{0.05in}

$G_3 =
    \left\{ \begin{array}{ccc}
        \mbox{$[ \{c_3\}, \{c_1\},\{c_2\} ]$} ,& &  \ \mbox{$n_3 = 0$  mod $3$} \\
        \mbox{$[ \{c_3\}, \{c_2\},\{c_1\} ]$} ,& &  \ \mbox{$n_3 = 2$  mod $3$} \\
    \end{array}
    \right.$

\vspace*{0.2in}

For $l=3$,

$G_i = [ \{c_i\}, \{c_{i+1}\},\{c_{i+2}\} ]$

\vspace*{0.1in}

\item \emph{Case 2: ($l = 2$)}

For $l = 2$, we shall now come up with a protocol such that only
one node in the third path  encounters back-flow. Then, we show
that the DMT for this protocol is equal to the MISO bound. We
describe the coloring scheme for the protocol as follows.

$G_1 = [ \{ c_1\},\{ c_2\},\{ c_3\} ]$

$G_2 = [ \{ c_2\}, \{c_3\},\{ c_1\} ]$

$G_3 = [ \{ c_3 \},\{ c_1\}, \{c_2\} ]$

\vspace*{0.1in}

After this assignment, we make the following modifications to
$A_{ij}$:

\vspace*{0.1in}

$A_{3(n_3)} = \{c_3\}$

$A_{3(n_3-1)} = \{c_1\}$, if $n_3 = 2$ mod $3$

\vspace*{0.1in}

One can check that this will lead to back-flow at only one node,
say $R_{ij}$, in the third path, whose position will depend on
whether $n_3 = 0$ (mod $3$) or $n_3  =  2$ (mod $3$).

%Next, we consider a single source--single sink channel with
%only one path, with interference at one node. Such a network is
%shown in Fig.~\ref{fig:int_3}.

For the given protocol, there is no back-flow in paths $P_1$ and
$P_2$, and therefore,

\beqa H_i & = & {g_i}{I_m} \ \text{for} \ i = 1, 2. \eeqa

For path $P_3$, the channel matrix is no longer diagonal because
there is back-flow, rather the matrix is lower triangular. But
according to Lemma~\ref{lem:backflow_upper_bound},
 \vspace*{0.1in}

det $(I_m + \rho {H_3}{H_3}^{{\dagger}})$ $=$ det $(I_m + \rho
H_3^{'}H_3^{'{\dagger}})$, where $H_3^{'} = g_3 I_m $.

\vspace*{0.1in}

Therefore, the DMT of the proposed protocol is the same as the
case when $H_3$ is a diagonal matrix, and hence, would achieve the
MISO bound by Corollary \ref{cor:dmt_equal}.\eit \epf

\bthm \label{thm:k2_max_rate} For $K=2$ and $n_i>1$, the maximum
 achievable rate for any orthogonal protocol is given by   \beqa R_{max} & \leq &
    \left\{ \begin{array}{ccc}
        1 ,& &  \ \mbox{$n_1 + n_2 = 0$ \ mod \ $2$} \\
        \frac{2n_2 - 1}{2n_2} ,& &  \ \mbox{$n_1 + n_2 = 1$ \ mod \ $2$}
    \end{array}
    \right.  \label{eqref:k2_rmax}
    \eeqa
    where $n_1 \leq n_2$.
\ethm

\bpf By Prop~\ref{prop:coloring}, any orthogonal protocol
corresponds to a coloring of the edges, described by the map
$\psi$.

For $K=2$, we consider the network as a cycle with edges
$l_1,l_2,...,l_{n_1+n_2}$ with associated sets of colors
$D_1,D_2,...,D_{n_1+n_2}$ respectively. Here,    \beqan l_j & = &
    \left\{ \begin{array}{ccc}
        e_{1j} \ ,& &  \ \mbox{$j \leq n_1$} \\
        e_{2(n_2+n_1+1-j)} \ ,& &  \ \mbox{$n_1 < j \leq n_1+n_2$}
    \end{array}
    \right.
\eeqan \beqan D_j & = &
    \left\{ \begin{array}{ccc}
        A_{1j} \ ,& &  \ \mbox{$j \leq n_1$} \\
        A_{2(n_2+n_1+1-j)} \ ,& &  \ \mbox{$n_1 < j \leq n_1+n_2$}
    \end{array}
    \right.
\eeqan with a single constraint, \beqa D_j \cap D_{(j+1) \ mod \
(n_1+n_2)} = \phi \label{eqref:con0} \eeqa

Now suppose we have a coloring scheme with N colors. Then each
color can be an element of the sets of colors corresponding to at
most $\lfloor\frac{n_1+n_2}{2}\rfloor$ edges. This is because, if
there are more colors, then the half duplex constraint must be
violated. So we have,

 \beqa
  \sum_{i = 1}^{2} \ \sum_{j = 1}^{n_i} \ |A_{ij}| & \leq & \left \lfloor\frac{n_1+n_2}{2} \right \rfloor N \nonumber \\
  \text{ie.}, n_1m_1 + n_2m_2 & \leq & \left \lfloor\frac{n_1+n_2}{2} \right \rfloor N \label{eqref:con1}
  \eeqa

Since $n_i \geq 2$ in each of the paths, the constraint
(\ref{eqref:con0}) also implies that,

 \beqa
  2{m_1} \leq {N} \label{eqref:con2}\\
  2{m_2} \leq {N} \label{eqref:con3}
  \eeqa

To find the maximum rate, we pose the maximization problem:

Maximise $(\frac{m_1}{N}+\frac{m_2}{N})$ subject to
(\ref{eqref:con1}), (\ref{eqref:con2}), and (\ref{eqref:con3}).

This is easily solved to be,

\beqan
  \frac{m_1}{N} & = & 0.5 \\
  \frac{m_2}{N} & = & \frac{1}{n_2}\left \lfloor\frac{n_1+n_2}{2} \right\rfloor \ - \
  \frac{n_1}{2n_2}.
\eeqan

So the maximum rate of the protocol is given by,

   \beqa R_{max} & \leq &
    \left\{ \begin{array}{ccc}
        1 ,& &  \ \mbox{$n_1 + n_2 = 0$ \ mod \ $2$} \\
        \frac{2n_2 - 1}{2n_2} ,& &  \ \mbox{$n_1 + n_2 = 1$ \ mod \ $2$}
    \end{array}
    \right.  \label{eqref:kequals2}
    \eeqa
    where $n_1 \leq n_2$.

\epf

\bconstr \label{constr:k2_max_rate} This construction establishes
an orthogonal protocol for $K = 2$ which achieves maximum rate. By
Prop~\ref{prop:coloring}, it is sufficient to establish a coloring
of the edges. We will give the map $\psi$ explicitly for the given
network by specifying $A_{ij} \ \forall i,j$.

We consider the network as a cycle with edges
$l_1,l_2,...,l_{n_1+n_2}$ with associated sets of colors
$D_1,D_2,...,D_{n_1+n_2}$, as in the proof of Theorem
\ref{thm:k2_max_rate}

For $K=2$,  respectively. Here,    \beqan l_j & = &
    \left\{ \begin{array}{ccc}
        e_{1j} \ ,& &  \ \mbox{$j \leq n_1$} \\
        e_{2(n_2+n_1+1-j)} \ ,& &  \ \mbox{$n_1 < j \leq n_1+n_2$}
    \end{array}
    \right.
\eeqan \beqan D_j & = &
    \left\{ \begin{array}{ccc}
        A_{1j} \ ,& &  \ \mbox{$j \leq n_1$} \\
        A_{2(n_2+n_1+1-j)} \ ,& &  \ \mbox{$n_1 < j \leq n_1+n_2$}
    \end{array}
    \right.
\eeqan with a single constraint, $D_j \cap D_{(j+1) \ mod \
(n_1+n_2)} = \phi$.

\item Case 1: $(n_1+n_2) = 0 \ mod \ 2$

We will have $C = \{c_1, c_2\}$. Define $\psi$ to be such that
\beqan D_j & = &
    \left\{ \begin{array}{ccc}
        \{c_1\} \ , & \ j = 1,3,...,n_1+n_2-1 \\
        \{c_2\} \ , & \ j = 2,4,...,n_1+n_2
    \end{array}
    \right.
\eeqan

\item Case 2: $(n_1+n_2) = 1 \ mod \ 2$ We have the set of colors
$C = \{c_1, c_2,...,c_N\}$, where $N=2n_2$. We will add colors to
$D_j$ using the following algorithm.

\ben \item Step 1: $D_j \leftarrow \phi \ \forall j \in
\{1,2,...,n_1+n_2\}$.

\item Step 2: Now we will add colors to each of the set $D_j$
using the following algorithm. In the algorithm, whenever we refer
to $D_j$, with $j > n_1+n_2$, we mean $D_j = D_{j \ mod \
(n_1+n_2)}$ and with $j = 0$, we mean $D_j = D_{n_1+n_2}$.

$\{$

$\hspace{0.5in}t \leftarrow 1;$

$\hspace{0.5in}For \ k = 1 \ to \ n_2 \ in \ steps \ of \ 1:$

$\hspace{0.5in}\{$

$\hspace{1in}For \ i = 1 \ to \ n_1+n_2-2 \ in \ steps \ of \ 2:$

$\hspace{1in}\{$

$\hspace{1.5in}D_{i-k+1} \leftarrow D_{i-k+1} \cup \{c_t\};$

$\hspace{1in}\}.$

$\hspace{1in}t \leftarrow t+1;$

$\hspace{0.5in}\}.$

$\hspace{0.5in}For \ k = 1 \ to \ n_2 \ in \ steps \ of \ 1:$

$\hspace{0.5in}\{$

$\hspace{1in}For \ i = 1 \ to \ n_1+n_2-2  \ in \ steps \ of \ 2:$

$\hspace{1in}\{$

$\hspace{1.5in}D_{(n_1+n_2)-(k-1)-i} \leftarrow D_{(
n_1+n_2)-(k-1)-i} \cup \{c_t\};$

$\hspace{1in}\}.$

$\hspace{1in}t \leftarrow t+1;$

$\hspace{0.5in}\}.$

$\}.$ \een \econstr

\bnote \label{note:achieves_max_rate} The orthogonal protocol
shown in construction (\ref{constr:k2_max_rate}) achieves maximum
rate given in Theorem ~\ref{thm:k2_max_rate}. In case 1, it is
clear that rate achieved is 1. In case 2, the number of colors
used are $2n_2$. In the first loop of the construction, out of the
$n_2$ colors used, $n_2-1$ colors are added to either $D_{n_1}$ or
$D_{n_1+1}$ and all the $n_2$ colors are added to either $D_1$ or
$D_{n_1+n_2}$. In the second loop of the construction, out of the
$n_2$ colors used, $n_2-1$ colors are added to either $D_1$ or
$D_{n_1+n_2}$ and all the $n_2$ colors are added to either
$D_{n_1}$ or $D_{n_1+1}$. So the rate of the protocol would be
$\frac{2n_2 - 1}{2n_2}$. \enote

\begin{figure}[!h]
  \centering
  \subfigure[Time slot 1]{\label{fig:illustration_1}\includegraphics[width=67mm,  height=50mm]{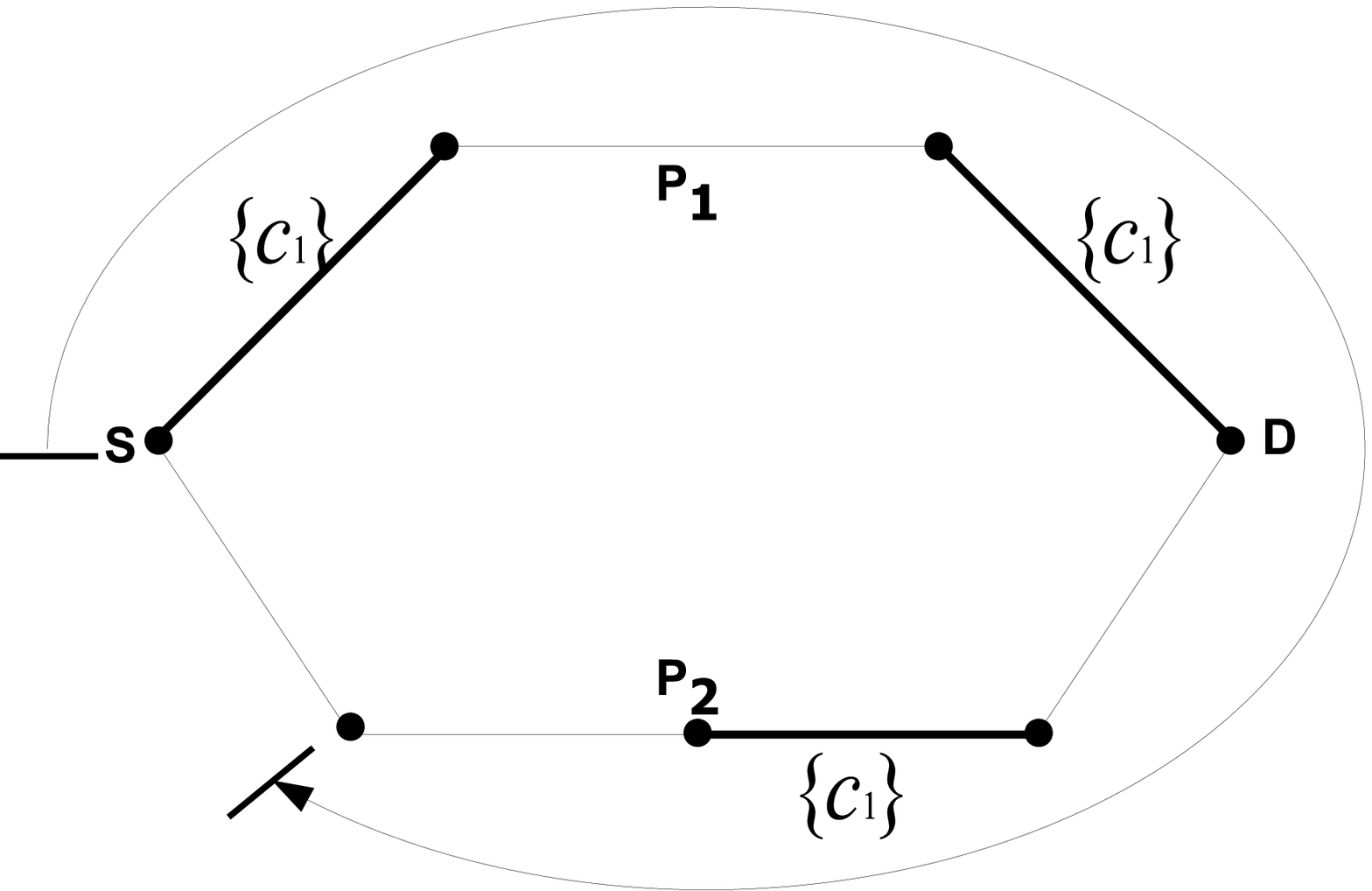}}
  \subfigure[Time slot 2]{\label{fig:illustration_2}\includegraphics[width=67mm, height=50mm]{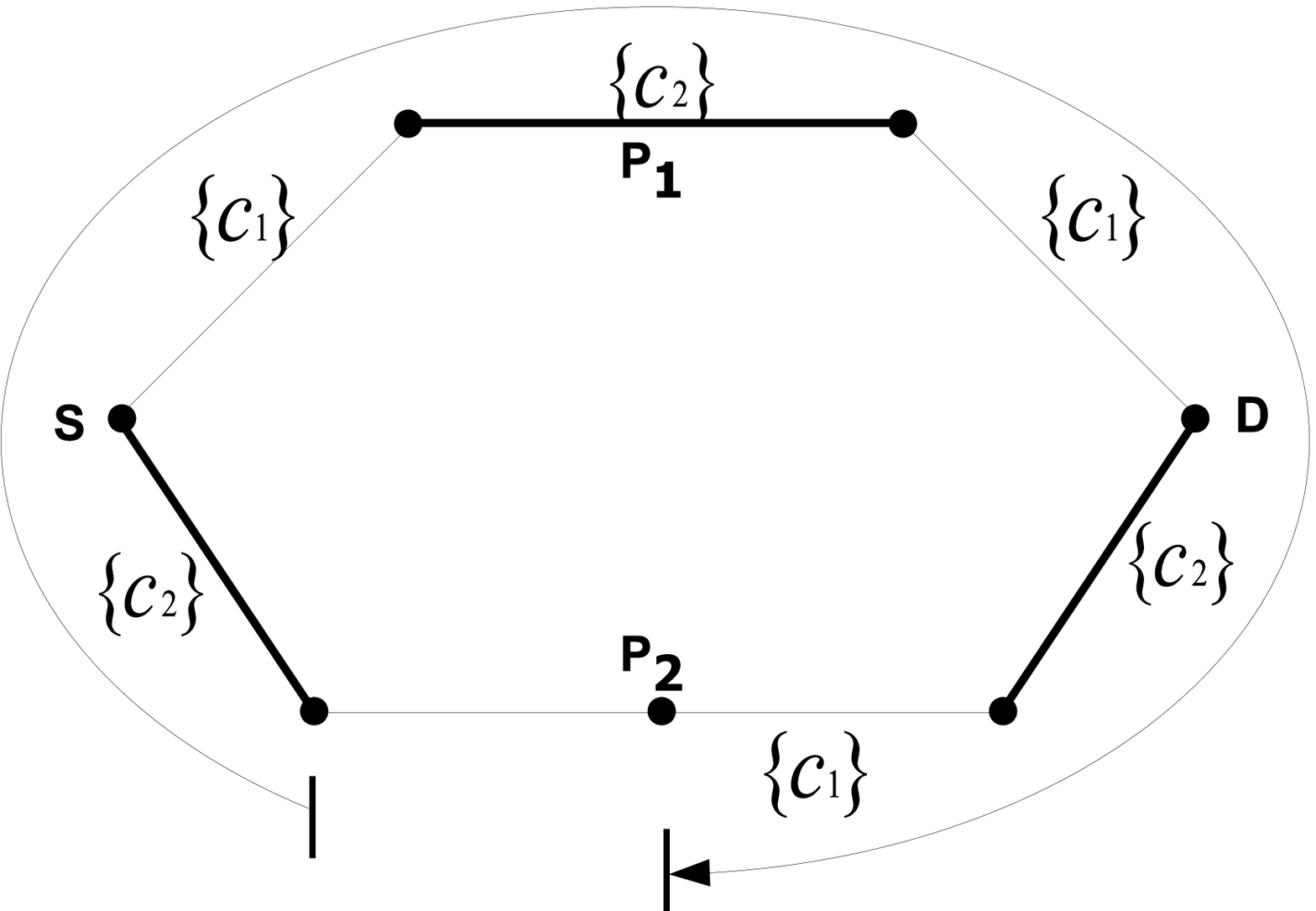}}
  \subfigure[Time slot 3]{\label{fig:illustration_3}\includegraphics[width=67mm, height=50mm]{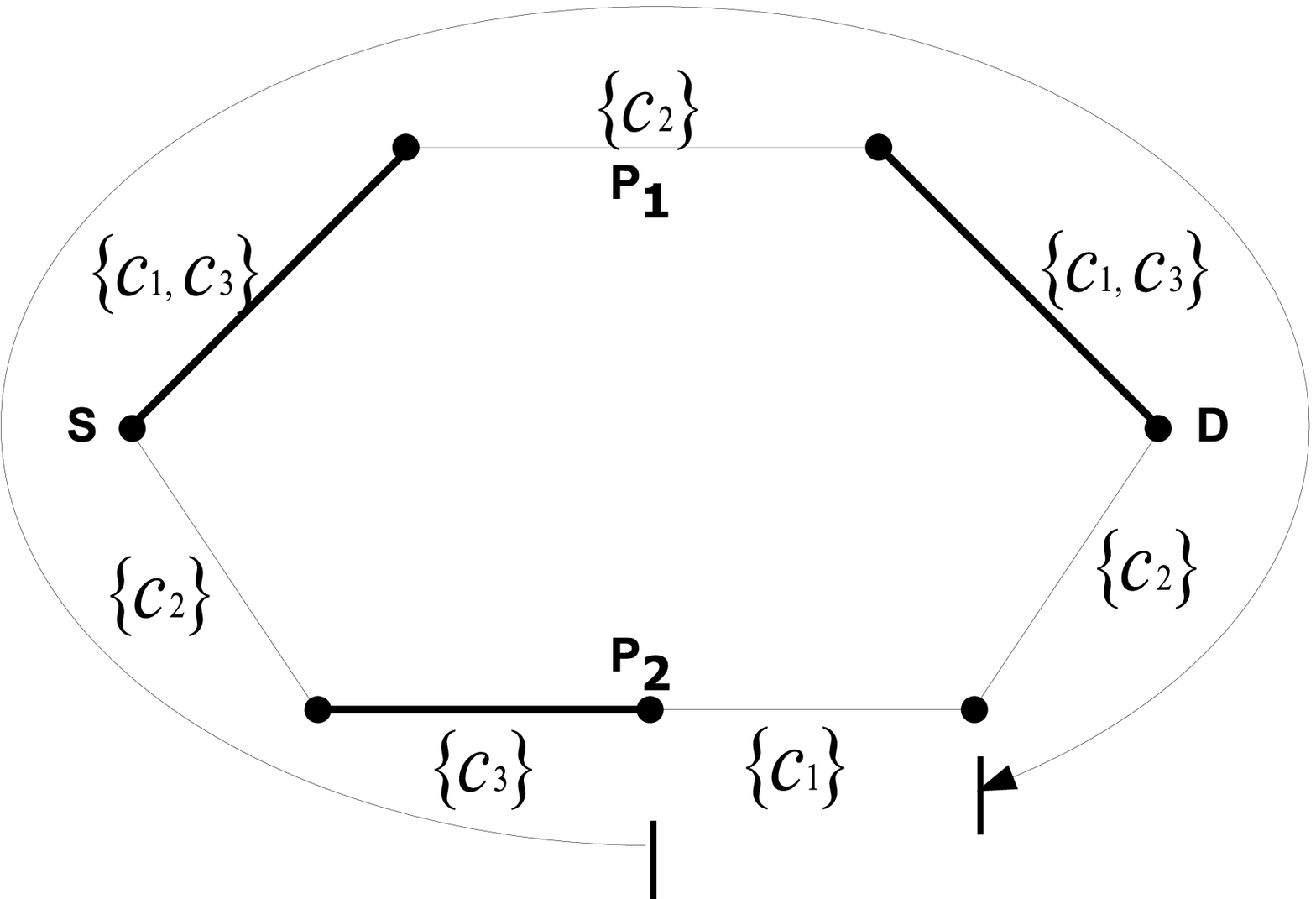}}
  \subfigure[Time slot 4]{\label{fig:illustration_4}\includegraphics[width=67mm, height=50mm]{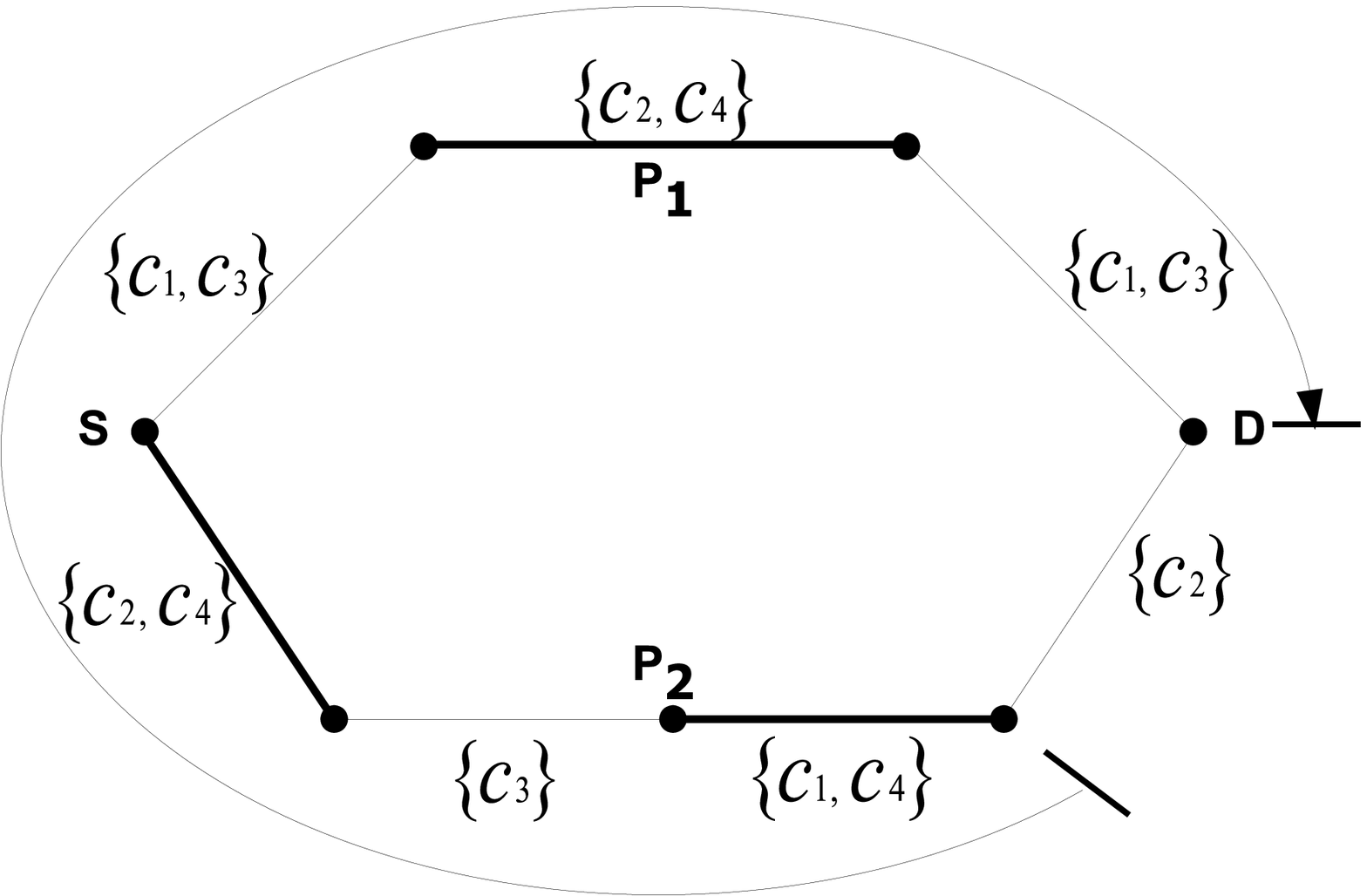}}
  \caption{Protocol Illustration: ($n_1$, $n_2$) = (3,4) [contd...]}
  \label{fig:illustration_set_1}
\end{figure}

\begin{figure}[!h]
  \centering
  \subfigure[Time slot 5]{\label{fig:illustration_5}\includegraphics[width=67mm, height=50mm]{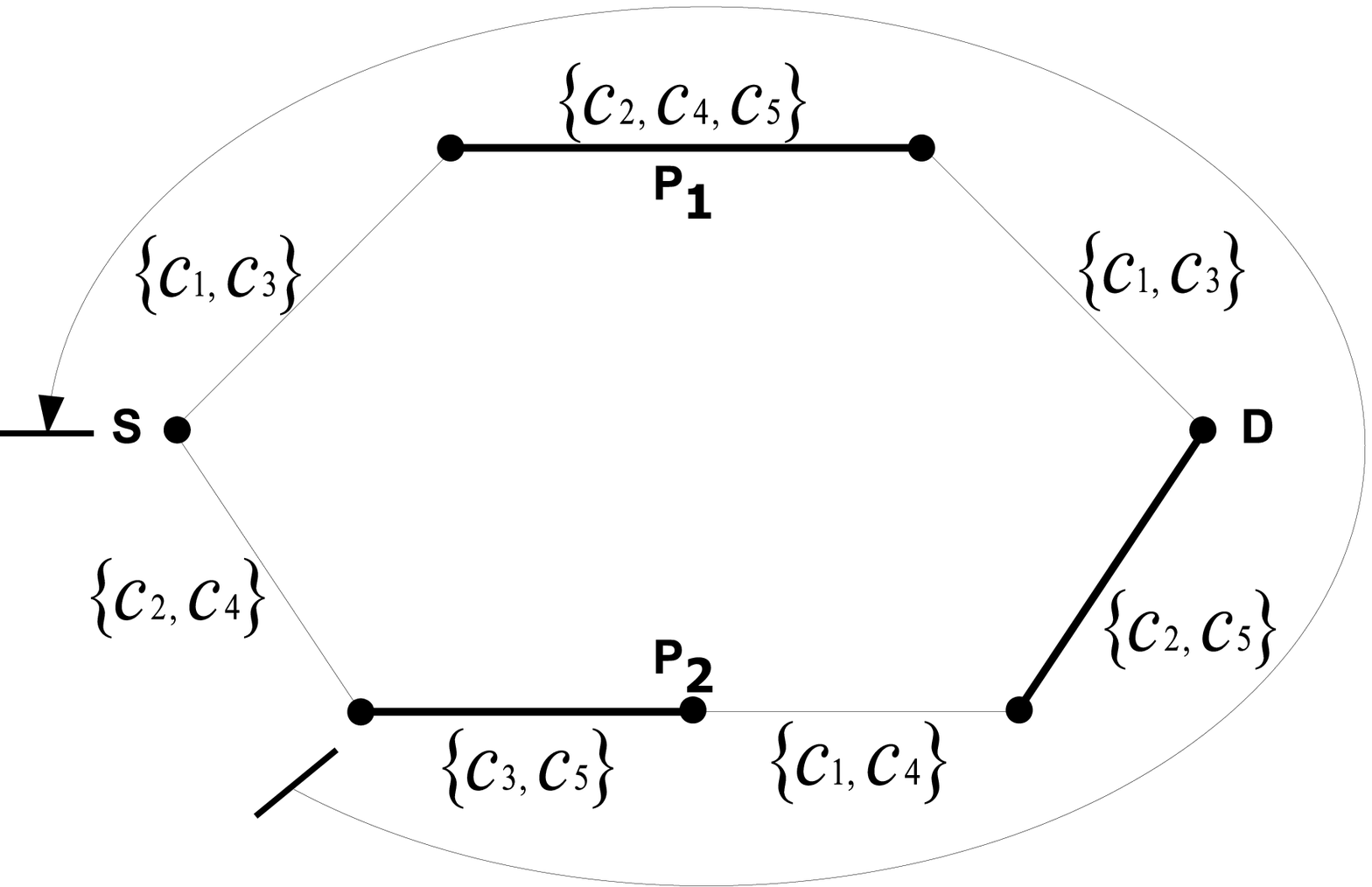}}
  \subfigure[Time slot 6]{\label{fig:illustration_6}\includegraphics[width=67mm, height=50mm]{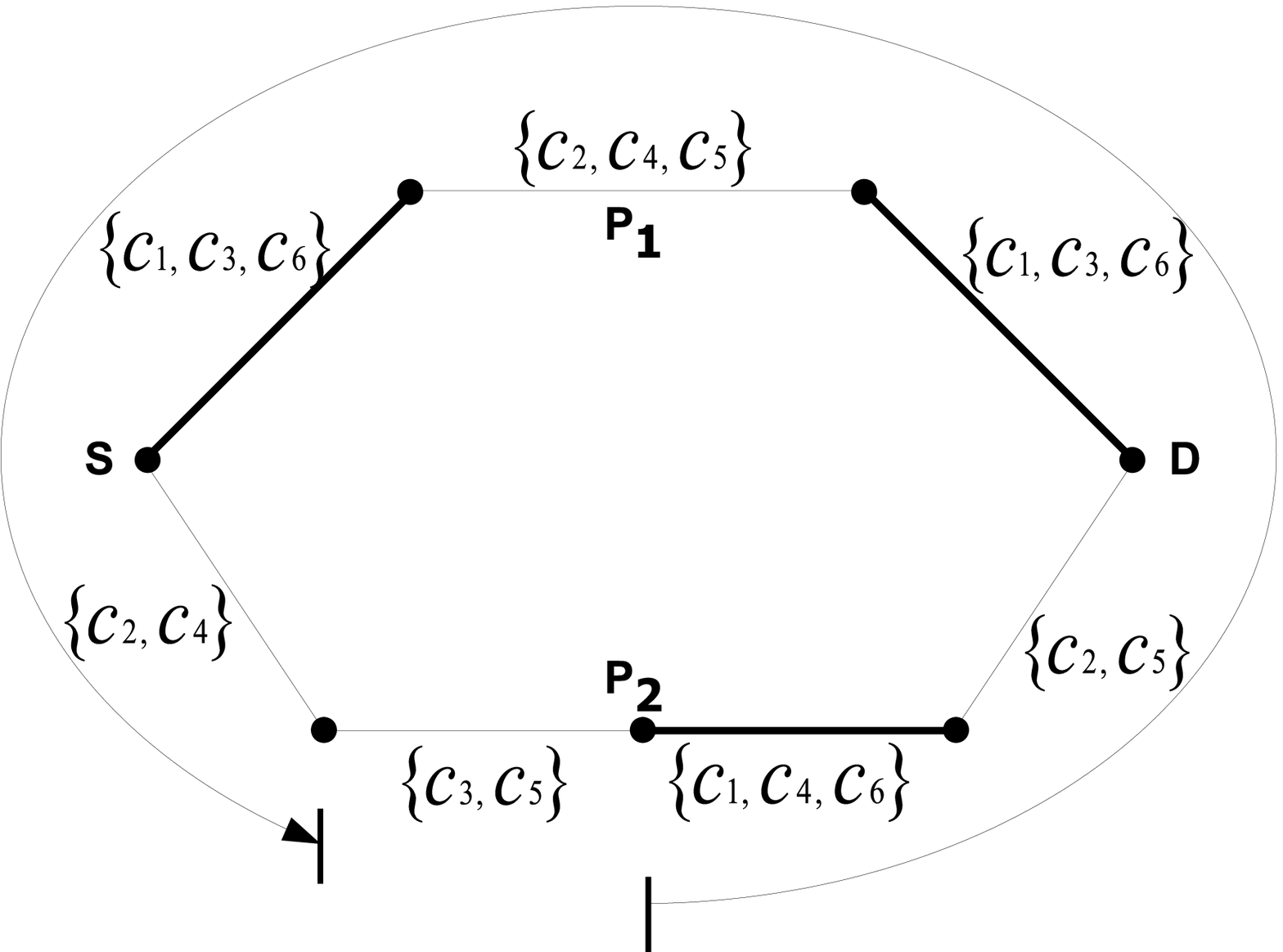}}
  \subfigure[Time slot 7]{\label{fig:illustration_7}\includegraphics[width=67mm, height=50mm]{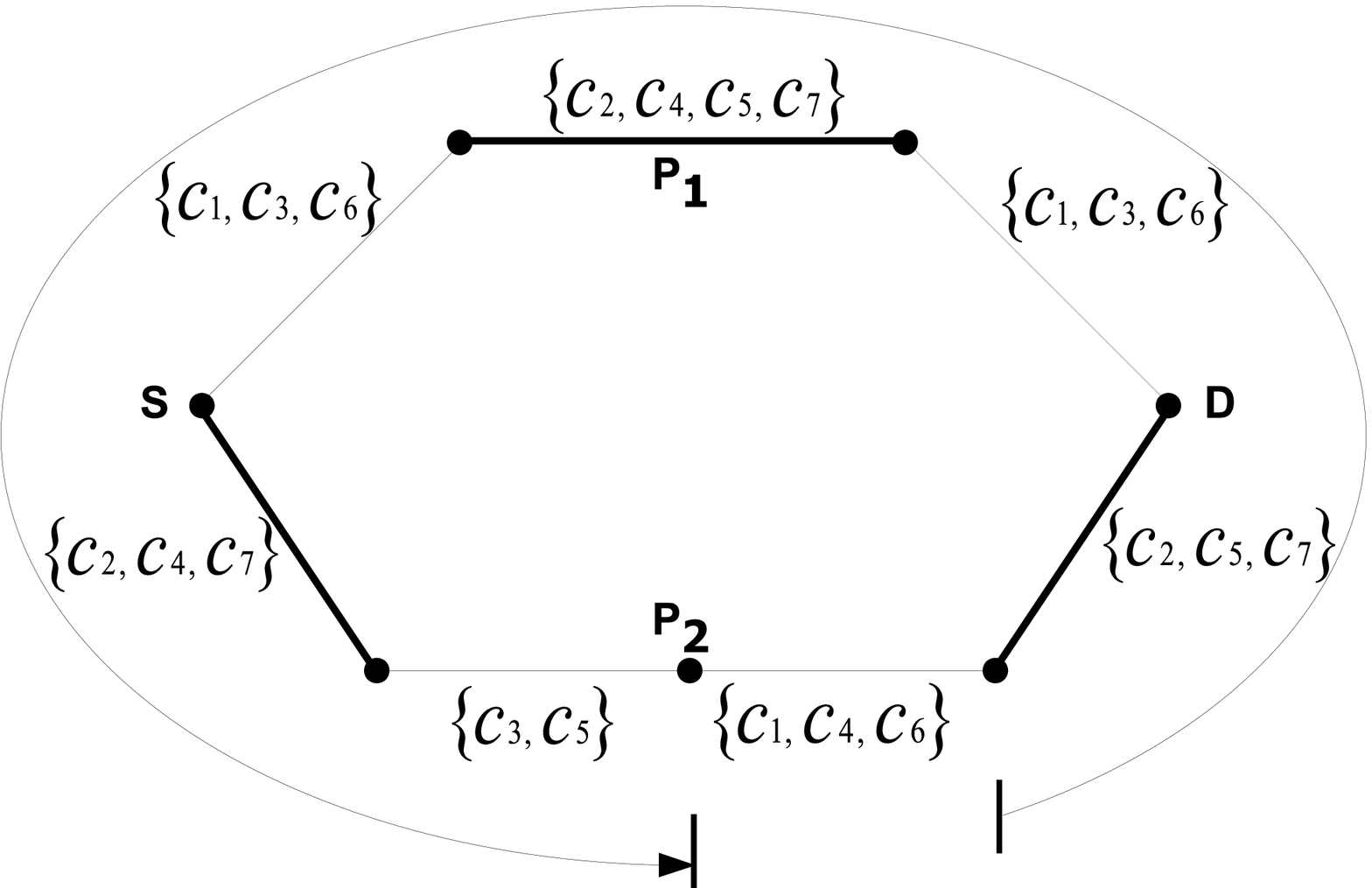}}
  \subfigure[Time slot 8]{\label{fig:illustration_8}\includegraphics[width=67mm, height=50mm]{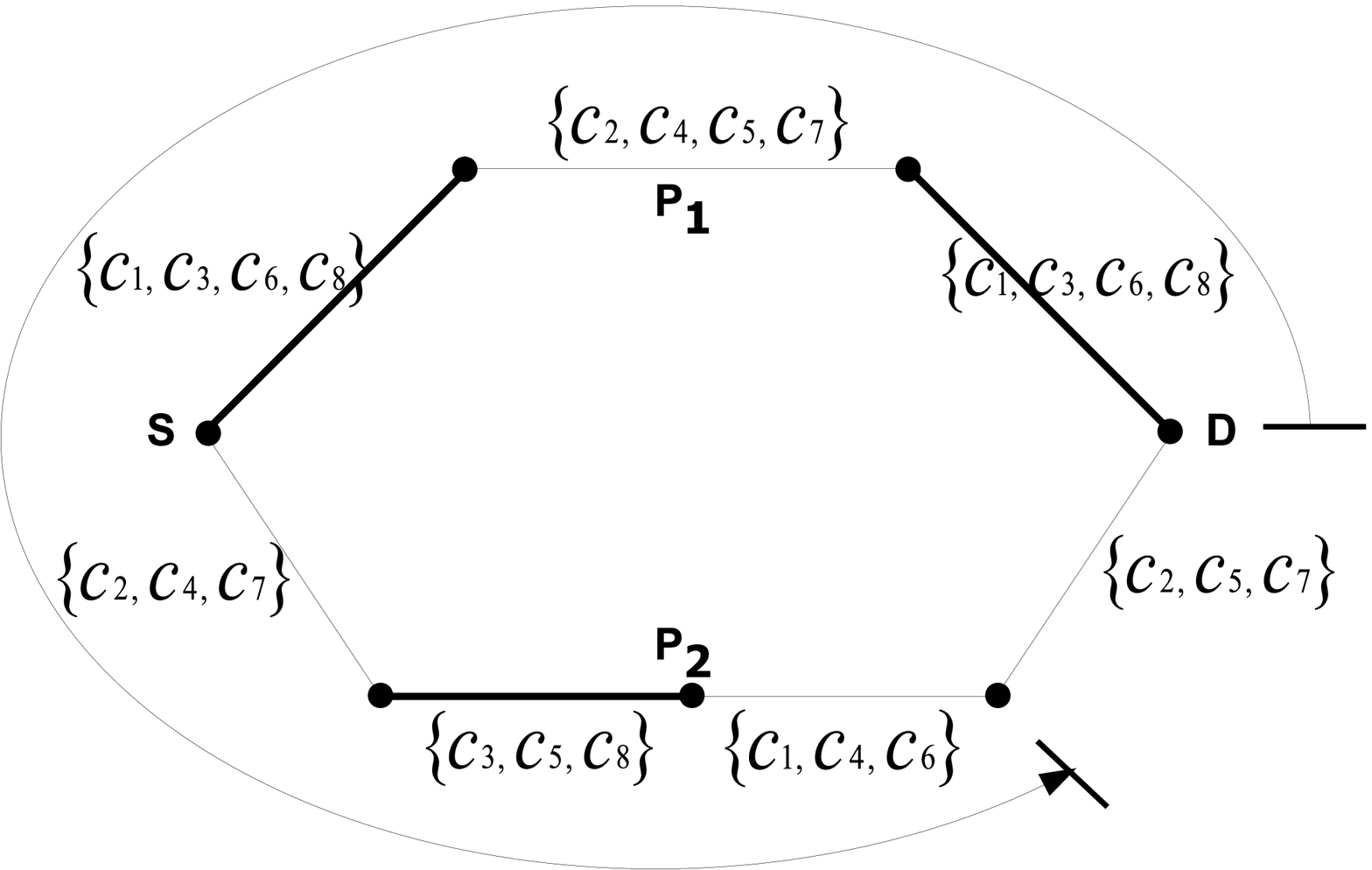}}
  \caption{Protocol Illustration: ($n_1$, $n_2$) = (3,4) [..contd.]}
  \label{fig:illustration_set_2}
\end{figure}

\subsubsection {Geometric Interpretation}

In this subsection, we interpret the protocol constructed by
Construction (\ref{constr:k2_max_rate}) in a geometric manner. We
assume $n_1 < n_2$ as in the previous section. As explained
earlier, at any given time instant a maximum of $\lfloor
\frac{n_1+n_2}{2} \rfloor$ edges can be active. Now $n_1 + n_2$ is
odd, and due to the half duplex constraint, only alternate edges
can be active. This means that, if we consider the entire network
at any time instant, every alternate edge will be colored except
for one place, where there will be two consecutive edges that are
not active. We will give the protocol by specifying at which two
consecutive places the edges will not be active, at every time
slot.

Consider the longer path and fix our pointer on the first edge
$e_{21}$ of the longer path $P_2$. Start a cycle from this edge
(consider the whole network as a cycle now), and activate
alternate edges beginning from the next edge following the pointer
in the clockwise direction, for the first time slot. This defines
the set of edges, which are active for the first time slot.
Hereafter, a set of edges which are simultaneously active at a
time slot will be referred to as the activation set for that time
slot. Now, move the pointer to the next edge $e_{22}$ of the
longer path $P_2$ and repeat the same procedure. Now the
activation set for the second time slot is defined. Continue the
procedure, moving the pointer to all of the edges $e_{2i}, \
i=1,2,...,n_2$. Thus the activation sets for the first $n_2$ time
slots of the protocol is specified. For the next $n_2$ time slots
of the protocol, the same procedure is followed, except that an
anti-clockwise cycle is used instead of clockwise cycle.

Thus the cycle length of the protocol equals $2{n_2}$. By using
this procedure, the edges in the shorter path $P_1$ always gets
activated every alternate time instant. So, each edge in the
shorter path gets $n_2$ colors. On the other hand, the edges on
the longer path $P_2$ also get activated alternately except that
they give up their transmission opportunity twice during the whole
duration of $2{n_2}$ time slots. So each edge in the longer path
$P_2$ gets $n_2 - 1$ colors.

\vspace{0.1in}

This illustrated with an example, $(n_1, n_2) = (3,4)$. In Fig.
\ref{fig:illustration_set_1}, activation sets for first $n_2$ time
slots of the protocol are defined. Here, we can observe that the
pointer moves in the clockwise direction. In Fig.
\ref{fig:illustration_set_2}, activation sets for the next $n_2$
time slots of the protocol are defined. Pointer is moved in the
clockwise direction in Fig. \ref{fig:illustration_set_1}; in
contrast, it is moved anti-clockwise in Fig.
\ref{fig:illustration_set_2}.

\bthm \label{thm:2PP_DMT} For a 2-PP network, if the two path
lengths are equal modulo $2$, then the DMT achieved by the
orthogonal protocol of Construction~\ref{constr:k2_max_rate} is
equal to the MISO bound, i.e., $d(r) = 2(1-r)^{+}$. \ethm

\bpf The proof follows from Lemma~\ref{lem:dmt_opt},
Lemma~\ref{lem:backflow_upper_bound} and
Theorem~\ref{thm:k2_max_rate}.\epf

\bthm \label{thm:KPP} For a KPP network, there exists an
orthogonal protocol achieving the MISO bound as long as $K \geq 3$
or $K=2$ and $n_1 = n_2 \text{ mod } 2$. \ethm

\bpf Clear by combining Theorem~\ref{thm:K_geq_4},
Theorem~\ref{thm:k3_miso} and Theorem~\ref{thm:2PP_DMT}. \epf

\subsection{KPP Networks with Direct Link}

\bthm \label{thm:KPP_D}For KPP(D) networks with half duplex
relays, single antenna nodes and with a direct link, the MISO
bound on DMT is achievable whenever there is an orthogonal
protocol avoiding back-flow that achieves the MISO bound in the
absence of direct link. \ethm

\bpf By hypothesis, the given KPP network with half duplex relays
and single antenna nodes, achieves optimal DMT in the absence of
direct link. We know by Theorem~\ref{thm:K_geq_4}, all KPP
networks with $K>3$ achieve optimal DMT.

Consider any KPP network with $K \geq 4$. We have also established
that there exists a protocol, $P$, with cycle length $K$,
achieving optimal DMT, in which the source sends one symbol each
through every path during one cycle. Now assume that a direct link
added between the source and the sink.

Define a protocol $P'$ as $P$ with a modification such that nodes
preceding the sink do not forward the symbols, but buffer them.
(Each node is assumed to have enough buffer length for this). The
protocol $P'$ is run for $D$ time slots on the network till all
the nodes preceding the sink have at least one symbol in their
buffer. Now switch back to the protocol $P$.

Up to and including $D$ time slots, the sink receives $D$ symbols
through the direct link. After $D$ time slots, the sink receives
one symbol through the direct link, and another through a relayed
path. By the definition of the protocol, each symbol transmitted
by the source reaches the sink node through the direct link, and
through exactly one relayed path. Note that each symbol arrives at
the sink through the direct link, and a relayed path with a delay
characteristic of the path. This is the same setting as in
Theorem~\ref{thm:main_theorem}, and we invoke the results from
there.

Let the total time slots elapsed be $M = mK + D$ for some positive
integer $m$. Then the lower bound for DMT, $d(.)$ is given by,

\beqan d(r) & \geq & d_D(r) + d_C(\frac{M}{M-D}r)) \\
\text{where,  } d_D(r) & = & (1-r)^+ \\
d_C(r) & = & K(1-r)^+ \\
\eeqan

As $m$ tends to infinity, the DMT lower bound coincides with the
cut-set bound, and thus the optimal DMT is achieved.

\epf

\section{Half Duplex KPP(I) networks \label{sec:half_duplex}}

In this section, we consider KPP networks in the presence of
interference links between paths, i.e., KPP(I) networks. There is
no direct link is KPP(I) networks as per the definition. We prove
that the MISO bound is achievable even in KPP(I) networks.

The basic idea here is to consider the backbone KPP network for
the given KPP(I) network. An orthogonal protocol is designed for
the backbone network. This protocol is run on the KPP(I) network.
It is obvious that there are now interference terms in the
transfer matrix. However, if the transfer matrix can be written as
a lower triangular matrix with the $K$ product coefficients on the
diagonal, then we can use Theorem~\ref{thm:main_theorem} and prove
that the MISO bound is achievable.

\subsection{Inteference does not impair DMT}
Next, we consider the case of causal interference, which we define
first.

\bdefn \label{def:Causal_Interference} Consider a KPP(I) network
with single antenna nodes. Let us operate the backbone KPP network
using an orthogonal protocol which induces an AF protocol on the
KPP(I) network. Let $H$ denote the channel matrix induced by the
AF protocol in the KPP(I) network and $H_1$ denote the diagonal
channel matrix induced by the orthogonal protocol in the backbone
KPP network. If the protocol is such that \bit \item $H$ is lower
triangular, \item Diagonal entries of $H$ are same as that of $H_1
$, \eit then the KPP(I) network is said to admit causal
interference under that protocol. \edefn

Now we prove a Lemma which asserts that the DMT of a KPP(I)
network with causal interference is same as that of the backbone
KPP network under the same protocol.

\blem \label{lem:causal_interference} Consider a KPP(I) network
with single antenna nodes, running on an AF protocol which admits
causal interference. Let the induced channel matrix be $H$, and
the diagonal part of $H$ be $H_d$. Then the DMT of $H$ is same as
that of $H_d$. \elem

\bpf  The presence of causal interference creates entries in the
strictly lower-triangular portion of the transfer matrix. Since
the DMT of a lower triangular matrix is lower bounded by the DMT
of the corresponding diagonal matrix, by
Theorem~\ref{thm:main_theorem}, $d_{H}(r) \geq d_{H_d}(r)$.

Now shall prove that $d_{H}(r) \leq d_{H_d}(r)$, which will
complete the proof of the lemma. Since an orthogonal protocol is
employed, all the entries in a row of the matrix $H$ will have a
common term $h_i$ corresponding to the fading coefficient of the
last link connecting to the sink. So with causal interference,
then the channel matrix would be \beqan H & = & \left[
\begin{array}{ccccc}
        h_1(g_{11})                 &   0               & \hdots    & &0\\
        h_2(g_{21})   & h_2(g_{22})               & 0     & &0\\
        h_3(g_{31})   & h_3(g_{32}) & h_3 (g_{33})   & &\\
        \vdots              &                   &\vdots &\ddots &\\
        h_n(g_{n1}) & \hdots        & \hdots      & &h_n (g_{nn})
        \end{array}
        \right],
\eeqan where every $g_{ij}$ is a polynomial function of Rayleigh
fading coefficients. Since the interference is causal, if the
network does not have interference links(i.e., in the backbone KPP
network), the same protocol would yield a channel matrix, \beqan
H_d & = & \left[
\begin{array}{ccccc}
        h_1(g_{11})                 &   0               & \hdots    & &0\\
        0   & h_2(g_{22})               & 0     & &0\\
        0   & 0 & h_3(g_{33})   & 0 &\\
        \vdots              &                   &\vdots &\ddots &\\
        0 & \hdots        & \hdots      & &h_n(g_{nn})
        \end{array}
        \right].
\eeqan

\beqan \text{Let   } H_1 & = & \left[
\begin{array}{ccccc}
        h_1                 &   0               & \hdots    & &0\\
        0   & h_2               & 0     & &0\\
        0   & 0 & h_3   & 0 &\\
        \vdots              &                   &\vdots &\ddots &\\
        0 & \hdots        & \hdots      & &h_n
        \end{array}
        \right].
\eeqan

$(I+\rho{H}{H}^\dagger)$ is a positive definite Hermitian matrix
and by invoking Theorem 16.8.2 of \cite{CovTho}, we have that the
determinant is upper bounded by the product of row-norms:

\beqa \det(I+\rho{H}{H}^\dagger) & < & (1 +
\rho|h_1|^2|g_{11}|^2)(1 +
\rho|h_2|^2|g_{22}|^2 + \rho|g_{21}|^2|h_2|^2)\cdots \nonumber \\
& & (1 + \rho|h_n|^2|g_{nn}|^2 + \rho|g_{n(n-1)}|^2|h_n|^2 +
\cdots
+ \rho|g_{n1}|^2|h_n|^2) \nonumber \\
& = & \prod_{i=1}^{n}(1 + \rho|h_i|^2(|g_{ii}|^2 + |g_{i(i-1)}|^2 + \cdots + |g_{i1}|^2)) \nonumber \\
& \doteq & \prod_{i=1}^{n}(1 + \rho|h_i|^2) \label{eq:scalar_bound_doteq1} \\
& = & \det(I+\rho{H_1}{H_1}^\dagger) \nonumber \eeqa

The dot equivalence \eqref{eq:scalar_bound_doteq1} follows from
equation \eqref{eq:rho} in the proof of
Lemma~\ref{lem:noise_white}.

\beqa \text{Now,  }\det(I+\rho{H}{H}^\dagger) & \dot \leq & \det(I+\rho{H_1}{H_1}^\dagger) \nonumber \\
 & \doteq  & \det(I+\rho{H_d}{H_d}^\dagger) \label{eq:h1_hd_equivalence} \\
\Rightarrow d_H(r) & \leq  & d_{H_d}(r) \nonumber \eeqa

Equation~\eqref{eq:h1_hd_equivalence} follows from the fact that
product of absolute value of Rayleigh random variables is
equivalent to a single Rayleigh random variable in the scale of
interest, as long as all the variables involved in the two
matrices $H_1$ and $H_d$ are independent. \epf

\subsection{Causal Interference}

By Lemma~\ref{lem:causal_interference}, it is clear that the
cut-set bound for a KPP(I) network can be attained if there is a
protocol that yields a lower triangular matrix with $K$
independent coefficients along its diagonal repeated periodically
(except  maybe the first $D$ time instants). Specifically if the
input-output relation can be written in the following form, then a
DMT of $d(r) = K(1-r)^+$ is achievable.

\beqa \left[ \begin{array}{c}
\bold{y_1}\\
\bold{y_2}\\
\vdots\\
\bold{y_K}\\
\end{array} \right]
 &=&  \left[\begin{array}{cccc}
    \bold{g_1} & & & \\
    * & \bold{g_2} & &\\
    * & * & \ddots &\\
    * & * & * & \bold{g_K} \\
    \end{array}\right]
    \left[ \begin{array}{c}
             \bold{x_1}\\
             \bold{x_2}\\
             \vdots\\
             \bold{x_K}\\
            \end{array} \right] + \bold{n} \label{eq:lower_triangular}
\eeqa

where $\bold{g_i} =\prod_{j=1}^{n_i} \bold{g_{ij}}$ and $*$
denotes any entry, either zero or non-zero.

This would be our aim in the rest of the section - to establish
when it is possible to find a protocol yielding such a channel
matrix. Let us first consider the KPP network without
interference, running on an orthogonal protocol. In this case, due
to the different delays on the different paths, an input-output
relation like Equation~\eqref{eq:lower_triangular} does not hold
immediately. In order to do so, first we consider a permutation of
the input for which it is possible to do so.

We consider symbols received by the sink from $n$th time instant
onwards, with $n$ sufficiently large enough, such that the sink
receives symbols from all the $K$ paths periodically. Consider the
received symbols $y_{n+1},y_{n+2},\ldots,y_{n+K}$, in $K$
consecutive time instants, each of the symbol traversing a
distinct path. Let the symbols received at time $n+i$ be $x_{m_i}$
and assume that the data comes through path $P_i$. Let us consider
the transfer matrix between $y_{n+1},y_{n+2},\ldots,y_{n+K}$ and
$x_{m_1},x_{m_2},\ldots,x_{m_K}$.

\beqa \left[ \begin{array}{c}
\bold{y_{n+1}}\\
\bold{y_{n+2}}\\
\vdots\\
\bold{y_{n+K}}\\
\end{array} \right]
 &=&  \left[\begin{array}{cccc}
    \bold{g_1} & & & \\
     & \bold{g_2} & &\\
     &  & \ddots &\\
     &  &  & \bold{g_K} \\
    \end{array}\right]
    \left[ \begin{array}{c}
             \bold{x_{m_1}}\\
             \bold{x_{m_2}}\\
             \vdots\\
             \bold{x_{m_K}}\\
            \end{array} \right] + \bold{n}  \label{eq:equiv_lower_triangular}
\eeqa

Now consider any KPP(I) network built on the above backbone KPP
network. We will give a sufficient condition on the interference
so that the channel matrix has a structure like
Equation~\eqref{eq:lower_triangular}.

\bprop \label{prop:Causal_Interference} If the interference in a
KPP network, running a particular protocol, has the following
property:

For each backbone path the following conditions are satisfied:
\bit \item \emph{Condition 1:} The delay experienced by data
travelling on any other path from the first node of the backbone
path should be no lesser than the delay on the backbone path from
the first node to the sink.
%There is no shorter path from the first node on the given path to the last node on any path.

\item \emph{Condition 2:} The unique shortest delay from the first
node on the given path to the last node on that path is through
the actual path from that node to the sink.
%The unique shortest path from the first node on the given path to the last node on that path is the actual path from that node to the sink.
\eit

Then the matrix connecting the output and a permuted version of
the input will be lower triangular with K independent coefficients
along its diagonal repeated periodically (except maybe the first
$D$ time instants).

\eprop

\bpf Consider the KPP network with interference. Reduce this to a
network without interference, i.e. assume that relays in different
paths are isolated from each other and write the input-output
transfer matrix as in Equation~\ref{eq:equiv_lower_triangular}.

Let us consider a given symbol $x_{m_i}$ transmitted from the
source. We are now looking for all possible ways in which this
data can reach the sink, since these contribute to the entries
other than the diagonal entries in the matrix that we are
interested in. We want to get a lower triangular matrix with the
$K$ product coefficients appearing on the diagonal.

A symbol from the source $x_{m_i}$ can get to a sink only after it
is passed through the first node on the actual path in which it
was intended to be sent if there were no interference. So we are
interested in all possible path delays from the first node on the
actual path to the sink.

If the data reaches through all other paths later than it does on
the backbone path, then the matrix is bound to be lower
triangular. This is ensured by \emph{Condition 1}. Now, we want
the coefficients on the diagonal to be equal to $g_i$.  This
requires that there is no path of same length splitting from a
path and merging back into the path with the same delay as the
actual path. This will add another coefficient to the $g_i$ which
might create a problem. To ensure that this does not occur, we
have \emph{Condition 2}.

More formally, since the network satisfies \emph{Condition 1} of
theorem above, we have that given that a symbol $x_{m_i}$
influences output $y_{n+i}$ through the shortest path, the same
symbol  $x_{m_i}$ will not influence any $y_j$, for $j<n+i$. Since
the network satisfies \emph{Condition 2} of theorem above, we have
that the symbol $x_{m_i}$ is coupled to $y_{n_i}$ through $g_i$,
since there is no other coefficient that sums to this.

This means that in the representation given by
Equation~\ref{eq:equiv_lower_triangular}, a given column
corresponding to the input $x_{m_i}$ will look like:
$\text{Column}_i = [0 \ 0 \ \hdots \  g_i \  * \ * \ \hdots \
*]^{\text{T}}$, where $*$ denotes some entry (zero or non-zero).

%
%\beqa \text{Column}_i & = & \left[ \begin{array}{c}
%0\\
%0\\
%\vdots\\
%g_i\\
%*\\
%*\\
%\vdots\\
%*\\
%\end{array} \right]  \label{eq:column}
%\eeqa

This clearly means that the matrix representation is lower
triangular with $g_i$ on the diagonal repeating periodically.
i.e., it is of the form \eqref{eq:lower_triangular} and therefore,
by Theorem~\ref{thm:main_theorem}, the upper bound on DMT is
achievable: $d(r) = K(1-r)$.

\epf

\bnote \label{remark:connection_to_protocol} The conditions in
this proposition depend on the actual delays experienced by the
data travelling through various paths. However, the actual delays
depend on the protocol used.  To simplify the criterion in terms
of characteristics of network topology, we define a class of
protocols with ``almost continuous activation'' in the next
section. This modified criterion can be computed by a simple
examination of the network.\enote

\subsection{Protocols with Almost Continuous Activation\label{sec:cont_activation}}

In this section, we define a class of protocols with ``almost
continuous activation`` where in conditions in
Proposition~\ref{prop:Causal_Interference} can be reduced to
conditions on the path lengths of the network.

\bdefn \label{defn:cont_activation_node} An orthogonal protocol
for a KPP network is said to have continuous activation at a relay
node if the node transmits whatever it receives from the incoming
edge in the last instant in the immediately next time
instant.\edefn

\bdefn \label{defn:cont_activation} An orthogonal protocol for a
KPP network is said to have continuous activation if the protocol
has continuous activation at \emph{all} relay nodes.\edefn

\bdefn \label{defn:almost_cont_activation} An orthogonal protocol
for a KPP network is said to have almost continuous activation if
the protocol has continuous activation at all relay nodes
\emph{except} possibly the first hop node on each parallel
path.\edefn

Protocols with almost continuous activation will be used in the
future sections to establish a sufficient condition for
achievability of DMT upper bound. Protocols with almost continuous
activation have the property that the data passes continuously
through the edges of the backbone paths of the KPP network in
successive instants after the first hop.

\bthm \label{thm:KPP_almost_cont} For a KPP network without
interference, there exists a protocol with almost continuous
activation whenever $K \geq 3$. \ethm

\bpf Let us assume without loss of generality that the paths are
ordered in ascending order of their sizes ordered modulo $K$. Let
us consider a given path $P_i$. Let us fix the color on the first
edge to be $c_i$, i.e., $A_{i1}  = c_i$.

The next edge can be anything other than $c_i$ in order to satisfy
the half duplex constraint. Once the color on the next edge is
fixed, the colors on the rest of the edges are known because the
protocol must have almost continuous activation. Let the next edge
have color $c_m$. $A_{i2} = c_m$ and we know that $m \neq i$. So
we must color the remaining edges consecutively: $A_{ij} =
c_{m+j-2}, j \geq 2$.

We have $K-1$ choices for $m$ and therefore these will lead to
$K-1$ different colors for the last edge $e_{i{n_i}}$. These are
all possible colors $c_1,c_2,...,c_K$ except the one color that
will appear on the last edge if $A_{i2}  = c_i$. Let us try to
determine the one color that \emph{can not} appear on the last
edge, because if it does, then the half duplex constraint will be
violated.

Let $n_i = a \text{ mod } K$. Then if $A_{i2} = c_i$, then
$A_{i{n_i}} = c_{ (i+a-2) \text{ mod }} K$.

This means that if the starting color is $c_i$, then there are
$K-1$ colors allowed except the one stated here: $A_{i{n_i}} \neq
c_{ (i+n_i-2) \text{ mod } K }$. Let $S_i = C \setminus  \{ c_{
(i+n_i-2) \text{ mod } K } \}$. Therefore, $S_i$ is the set of all
allowed colors on the last edge in path $P_i$. We represent this
symbolically by $c_i  \longleftrightarrow c_j, \forall c_j \in
S_i$, where $\longleftrightarrow$ denotes the terminal edge
compatibility relation.

Now we have a set of starting colors $A = \{ c_i, i=1,2,...,K \}$.
The set of ending colors (i.e., the colors on the ending edges)
should also be the set $B = \{ c_i, i=1,2,...,K \}$ since we want
a rate one protocol. Now visualize a bipartite graph $\mathcal{G}$
between the sets $A$ and $B$. Where $c_i$ in $A$ is connected to
$c_j$ in $B$ if $c_j \in S_i$.

\bdefn A complete matching on this bipartite graph $\mathcal{G}$
is a subgraph of $\mathcal{G}$ where every node in $A$ is
connected to exactly one node in $B$ and these nodes in $B$ are
distinct. \edefn

Any complete matching on $\mathcal{G}$ specifies a protocol with
almost continuous activation and vice versa, since a protocol with
almost continuous activation is specified by just the starting and
the ending colors. From the theory of bipartite matching
\cite{LinWil}, we have the following proposition:

\bprop \label{prop:matching} Let $\mathcal{G}$ be a bi-partite
graph from set $A$ to set $B$. Let $X \subset A$ be any subset of
$A$. A complete matching from $A$ to $B$ exists iff \beqa
|\Gamma(X)| \geq |X|, \forall X \subset A \label{eq:matching}
\eeqa where $\Gamma(X)$ denotes the set of all nodes that are
adjacent to any node in $X$ on the graph $\mathcal{G}$. \eprop

\bprop \label{prop:our_matching} The bipartite graph $\mathcal{G}$
has a complete matching whenever $K \geq 3$ \eprop

\bpf The bipartite graph $\mathcal{G}$ has a complete matching iff
$ |\Gamma(X)| \geq |X|, \forall X \subset A$.

Since each element in $A$ is connected to $K-1$ nodes in the set
$B$, we have that $|\Gamma(X)| \geq K-1,\forall X \subset A, X
\neq \Phi $. This means that the condition is satisfied
automatically for the sets for which $0 < |X| \leq K-1$.

Now the only condition to check is when $|X| = K$. In this case
the condition \eqref{eq:matching} reduces to \beqa \cup_{i=1}^{K}
S_i = C \eeqa

 This condition is violated $\iff$ all the
$S_i$ are equal.

$\iff$ All the $c_{ (i+n_i-2) \text{ mod } K }$ are equal.

$\iff$ All the $ (i+n_i-2) \text{ mod } K$ are equal.

$\iff$ All the $ (i+n_i) \text{ mod } K$ are equal to $M$ (say).

%$\iff$ All the $ (n_i) = M-i \text{ mod } K$ are equal.

$\iff$ All the $n_i$ are distinct modulo $K$ and $(i+n_i) \text{
mod } K$ are equal to $M$, for $i=1,2$.

Now since, all the $n_i$ are distinct modulo $K$ and the paths are
ordered in ascending order of their sizes ordered modulo $K$, we
have $n_i = i-1 \text{ mod } K$.

$\Rightarrow$ All the $n_i$ are distinct modulo $K$ and $(1 + 0)
\text{ mod } K = (2+1) \text{ mod } K $.

$\iff$ all the $n_i$ are distinct modulo $K$ and $0 = 2 \text{ mod
} K $.

$\Rightarrow$ $K \leq 2$.

Therefore there is \emph{no} complete matching on the bipartite
graph $\Rightarrow$ $K \leq 2$. The contra-positive of this
statement is that,

$K > 2$ $\Rightarrow$ There is a complete matching on the
bipartite graph.

Therefore a complete matching exists whenever $K \geq 3$. This
proves the proposition. \epf

Since a protocol with almost continuous activation exists whenever
a complete matching on the corresponding bipartite graph exists,
we have that protocols with almost continuous activation exist
whenever $K \geq 3$. Hence the theorem \epf

Now, we can translate conditions on the delay in
Proposition~\ref{prop:Causal_Interference} into conditions on path
lengths while using protocols with almost continuous activation.
This is formalized in the following proposition:

\bprop \label{prop:Modified_Causal_Interference} If the
interference in a KPP network, running a protocol with almost
continuous activation, has the following two properties, then the
matrix connecting the output and a permuted version of the input
will be lower triangular with K independent coefficients along its
diagonal repeated periodically (except maybe the first $D$ time
instants). For each backbone path,

\bit \item \emph{Condition 1:} The length of any other path from
the first node should be no lesser than the delay on the backbone
path from the first node to the sink.

\item \emph{Condition 2:} The unique shortest path from the first
node on the given path to the last node on that path is through
the backbone path from that node to the sink. \eit \eprop

\bpf This follows directly from
Proposition~\ref{prop:Causal_Interference} and
Remark~\ref{remark:connection_to_protocol}. \epf

\subsubsection{Optimal DMT for regular networks \label{sec:regular_dmt}}
Now we show that the MISO bound is achievable for regular
networks.

\bthm \label{thm:knregular_dmt} The optimal DMT $d(r)=L(1-r)^{+}$
of (K,L) Regular networks is achievable. \ethm

\bpf  Consider a (K,L) regular network. It can be treated as a
KPP(I) network and therefore the back-bone KPP network can be run
using an orthogonal protocol with almost continuous activation.
Consider the following protocol with almost continuous activation.
Let the colors be $c_1,c_2,...,c_K$, and assume $c_0 = c_K$ and
$c_{\ell} = c_{l \text{ mod } K}$.

$A_{ij} = c_{i+(j-1)}, i=1,2,..,K, j=1,2,..,L+1$.

With this protocol it can be seen that interference is causal,
i.e., interference satisfies the conditions of
Prop.~\ref{prop:Causal_Interference}. Therefore, the optimal DMT
of $L(1-r)^{+}$ is achievable for these networks. \epf

\bcor \label{cor:two_path} For a (2,L) layered network, a lower
triangular transfer matrix which contains the two product
coefficients corresponding to the two parallel paths alternately
on the diagonal can be obtained using the protocol with almost
continuous activation. \ecor

\bcor \label{cor:two_hop_DMT} For the two-hop relay network
without direct link, the optimal DMT is achieved. \ecor

\bpf The two-hop relay network without the direct link is a (K,1)
regular network, where $K$ denotes the number of relays in the
network. Thus Theorem~\ref{thm:knregular_dmt} implies this
corollary. \epf

\bnote The result in Corollary~\ref{cor:two_hop_DMT} was also
proved in an independent work \cite{GhaBayKha}. The protocol used
in this paper and in \cite{GhaBayKha} are essentially the same as
the SAF protocol \cite{YanBelSaf}, except that it is used in a
network without direct link. However, the proof techniques used
here and in \cite{GhaBayKha} are very different. \enote

\subsubsection{Optimal DMT for KPP(I) networks}

In this section, we prove that the MISO bound can be achieved on
\emph{all} KPP(I) networks, with $K \geq 3$.

 In Prop.~\ref{prop:Causal_Interference}, we gave a
sufficient condition to establish when a network can be used along
with a given protocol in order to achieve the optimal DMT. Later
in Prop.~\ref{prop:Modified_Causal_Interference}, we gave a
sufficient condition on path lengths in a network such that the
network can be used along with a protocol with almost continuous
activation to get the optimal DMT.  Suppose the network does not
meet the sufficient condition given in
Prop.~\ref{prop:Modified_Causal_Interference}. It is possible that
the protocol can be modified to make the network meet the
sufficient condition of Prop.~\ref{prop:Causal_Interference}. We
do so here by adding delays to internal nodes of the network such
that, even though the path lengths do not satisfy the constraints,
the delays do. By appropriately choosing a protocol and adding
delays, we can make the network and the protocol jointly satisfy
the conditions of Prop.~\ref{prop:Causal_Interference}. This leads
us to the following Theorem:

\bthm \label{thm:KPP_2_3_Causal} Consider a KPP(I) network with
$K=3$. There exists a set of delays which when added appropriately
to various nodes in the networks, and when used along with the
protocol with almost continuous activation, satisfies the
conditions of Prop.~\ref{prop:Causal_Interference}. \ethm

\bpf The proof is omitted here for brevity. The proof makes use of
decomposing the given network into various layers, each of which
can be balanced individually and the layers can put together to
give a solution for the entire network. \epf

\bthm \label{thm:KPP_I} Consider a KPP(I) network with $K \geq 3$.
The cut-set bound on the DMT $d(r) = K(1-r)^+$ is achievable.
\ethm

\bpf For $K=3$, it follows from Theorem~\ref{thm:KPP_2_3_Causal}.

Now, we will consider the case when $K > 3$. Consider a $3$
parallel path sub-network of the original network. By
Theorem~\ref{thm:KPP_2_3_Causal}, we can get a matrix with these
three product coefficients along the diagonal. There are now
$^KC_3$ possible $3$PP subnetworks. If each of these subnetworks
is activated in succession, it would yield a lower triangular
matrix with all the $K$ product coefficient $g_i$ repeated thrice
$K$ choose $3$ times on the diagonal. By
Theorem~\ref{thm:main_theorem}, the DMT of this matrix is better
than that of the diagonal matrix alone. The diagonal matrix has a
DMT equal to $K(1-r)^+$. Therefore a DMT of $d(r) \geq K(1-r)^{+}$
can be obtained. However, since $d(r) \leq K(1-r)^{+}$ by cutset
bound, we have $d(r) = K(1-r)^{+}$. \epf

\section{Layered Networks \label{sec:layered}}

\blem \label{lem:product_channel_DMT} Let $\mathcal{H} \subset \{
h_{11},h_{12},...,h_{1M_1} \} \times \{ h_{21},h_{22},...,h_{2M_2} \} \times ...
\times \{ h_{K1},h_{K2},...,h_{KM_K} \}$. Let $|\mathcal{H}| = N$. Let each $h_{ij}$
appear in $N_i$ of the terms in $\mathcal{H}$ irrespective of $j$. Then $N_iM_i =
N$. Let $N_{\text{max}} := \max_{i=1}^{N} N_i$ and $M_{\text{min}} := \min_{i=1}^{K}
M_i$.

Let ${h}_i, \ i=1,2,..,N$ be the elements of $\mathcal{H}$.

Let $\psi: \mathcal{H} \rightarrow G$ be a map such that $\psi( (
a_1, a_2,...,a_K ) ) = \Pi_{j=1}^{K} {a}_{i}$. Now let $g_i =
\psi(H_i), \ i=1,2,...,N$. Then each $g_i$ is of the form
$\Pi_{k=1}^{K} h_{kl(i,k)} $, where $l(i,k)$ is a map from $[N]
\rightarrow [M_k]$ for a fixed $k \in [K]$.

Let $H$ be a $N \times N$ diagonal matrix with the diagonal
elements given by $H_{ii}=g_i$.

The DMT of the parallel channel $H$ is a linear DMT between a diversity of
$\frac{N}{N_\text{max}}$ and a multiplexing gain of $N$: \beqa d(r) & = &
\frac{(N-r)^{+}}{N_\text{max}} \eeqa \elem

\bpf

Let us assume without loss of generality that $N_1 \geq N_2 \geq ... N_K$.

%Define a variable transformation: $\alpha_{ij}$ : $\rho^{-\alpha_{ij} = |h_ij|^2$.
%Now a distribution is induced on $\alpha_{ij}$.
$H = $ diag $(H_{ii})$. $H_{ii} = \Pi_{k=1}^{K} h_{kl(i,k)} $.

Consider a variable transformation where $\alpha_{kj}$ is defined
such that $\rho^{-\alpha_{kj}} = |h_{kj}|^2$.

Now the DMT $d(r)$ is given by the following defining equation:
\beqa \rho^{-d(r)} & = & Pr\{\log\det(I+\rho{H}{H}^\dagger) \leq r\log\rho\}  \nonumber \\
& = & Pr\{\det(I+\rho{H}{H}^\dagger) \leq \rho^r \}  \nonumber \\
& = & Pr\{\Pi_{i=1}^{N} (1+\rho{|H_{ii}|}^2) \leq \rho^r \}  \nonumber \\
& = & Pr\{\Pi_{i=1}^{N} (1+\rho{ \Pi_{k=1}^{K} |h_{kl(i,k)}|^2 }) \leq \rho^r \}  \nonumber \\
& = & Pr\{\Pi_{i=1}^{N} (1+\rho{ \Pi_{k=1}^{K} \rho^{-\alpha_{kl(i,k)}} }) \leq
\rho^r \}  \nonumber \\
& = & Pr\{\Pi_{i=1}^{N} (1+{  \rho^{1-\sum_{k=1}^{K} \alpha_{kl(i,k)}} }) \leq
\rho^r \}  \nonumber \\
& \doteq & Pr\{\Pi_{i=1}^{N} {  \rho^{(1-\sum_{k=1}^{K} \alpha_{kl(i,k)})^{+}} }
\leq \rho^r \} \nonumber  \\
& = & Pr\{\sum_{i=1}^{N} {  {(1-\sum_{k=1}^{K} \alpha_{kl(i,k)})^{+}} } \leq r \} \label{eq:exact_layered}  \\
&  \leq & Pr\{\sum_{i=1}^{N} {  {(1-\sum_{k=1}^{K} \alpha_{kl(i,k)})} } \leq r \}  \label{eq:approx_layered} \\
& = & Pr\{ N - \sum_{k=1}^{K} { N_k \sum_{j=1}^{M_k} \alpha_{kj} } \leq r \}
\nonumber \eeqa

The last equality follows since each $|h_{ij}|^2$ appear in $N_i$ of the terms in
$\mathcal{H}$ irrespective of $j$ and so do the corresponding $\alpha_{ij}$. Let
$d_1(r)$ be defined as the SNR exponent of the RHS in the last equation above, i.e.,

\beqa Pr\{ N - \sum_{k=1}^{K} { N_k \sum_{j=1}^{M_k} \alpha_{kj} } \leq r \} & = &
\rho^{-d_1(r)} \eeqa

Now, \beqa d(r) & \geq & d_1(r) \\
& = & \inf_{ \{ N - \sum_{k=1}^{K} { N_k \sum_{j=1}^{M_k} \alpha_{kj} } \leq r  \ ,
\ \alpha_{kj} \geq 0 \} } { \ \ \sum_{k=1}^{K}  \sum_{j=1}^{M_k} \alpha_{kj} } \eeqa

Define \beqa \alpha_k & := & \sum_{j=1}^{M_k} \alpha_{kj} \eeqa

\beqa d_1(r) & = &  \inf_{ \{ N - \sum_{k=1}^{K} { N_k  \alpha_{k} } \leq r \ , \
\alpha_{k} \geq 0 \} } { \
\ \sum_{k=1}^{K}  \alpha_{k} } \\
& = &  \inf_{ \{ \sum_{k=1}^{K} { N_k \alpha_{k} } \geq N - r \ , \  \alpha_{k} \geq
0 \} } { \ \ \sum_{k=1}^{K} \alpha_{k} } \eeqa

\emph{Claim:} The infimum of $\sum_{k=1}^K \alpha_k$ under the
constraint $\{ \sum_{k=1}^{K} { N_k \alpha_{k} } \geq N - r \} \ ,
\ \alpha_{k} \geq 0$ is attained by $ \alpha_1 = \frac{N -
r}{N_1}, \ \alpha_i = 0, \forall i = 2,...,N $ and the value of
the infimum is $\frac{N - r}{N_1}$.

\emph{Proof:} The proof is simple and is skipped here.

%Since the objective function is convex, local minimum is the same as global minimum.
%It is sufficient to prove that the stated $\{ \alpha_i \}$ is a local minimum. Let
%us assume that $\alpha_i^{'} = \delta_i \geq 0, i=2,...,K$. Then
%
%
%\beqan \sum_{k=1}^{K} { N_k \alpha_{k}^{'} } & \geq & N - r \\
% { N_1 \alpha_{1}^{'} } & \geq & N - r - \sum_{k=2}^{K} { N_k \alpha_{k}^{'} } \\
% N_1 \sum_{k=1}^K { \alpha_{k}^{'} } & \geq & N - r + \sum_{k=2}^{K} {(N_1 - N_k)  \delta_{k} } \\
% \sum_{k=1}^K { \alpha_{k}^{'} } & \geq & \frac{N - r}{N_1} + \sum_{k=2}^{K} {\frac{N_1 - N_k}{N_1} \delta_{k}} \\
%\sum_{k=1}^K { \alpha_{k}^{'} } & \geq & \frac{N - r}{N_1}  \eeqan
%
%The last equation follows since $\sum_{k=2}^{K} \frac{N_1-N_k}{N_1} \delta_{k} \
%\geq \ 0 $ because $N_1 \geq N_k, k=2,..,N$ and $\delta_k \geq 0$, for $k=2,..,N$.
%
%Therefore $ \alpha_1 = \frac{N - r}{N_1}, \ \alpha_i = 0, \forall i = 2,...,N $ is a
%local minimum, and thereby a global minimum.
%
This claim implies that $d(r) \geq d_1(r) = \frac{N - r}{N_1}$.

Now we will check that this lower bound is infact equal to the DMT
of the channel. Let us consider an assignment of $\alpha_{kj}$
suggested by the claim above: $\alpha_{1j} = \frac{N - r}{N_1 M_1}
= \frac{N-r}{N}, j=1,2,..,M_1$.

From \eqref{eq:exact_layered}, we know that \beqa d(r) & = &
\inf_{ \{ \sum_{i=1}^{N} { {(1-\sum_{k=1}^{K}
\alpha_{kl(i,k)})^{+}} } \leq r  \ , \ \alpha_{kj} \geq 0 \} } { \
\ \sum_{k=1}^{K} \sum_{j=1}^{M_k} \alpha_{kj} } \eeqa

We have to verify that this assignment yields the infimum under
the constraint stated here.

\emph{Claim:} The infimum of $\sum_{k=1}^{K} { \sum_{j=1}^{M_k}
\alpha_{kj}}$ under the constraint $\{\sum_{i=1}^{N} {
{(1-\sum_{k=1}^{K} \alpha_{kl(i,k)})^{+}} } \leq r  \ , \
\alpha_{kj} \geq 0 \}$ is attained by $\alpha_{1j} = \frac{N -
r}{N}, j=1,2,..,M_1$, $\alpha_{kj} = 0, \forall k>1$ and the value
of the infimum is $\frac{N - r}{N_1}$.

\emph{Proof:} Since the objective function is convex, local
minimum is the same as global minimum. It is sufficient to prove
that the stated $\{ \alpha_{kl} \}$ is a local minimum. To prove
that, we show that the objective function does not decrease in a
neighbourhood of the claimed optimal point. Let us assume that
$\alpha_{kl}^{'} = \delta_{kl} \geq 0, i=2,...,K$.

Since $\alpha_{1j} = \frac{N - r}{N} \leq 1$, we have that all
terms in the summation $\sum_{i=1}^{N} { {(1-\sum_{k=1}^{K}
\alpha_{kl(i,k)})^{+}} }$ are non-zero. By choosing $\delta_{kl}$
small enough, we can ensure that all terms in the summation are
non-zero.
\beqan \sum_{i=1}^{N} { {(1-\sum_{k=1}^{K} \alpha^{'}_{kl(i,k)})^{+}} } & \leq & r \\
 \sum_{i=1}^{N} { {(1-\sum_{k=1}^{K} \alpha^{'}_{kl(i,k)})} } & \leq & r \\
 N - \sum_{k=1}^{K} { N_k \sum_{j=1}^{M_k} \alpha_{kj}^{'} } & \leq & r \\
 \sum_{k=1}^{K} { N_k \sum_{j=1}^{M_k} \alpha_{kj}^{'} } & \geq & N - r \\
 { N_1 \sum_{j=1}^{M_k} \alpha_{1j}^{'} } & \geq & N
- r - \sum_{k=2}^{K} {  N_k \sum_{j=1}^{M_k} \alpha_{kj}^{'} } \\
N_1 \sum_{k=1}^{K} {  \sum_{j=1}^{M_k} \alpha_{kj}^{'} } & \geq &
N - r + \sum_{k=2}^{K} { (N_1 - N_k) \sum_{j=1}^{M_k}
\alpha_{kj}^{'} } \\
\sum_{k=1}^{K} { \sum_{j=1}^{M_k} \alpha_{kj}^{'} } & \geq
& \frac{N - r}{N_1} + \sum_{k=2}^{K} { \frac{N_1 - N_k}{N_1} \sum_{j=1}^{M_k} \delta_{kj}  } \\
\sum_{k=1}^{K} { \sum_{j=1}^{M_k} \alpha_{kj}^{'} } & \geq &
\frac{N - r}{N_1} \eeqan

The last equation follows since ${N_1-N_k} \geq \ 0, \ k \geq 2 $ and $\delta_{kj}
\geq 0$.

Therefore $\alpha_{1j} = \frac{N - r}{N}, j=1,2,..,M_1$,
$\alpha_{kj} = 0, \forall k>1$ is a local minimum, and thereby a
global minimum. This yields a DMT of

\beqan d(r) & = & \sum_{k=1}^{K} {\sum_{j=1}^{M_k} \alpha_{kj}}  =  {\sum_{j=1}^{M_1} \alpha_{1j}}\\
d(r) & = &  {\sum_{j=1}^{M_1} \frac{N - r}{N}}  = { \frac{N -
r}{N_1}} \eeqan

Thus $d(r) = d_1(r) = \frac{N - r}{N_1}$ is indeed the DMT of the channel described.

\epf

\bdefn \label{def:Bipartite_Graph} Given a set of paths $P$ in a
layered network, the bipartite graph corresponding to the path set
$P$ is defined as follows: \bit \item Construct a bi-partite graph
with vertices $P$ on the left and vertices $P$ again on the right.
\item Connect an element $P_i$ on the left to $P_j$ on the right
if the two paths are node disjoint. \eit \edefn

\blem \label{lem:Bipartite_Graph} Consider a set of paths $A := \{
a_i,i=1,2,...,N \} $ in a given layered network. Let the product
of the fading coefficient on the $i$-th edge disjoint path $a_i$
be $g_i$. Construct the bi-partite graph corresponding to $A$
according to Definition.~\ref{def:Bipartite_Graph}. If there
exists a complete matching in this bi-partite graph, then these
edges can be activated in such a way that the DMT of this protocol
is greater than or equal to the DMT of a parallel channel with
fading coefficients $g_i,i=1,2,...,N$ with the rate reduced by a
factor of $N$, i.e., $d(r) \geq d_{H_d}(Nr)$, where $H_d =
diag(g_1,g_2,...,g_N)$ \elem

\bpf

Suppose there is a complete matching $\pi$ on the graph
constructed as above. The complete matching specifies for every
edge disjoint path on the left $a_i$, a partner on the right
$a_{\pi_i}$. The length of each path and therefore the delay is
equal to $D:=L+1$.

Step - 1 : Activate path $a_1$ along with path $a_{n_1}$  for a
period $2T$, where $T>D$: treating these two paths as a $2-PP$
Network, since these two paths are node disjoint.This network
potentially has interference, but no direct link. Since this
network is a subnetwork of a layered network, this 2-PP network
has both the edges to be of the same length and causal
interference and therefore rate-1 can be achieved on this network
by Corollary~\ref{cor:two_path}. So the technique used in
Section.~\ref{sec:regular_dmt} can be used on this network to get
a  matrix, with zeros on the first $D$ rows. After deleting these
$D$ rows, the matrix will be lower triangular due to causal
interference and the diagonal in the matrix comprised of
coefficients equal to $g_1$ and $g_{n_1}$ alternately for $T-D$
durations each. After this is done, the various nodes in the
network store the data that have not yet been passed to the sink.
This data will be used in the future when this path is activated
again.

Step - 2 : Repeat Step - 1 for all the paths $a_1,...,a_N$. The net transfer matrix
will comprise $ND$ zero rows, which effectively signifies a rate loss.

On removing these zero rows we get a transfer matrix, $H$. The DMT
of the protocol is $d(r) = d_{H}(2NTr)$. By using
Theorem~\ref{thm:main_theorem}, we get that $d_{H}(r) \geq
d_{H_1}(r)$, where $H_1$ is the diagonal matrix corresponding to
the matrix $H$. But $H_1$ contains $2T-D$ entries each of $g_i$,
therefore this matrix DMT is given by $d_{H_1}(r) =
d_{H_d}(\frac{1}{2T-D}r)$  where $H_d = diag (g_1,...,g_N)$.
$\Rightarrow d(r) = d_H(2NTr) \geq d_{H_1}(2NTr) = d_{H_d}(
N\frac{2T}{2T-D}r)$.

For $T$ tending to infinity, we get $d(r) \geq d_{H_d}(Nr)$.

\epf

\bnote This activation can also be done in a cyclic way in order to reduce the delay
of data transfer. In the modified scheme, the method used above can be repeated for
$L$ cycles. Now, instead of letting $T$ going to infinity, we can tend $L$ to
infinity to get the same DMT as above. \enote

A sufficient condition that guarantees that a linear DMT between
the maximum diversity and multiplexing gain on a general layered
network is given in Lemma~\ref{lem:General_layered_network}.

\blem \label{lem:General_layered_network} For a general layered
network, a linear diversity multiplexing tradeoff of $d(r) =
d_\text{max}(1-r)^{+}$ between the maximum diversity gain
$d_\text{max}$ and the maximum multiplexing gain $1$ is achievable
whenever the bipartite graph corresponding to the set of edge
disjoint paths $e_i$, $i=1,2,...,d_\text{max}$ from the source to
the sink has a complete matching.\elem

\bpf

By using Lemma~\ref{lem:Bipartite_Graph} we will be able to get a
DMT of $d(r) = d_{H_d}(d_\text{max}r)$. But since the paths are
edge disjoint, the fading coefficients are independent, we get
$d_{H_d}(r) = ( d_\text{max} - r )^{+}$. Therefore, we get, $d(r)
= d_{\text{max}}(1-r)^{+}$ \epf

\bdefn \label{def:FD_paths} A path from a source to sink in a
layered network is said to be \emph{forward-directed} if all the
edges in the path are directed from one layer to the next layer
towards the sink (i.e., no edge in the path goes from one layer to
the previous layer and there is no edge which starts and ends in
the same layer.) \edefn

\blem \label{lem:complete_matching} Let $P_1,..,P_N$ be the set of
all forward directed paths in a fully connected layered network.
Then the bipartite graph of the path set $P$ has a complete
matching. \elem

\bpf We will prove this by producing an explicit complete matching
on the bipartite graph. Let the layered network have $L$ layers.
Let there be $R_i$ relays in the $i$-th layer. Let us fix an
(arbitrary) ordering on the relays in each hop. Let the relays in
the $j$-th hop be indexed $0,1,...,R_j-1$. The number of paths is
given to be equal to $N$.

A forward-directed path $P_i$ is specified completely if all the
relays through which the path passes. This is denoted by the $L$
tuple $B_i=(b_{i1},...,b_{iL})$, where $b_{ij}$ denotes the index
of the relay in the $j$-th hop through which path $P_i$ passes.
Each $L$-tuple specifies a path from source to sink, since the
layered network is fully connected. Now in this notation, two
forward-directed paths $P_i$ and $P_j$ are node-disjoint if the
tuples $B_i$ and $B_j$ are distinct in all the $L$ positions.

Consider a map $\alpha:P \rightarrow P$, where \beqan \alpha(P_i)
= \alpha(B_i) = \alpha(b_{i1},b_{i2},...,b_{iL}) & = & (b_{i1}+1
\text{ mod } R_1,b_{i2}+1 \text{ mod } R_2,...,b_{iL} \text{ mod }
R_L). \eeqan

It can be checked that this map is a bijection from $P$ to $P$.
Since $R_i > 1 \ \forall i$, $B_i$ and $\alpha(B_i)$ are
point-wise distinct, and thereby the paths $P_i$ and $\alpha(P_i)$
are node disjoint. Therefore the map $\alpha$ defines a complete
matching on the graph. \epf

\bthm \label{thm:fully_connected_layered} For a fully-connected
layered network, a linear DMT between maximum diversity and
maximum multiplexing gain of $1$ is achievable. \ethm

\bpf Consider a fully connected layered network with $L$ layers.
Let there be $R_i$ relays in the $i$-th layer for $i=0,1,...,L+1$.
Let $R_0=R_{L+1}=1$ since there is one source and one sink and
$M_i := R_{i-1}R_i, i=1,2,...,L+1$ be the number of fading
coefficients in the $i$-th hop. Let $h_{ij}, j=1,2,..,M_i$ be the
fading coefficients on the $i$-th hop for $ i=1,2,...,L+1$. Let
$N$ be the total number of forward-directed paths from source to
sink, and $P_i,i \in [N]$ be the various forward-directed paths.
Let $P$ denote the set of all these forward-directed paths. Then
$|P| = N = \Pi_{i=1}^{L} R_i$. Let $g_i$ be the product fading
coefficient on path $P_i$.

Let $M_\text{min} = \min_{i=1}^{L+1} M_i$. Then $d_\text{max} =
M_\text{min}$ by Theorem~\ref{thm:mincut}.

By Lemma~\ref{lem:complete_matching}, the bipartite graph
corresponding to $P$ has a complete matching. $P$ satisfies the
criterion of Lemma~\ref{lem:Bipartite_Graph} and therefore, we can
obtain a DMT of $d(r) \geq d_{H_d}(Nr)$. Now, we need to compute
$d_{H_d}(r)$. To that effect, we make the following observations,
which will enable us utilize Lemma~\ref{lem:product_channel_DMT}.

A given path $P_i$ can be alternately represented as the set $G_i
= (h_{1l(i,1)}, h_{2l(i,2)},...,h_{(L+1)l(i,{L+1})})$  of fading
coefficients on that path. Consider the set of all $G_i$, i.e., $G
= \{ G_i,i \in [N] \}$.

Now let $g_k, k \in [N]$ be the product fading coefficient on path
$G$. Now clearly \beqan G & \subset & \{
h_{11},h_{12},...,h_{1M_1} \} \times \{ h_{21},h_{22},...,h_{2M_2}
\} \times ... \times \{ h_{(L+1)1},h_{(L+1)2},...,h_{(L+1)M_{L+1}}
\} \eeqan

Now each $h_{ij}$ appears in the same number $N_i$ of terms in $G$
irrespective of $j$, where $N_i = \frac{N}{M_i}$ and
$N_\text{max}= \max_{i=1}^{L+1} N_i$.

If $\psi$ is defined as in Lemma~\ref{lem:product_channel_DMT},
then $g_i = \psi(G_i)$. Now we have satisfied all the conditions
of Lemma~\ref{lem:product_channel_DMT} and therefore, $d_{H_d}(r)
= \frac{N-r}{N_\text{max}}$.

Now \beqan d(r) & \geq & d_{H_d}(Nr) \\
& = & \frac{(N-Nr)^{+}}{N_\text{max}} \\
& = & M_\text{min}(1-r)^{+} \\
\Rightarrow d(r)& \geq & d_\text{max}(1-r)^{+} \eeqan \epf

For fully connected layered networks with $L < 4$, the min-cut is either at the
source side or at the sink side, and hence we have the following corollary:

\bcor \label{cor:optimal_layered} For a fully connected layered network with $L <
4$, the optimal DMT is achievable.\ecor

\bpf Consider a layered network with $L=1$, i.e., there is only
one layer. Let there be $n_1$ relay antennas in the relaying
layer. The DMT upper bound is $n_1(1-r)^{+}$ from the cut-set
bound, which is achieved.

Let $L=2$ and there be $n_1$ and $n_2$ relays in layers $1$ and $2$. Then the cutset
bound on DMT is $\min \{n_1,n_2 \} \ (1-r)^{+}$, which is achieved.

Let $L=3$ and there be $n_1,n_2,n_3$ relay antennas in the corresponding layer. It
can be seen that $d_{\text{max}} = \min \{n_1,n_2\}$ and that the DMT upper bound is
$\min \{n_1,n_2\} \ (1-r)^{+}$, which is indeed achieved. \epf

\section{Networks with Multiple Antenna Nodes\label{sec:half_duplex_multiple}}

In this section we consider families of single source single sink
networks with potentially all nodes having multiple antennas. We
consider KPP networks with interference and Layered networks under
both half duplex and full duplex constraint.

\subsection{Achievable DMT for Certain Networks with Multiple antenna nodes\label{sec:multiple_antenna}}

\subsubsection{Full Duplex Layered Networks\label{sec:layered_multiple}}

\begin{figure}[h!]
\centering
\includegraphics[height=65mm]{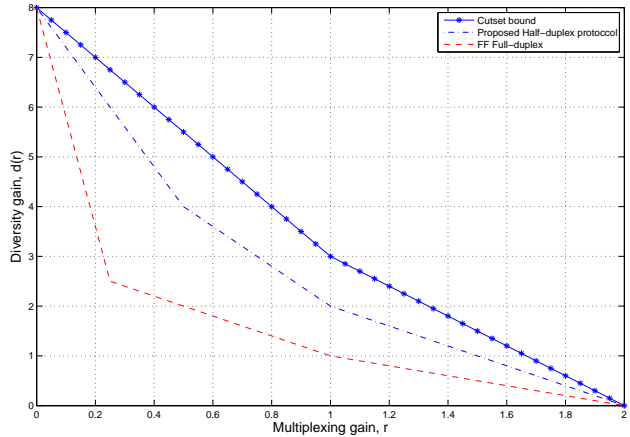}
\caption{Comparison of various protocols for $(2, 4, 2)$
network\label{fig:multiple_ant_dmt_curve}}
\end{figure}

We consider layered networks with multiple antennas at the source and the sink.
Multiple antennas at relays can be handled by replacing the relay with multiple
single-antenna relays in the same layer. We do not assume directed antennas and
consider undirected edges. However this creates a back-flow, which induces a lower
triangular matrix, that we handle using Theorem.~\ref{thm:main_theorem}.

\bdefn A single source single sink layered network with multiple antennas at the
source and the sink is referred to as an $(n_0, n_1, \ldots, n_L, n_{L+1})$ network
if the network has $L$ layers, with the source having $n_0$ antennas, the sink
having $n_{L+1}$ antennas, and the $i$-th layer of relays having $n_i$ nodes with
single antennas. \edefn

In \cite{YanBelNew}, parallel AF and flip-and-forward (FF)
protocols have been proposed for the $(n_0, n_1, \ldots, n_{L+1})$
network with full duplex operation and directed antennas, so that
back-flow is avoided. The parallel AF protocol aims to achieve the
full diversity for the network, whereas FF achieves the extreme
points of full multiplexing gain and the full diversity gain. In
\cite{YanBelNew},it has been proved that FF achieves a better DMT
than AF. However, the DMT curves of both these protocols lie far
away from the cut-set DMT bound. We propose a protocol with
achievable DMT better than the existing protocols for a $(n_0,
n_1, \ldots, n_{L+1})$ network under the full-duplex constraint.

In parallel AF and FF, the key idea is to partition the relay nodes in each layer
into subsets of nodes called super nodes. A sequence of consecutive super nodes from
source to sink form an AF path, and a set of AF paths is defined as a parallel
partition in \cite{YanBelNew}. An independent parallel partition is defined as a
parallel partition where any two different AF paths do not share common
edges\cite{YanBelNew}.

We propose a protocol which uses different partitioning depending
upon the multiplexing gain $r$ (we will refer to $r$ as the rate
by abuse of notation). \footnote{The idea of varying the protocol
parameters depending on $r$ was used in \cite{EliVinAnaKum} for
the NSDF protocol.} The basic intuition is that, at lower rates,
we can exploit the diversity of the network by creating more
parallel AF paths. At higher rates, super nodes are to be chosen
such that each AF path has enough degrees of freedom.

Let $P_i$ be the number of partitions in layer $i$. Let ${P}$
denote a particular partitioning which is specified by the vector
of $(P_0,P_1,P_2,...,P_{L+1})$ and let $\mathcal{P}$ denote all
possible partitionings.

Given that the layer $i$ has $P_i$ partitions, the number of
independent AF paths is \beqan N & = & \min_{\{i=0,1,2,...,L\}}
P_i P_{i+1} \eeqan

The protocol is as follows: Activate all the $N$ parallel paths
successively so that each path is activated for $T$ time instants.
During the activation of $i$th path, we will get a transfer matrix
that is block lower-triangular with $H_i$, the product matrix for
the $i$-th path on the diagonal. Since the matrix is lower
triangular, the DMT of this matrix is better than the DMT of
$H_i$. Let $d_i(r)$ be the DMT of this matrix, which can be
computed using the techniques for computing the DMT of product
Rayleigh matrices in \cite{YanBelNew}. Now the DMT of this induced
channel can be given using Theorem~\ref{thm:main_theorem} and the
parallel channel DMT in Lemma~\ref{lem:parallel_channel}:
\vspace{-0.15in} \beqan d_H(r) & \geq & \sup_{\{P \in
\mathcal{P}\}} \ \inf_{\{(r_1,r_2,\cdots,r_N): \ \sum_{i=1}^{N}
r_i = r\}} \ \sum_{i=1}^{N} {d_i(r_i)} \eeqan

The DMT of the protocol can be given as $d(r) = d_H(Nr)$.

Since the optimization is over the set of all possible partitions,
it might be difficult to compute the DMT in general. So we
consider a restricted case when the source and sink are
unpartitioned, and all the relay layers are partitioned into the
same size, $P$. Under this assumption, we have that $1 \leq P \leq
n_\text{min}$. Let $d_{(n_0,n_1,...,n_{L+1})}(r)$ denote the DMT
of a product channel $(n_0,n_1,...,n_{L+1})$, which we can compute
using the technique given in \cite{YanBelNew}. Let $n^P_i := \left
\lfloor \frac{n_i}{P} \right \rfloor, i=1,2,...,L$. When the relay
layer $i$ is partitioned into $P_i$ partitions, each partition
contains at-least $n^P_i$ relays. If it contains more, the
remaining relays are requested to be silent. This is done for
simplicity of computing the DMT.

The strategy of Theorem~\ref{thm:FD_No_Direct_Path} can be used to
obtain a DMT of $d_\text{max}(1-r)^{+}$ for a layered network (see
Corollory~\ref{cor:FD_KPP_Layered}). By combining this strategy
with the aforementioned strategy and chosing the one with the
better DMT based on $r$, we get a DMT of
\beqa d(r) & \geq & \max \{d_\text{max}(1-r)^{+}, \nonumber \\
&& \sup_{\{P \in [n_\text{min}]\}} \ \ P \
d_{(n_0,n^P_1,...,n^P_L,n_{L+1})}(r) , \}
\label{eq:DMT_MA_FD}\eeqa

The proposed protocol is essentially the same as \cite{YanBelNew} except for the
following differences: \bit \item We consider un-directed graph which gives rise to
back-flow. We are able to handle back-flow by using Theorem~\ref{thm:main_theorem}.
 \item We consider partitions of arbitrary size. Evaluating the DMT
with arbitrary sized partitions is made possible because of the parallel channel DMT
in Lemma~\ref{lem:parallel_channel}. \item The size of the partition is made
variable with respect to the rate. \footnote{However, the fact that FF protocol does
not depend on $r$ can make practical implementation simpler} \item We will show that
this result can be extended to half-duplex networks under the assumption that all
partitions are of equal size with $P_i>1$. \item It can be shown that the DMT of the
RHS in \eqref{eq:DMT_MA_FD} is strictly better than that of the FF protocol\eit

\emph{Example 1 :} Consider a $(2, 4, 2)$ layered network. The
achievable DMT curve using the FF protocol, the proposed protocol
and the cut-set bound are plotted in the
Figure~\ref{fig:multiple_ant_dmt_curve}.

\subsubsection{Half-Duplex Layered Networks}
We consider multi-antenna Layered networks with the additional
constraint of half-duplex relay nodes. We prove that the methods
provided above for full duplex networks can be generalized for the
half duplex network with bidirectional links.

Consider the partitioning method stated for full-duplex layered
networks, with $P_i = P, \forall i=1,2,...,L$, i.e., the relaying
layers are partitioned into equal number of partitions. Let the
source and sink be un-partitioned. When the relay layer $i$ is
partitioned into $P_i$ partitions, each partition contains
at-least $n^P_i := \left \lfloor {\frac{n_i}{P_i}} \right \rfloor$
relays. If it contains more, the remaining relays are requested to
be silent, as in the full duplex case.

The following observations are in place: Once we replace the nodes
corresponding to the same partition by a super-node, this virtual
network forms a regular network. This is because each relaying
layer has the same number of partitions and therefore the same
number of super-nodes. Therefore, this network can be treated as a
KPP networks with paths having equal lengths if $P > 1$. We use a
protocol with continuous activation on this regular network. Since
the paths are of equal length, the interference is causal making
the induced channel matrix lower triangular. This has better DMT
than the corresponding diagonal matrix by
Theorem~\ref{thm:main_theorem}. This yields the same lower bound
on DMT as in the full duplex case. Thus the DMT of the half duplex
network with the protocol is better than using the network with a
full duplex protocol and using the same partitioning. So we get:
\vspace{-0.1in}

\beqa d(r) & \geq & \max \{d_\text{max}(1-r)^{+}, \nonumber \\
&& \sup_{P \in \{2,3,..,n_\text{min} \}} \ \ P \
d_{(n_0,n^P_1,...,n^P_L,n_{L+1})}(r) , \}
\label{eq:DMT_MA_HD}\eeqa

\emph{Example 2:} For the case of $(2, 4, 2)$ network with half-duplex constraint,
the proposed protocol achieves the same DMT as the full duplex case of
$\emph{Example 1}$. However, the FF protocol used naively for a half-duplex system
will entail multiplexing gain loss by a factor of $\frac{1}{2}$.

\subsection{KPP(I) Networks}

Consider KPP(I) networks with multiple antennas at the source and
sink and potentially at all intermediate nodes.

\subsubsection{Full duplex KPP(I) Networks\label{sec:multiple_fullduplex}}

We consider full-duplex KPP(I) networks with multiple antenna
nodes. Given an underlying path $P_i$, we activate all edges in
the $P_i$ simultaneously. Let us call this process as activating
the path $P_i$ and the fading matrix thus obtained as $G_i$. So
$G_i = \Pi_{i=1}^{K} H_{ij}$. Let the DMT corresponding to this
product matrix be $d_i(r)$, which depends only on the number of
the antennas on the path $P_i$ and can be computed according to
formulae given in \cite{YanBelNew}.

Since activating different paths can potentially have different
DMTs, it is not optimal in general to use all paths equally.

When one is operating at a higher multiplexing gain, one might
want to use a path with higher multiplexing gain more frequently
in order to get greater average rate. While operating at a low
rate, all the paths must be used in order to get maximum
diversity. We consider a generic case where path $i$ is activated
for a fraction $f_i$ of the duration. These fractions can be
chosen depending on $r$ in order to maximize $d(r)$.

By so doing, we will get a parallel channel with repeated
coefficients. The DMT of such a channel was evaluated in
Lemma~\ref{lem:parallel_correlated_channel}. The conversion
however entails a loss factor, which is equal to the total number
of time instants for which the channel was used. After making this
rate correction, we get the following formula by modifying
equation \eqref{eq:parallel_repeated_coeffs}. So the achievable
DMT is given by, \vspace{-0.2in} \beqa d(r) & \geq &
\sup_{(f_1,f_2,\cdots,f_K)} \ \ \inf_{(r_1,r_2,\cdots,r_K): \
\sum_{i=1}^{K} \ f_i r_i = r }\ \sum_{i=1}^{K} {d_i(r_i)}
\nonumber \\ && \label{eq:multiple_antenna_KPP_FD} \eeqa
\vspace{-0.2in}

\subsubsection{Half Duplex KPP(I) Networks\label{sec:multiple_halfduplex}}

From Section~\ref{sec:half_duplex}, we know that under the half
duplex constraint, there exists a protocol activating the $K$
paths equally for KPP(I) networks with $K \geq 3$ causing only
causal interference. We can use the same protocol notwithstanding
the fact that the relays contain multiple antennas. By doing so,
we will get a transfer matrix which will be lower triangular.
Also, the diagonal entries of this channel matrix would remain the
same as though the relay nodes operate under full-duplex mode. By
Theorem~\ref{thm:main_theorem}, this gives a lower bound on the
DMT, and it is equal to DMT lower bound of the full duplex network
in \eqref{eq:multiple_antenna_KPP_FD}. Therefore even when there
is half duplex constraint, we can achieve the same DMT given by
the \eqref{eq:multiple_antenna_KPP_FD} with $f_i = \frac{1}{K}$
instead of the supremum.

If we want to achieve different fractions of activation for
different parallel paths, then we can follow a different trick for
$K \geq 4$. In this case, we can use the $^KC_3$ 3-parallel path
networks, but activate each $3$-parallel-path network for a
different fraction of time. Using this strategy, we can show that,
for $K \geq 4$, all time fractions $f_i$ for the parallel path
$P_i$ can be obtained as long as $(f_1,f_2,...,f_K) \in
\mathcal{F}$ where \beqan \mathcal{F} := \{(f_1,f_2,...,f_K):
\sum_{i=1}^{K} f_i = 1, \ \ 0 \leq f_i \leq \frac{1}{3} \} \eeqan

For $K \geq 4$, this yields a DMT of \vspace{-0.2in} \beqa d(r) &
\geq & \sup_{(f_1,f_2,\cdots,f_K)\in \mathcal{F}} \ \
\inf_{(r_1,r_2,\cdots,r_K): \ \sum_{i=1}^{K} \ f_i r_i = r }\
\sum_{i=1}^{K} {d_i(r_i)} \nonumber \\ &&
\label{eq:multiple_antenna_KPP_HD} \eeqa \vspace{-0.2in}

This is the same as the lower bound on the DMT for the full duplex
case, except that we are constrained to have all activation
fractions $f_i$ to be lesser than one-third.

\section{Code Design}
\subsection{Design of DMT achieving codes \label{sec:code_design}}

Consider any network and protocol described above, and let us say
the network is operated for $M$ slots. Let $L$ be the period of
the protocol and let us assume $M = mL + D$ for simplicity. We
will assume that after $D$ time instants the KPP network comes to
steady state, and we will neglect the first $D$ time instants.
Even though there is a rate loss of $\frac{M}{M-D}$ associated
with that, we can make this loss arbitrarily small by making $M$
large enough.

The induced channel is given by $Y=HX+W$ where $X,Y,W$ is a $M
\times 1$ vector and $H$ is a $M \times M$ matrix. However, to
design an optimal code for this channel, we need to use a space
time code matrix $X$. In order to obtain an induced channel with
$X$ being a $M \times T$ matrix, we do the following. Instead of
transmitting a single symbol, each node transmits a row vector
comprising of $T$ symbols during each activation. Then the induced
channel matrix takes the form: $Y=HX+W$, with $X,Y,W$ being $M
\times T$ matrices and $H$ the same $M \times M$ matrix as
earlier.

So there are totally $MT$ symbols transmitted. In the matrix $X$,
let us call the row vector of $T$ symbols in slot $i$ as $x_i$. To
address a specific symbol: the $j$-th symbol in slot $i$, we use
the notation $x_{ij}$. Let us use similar notation for the output:
$y_{ij}$ denotes the $j$-th symbol received in the $i$-th time
slot, and $y_i$ denotes the row vector of $T$ symbols received in
the $i$-th time slot.

Now from \cite{TavVis}, we know that if we use an approximately
universal code for $X$, then it will achieve the optimal DMT of
the channel matrix $H$ irrespective of the statistics of the
channel. Explicit minimal delay approximately universal codes for
the case when $T=M$ are given in \cite{EliRajPawKumLu},
constructed based on appropriate cyclic division algebras
\cite{SetRajSas}. These codes can be used here to achieve the
optimal DMT of the induced channel matrix.

\subsubsection{Short DMT Optimal Code Design}
The code construction provided above affords a code length of $TM
= M^2$. Also we need $M$ very large for the initial delay overhead
to be minimal. This entails a very large block length, and indeed
very high decoding complexity. Now a natural question is whether
optimal DMT performance can be achieved with shorter block
lengths. We answer this question for KPP networks by constructing
DMT optimal codes that have $T=L$ and a block length of $L^2$,
where $L$ is the period of the protocol used. We also provide a
DMT optimal decoding strategy that also requires only decoding a
$L \times L$ matrix at a time. This is a constant which does not
depend on $M$ and therefore, even if we make $M$ large, the delay
and decoding complexity are unaffected. This code construction can
be easily extended to other networks considered in this paper as
well.

After $D$ time instants, the KPP network attains steady state.
Consider the first $L$ inputs after attaining steady state
$x_{D+1},x_{D+2},...,x_{D+L}$. If the channel matrix is restricted
to these $L$ time slots alone, then channel matrix would be a
lower triangular matrix with the $L$ independent coefficients
$g_i$, $i=1,2,..,K$ repeated periodically. The DMT of this matrix,
after adjusting for rate, is $d_K(r) = K(1-r)^{+}$. So if we use a
$L \times L$ DMT optimal matrix as the input (this can be done by
setting $T=L$ and using a $L \times L$ approximately universal CDA
based code for the input), we will be able to obtain a DMT of
$d_K(r)$ for this subset of the data. This means that the
probability of error for this vector comprising of $T$ input
symbols will be of exponential order $P_e \doteq \rho^{-d_K(r)}$
if an ML decoder is used to decode the $L \times L$ matrix.

Let us assume that the first $L$ symbols has been decoded
independently. Let us now focus on the next $L$ received symbols
$y_{D+L+1},y_{D+L+2},...,y_{D+L+L}$. These symbols potentially
depend on the previous block of $L$ symbols and it is optimal to
decode all of these together. However we show that a Successive
Interference Cancellation (SIC) based method is DMT optimal as
well. After the first block of $L$ symbols are decoded, its effect
will be subtracted out from the remaining symbols, and then the
next block of $L$ symbols decoded independently. For the third
block, the effect of the first two blocks each of length $L$ will
be subtracted out and the third block decoded independently and so
on.

Let us evaluate the probability of error when this SIC based
method is used. Let us find the probability of error for $B$
blocks after the initial $D$ instants of silence. Let $E_i$ denote
the event that there is an error in any of the first $i$ blocks,
$F_i$ denote the event that there is an error in decoding the
$i$-th block. Proceeding by induction on the $i$-th statement
$P(E_i) =\rho^{-d_K(r)}$, we get
\beqan P (F_i) & = & P(F_i / E_{i-1}) P(E_{i-1}) + P(F_i / \overline{E_{i-1}}) P( \overline{E_{i-1}}) \\
& \leq &  P(E_{i-1}) + P(F_i / \overline{E_{i-1}}) \\
& \doteq & \rho^{-d_K(r)} + \rho^{-d_K(r)} \\
& \doteq & \rho^{-d_K(r)} \eeqan
\beqan \Rightarrow P (E_i) & = & P (\bigcup_{j=1}^{i} F_j)  \ \leq \  \sum_{j=1}^i P(F_j)\\
& \doteq & \sum_{j=1}^i \rho^{-d_K(r)} \doteq  \rho^{-d_K(r)}
\eeqan

Therefore, we have that the entire probability of error is of the
exponential order of $\rho^{-d_K(r)}$ and the scheme achieves the
optimal DMT of the $H$ matrix.

\subsection{Universal Full-Diversity Codes} \label{sec:univ_full_div}

Consider a input output equation of the form ${Y} ={HX} + {W}$
where $X,Y,H,W$ are $M \times M$ matrices.

Usually the code design criterion given for a input matrix to have
full diversity for rayleigh fading is that the difference of any
two possible input matrices be full rank. In this section we show
that such a criterion is sufficient to get full diversity on
\emph{any} channel matrix distribution. By full diversity here, we
mean that the code will attain a diversity equal to $d(0)$ for the
channel.

%Let us assume that the code matrix $X_n$ ($n$ for normalized) be
%such that each entry has unit variance. So at SNR $\rho$, we have
%$X = \sqrt{\rho} X_n$.
We quote the following theorem from the theory of approximately
universal codes (Theorem 3.1 in \cite{TavVis} ):

\begin{thm} \cite{TavVis} \label{thm:au_mimo}
A sequence of codes of rate $R(\rho):= r \log  \rho $ bits/symbol
is approximately universal over the MIMO channel if and only if,
for every pair of codewords, \beq\label{eq:mimo_univ_crit}
\lambda_1^2 \lambda_2^2\cdots \lambda_{n_{\text{min}}}^2  \geq
\frac{1}{2^{R(\rho)+o(\log \rho )}} = \frac{1}{{\rho^r} \
2^{o(\log \rho)}}, \eeq where $\lambda_1,\ldots
,\lambda_{n_{\text{min}}}$ are the smallest $n_{\text{min}}$
singular values of the normalized (by $\frac{1}{\sqrt{\rho}}$)
codeword difference matrix. A sequence of codes achieves the DMT
of any channel matrix if and only if it is approximately
universal. \end{thm}

Substituting $r=0$ corresponding to a multiplexing gain of $0$ in
Theorem~\ref{thm:au_mimo}, we get that the criterion is \beq
\lambda_1^2 \lambda_2^2\cdots \lambda_{n_{\text{min}}}^2  \geq
\frac{1}{ 2^{o(\log \rho)}}, \eeq

In particular, if a code satisfies, for all pairs of codewords,
the difference determinant is non-zero, i.e., \beq \lambda_1^2
\lambda_2^2\cdots \lambda_{n_{\text{min}}}^2 \geq L > 0, \eeq then
the code is approximately universal for a rate of $r=0$, and
therefore achieves, the $d(0)$ of any given channel matrix.

This criterion is the same as the criterion for full diversity on
a rayleigh channel. This means that all codes with full diversity
designed for the rayleigh fading MIMO channel are indeed full
diversity for a MIMO channel with any fading distribution.
Therefore we can use a full-diversity code designed for a rayleigh
fading MIMO channel to get full-diversity for any KPP or Layered
network, when used along with the corresponding protocol for these
networks.

\section*{Acknowledgment}
% optional entry into table of contents (if used)
%\addcontentsline{toc}{section}{Acknowledgment}
Thanks are due to K.~Vinodh and M.~Anand for useful discussions.

\end{document}